\documentclass[prd,aps,letterpaper,twocolumn,preprintnumbers,superscriptaddress,nofootinbib,floatfix]{revtex4}

\usepackage{amsmath}
\DeclareMathOperator\arccosh{arccosh}
\usepackage{amssymb}
\usepackage{color}
\usepackage{comment}
\usepackage{graphicx}
\usepackage{dcolumn}
\usepackage{bm}
\usepackage{hyperref}

\usepackage[normalem]{ulem}
\usepackage[utf8]{inputenc}
\usepackage{subfigure}
\usepackage{capt-of}

\newcommand{\nn}{\nonumber}

\newcommand{\df}{\mathrm{d}}
\newcommand{\img}{\mathrm{i}}

\newcommand{\cA}{\mathcal{A}}
\newcommand{\cO}{\mathcal{O}}

\newcommand{\bt}{\vec{b}_T}

\newcommand{\MS}{\overline{\mathrm{MS}}}
\newcommand{\RI}{\mathrm{RI}^\prime\mathrm{/MOM}}
\newcommand{\ns}{{\text{ns}}}
\newcommand\TMD{\mathrm{TMD}}

\newcommand{\mbraket}[3]{\left< #1 \vphantom{#2#3} \right|
 #2 \left| #3 \vphantom{#1#2} \right>} 

\DeclareRobustCommand{\eq}[1]{Eq.~(\ref{eq:#1})}
\DeclareRobustCommand{\eqs}[2]{Eqs.~(\ref{eq:#1}) and (\ref{eq:#2})}

\DeclareRobustCommand{\fig}[1]{Fig.~\ref{fig:#1}}

\begin{document}

\preprint{FERMILAB-PUB-20-102-T}
\preprint{MIT/CTP-5184}

\title{Collins-Soper Kernel for TMD Evolution from Lattice QCD}%

\author{Phiala Shanahan}
 \email{phiala@mit.edu}
 \affiliation{Center for Theoretical Physics, Massachusetts Institute of Technology, Cambridge, MA, USA 02139}%
\author{Michael Wagman}%
 \email{mwagman@fnal.gov}
 \affiliation{Center for Theoretical Physics, Massachusetts Institute of Technology, Cambridge, MA, USA 02139}%
  \affiliation{Fermi National Accelerator Laboratory, Batavia, IL 60510, USA}

\author{Yong Zhao}%
 \email{yzhao@bnl.gov}
 \affiliation{Center for Theoretical Physics, Massachusetts Institute of Technology, Cambridge, MA, USA 02139}%
 \affiliation{Physics Department, Brookhaven National Laboratory, Bldg. 510A, Upton, NY 11973, USA}

\begin{abstract}
   The Collins-Soper kernel relates transverse momentum-dependent parton distribution functions (TMDPDFs) at different energy scales. For small parton transverse momentum $q_T\sim \Lambda_\text{QCD}$, this kernel is non-perturbative and can only be determined with controlled uncertainties through experiment or first-principles calculations. This work presents the first exploratory determination of the Collins-Soper kernel using the lattice formulation of Quantum Chromodynamics. In a quenched calculation, the $N_f=0$ kernel is determined at scales in the range 250~MeV $< q_T < 2$~GeV, and an analysis of the remaining systematic uncertainties is undertaken. 
\end{abstract}

\maketitle

\section{Introduction}

Understanding the structure of matter has been a defining goal of physics for centuries. In the modern context, a primary objective is imaging the three-dimensional spatial and momentum structure of the proton, and of other hadrons. Some important aspects of this structure related to the transverse momentum of quarks and gluons in a hadron state are encoded in transverse-momentum-dependent parton distribution functions (TMDPDFs)~\cite{Collins:1981uk,Collins:1981va,Collins:1984kg}. Experimentally, these quantities can be constrained for the proton by Drell-Yan production and semi-inclusive deep inelastic scattering (SIDIS) of electrons off protons; the best current constraints are achieved via global fits to experimental data~\cite{Landry:1999an,Landry:2002ix,Konychev:2005iy,Su:2014wpa,DAlesio:2014mrz,Echevarria:2014xaa,Kang:2015msa,Bacchetta:2017gcc,Scimemi:2017etj,Bertone:2019nxa,Scimemi:2019cmh,Bacchetta:2019sam}, with improvements expected in the coming years from measurements at COMPASS~\cite{Gautheron:2010wva}, the Thomas Jefferson National Accelerator Facility~\cite{Dudek:2012vr}, RHIC~\cite{Aschenauer:2015eha}, and an Electron-Ion Collider~\cite{Accardi:2012qut}. Additional experimental information from dihadron production in $e^+ e^-$ collisions at Belle, and new ways of looking at hadrons inside jets~\cite{Buffing:2018ggv,Gutierrez-Reyes:2019vbx}, may also help constrain these fits.

Key to global fits of TMDPDFs is the ability to relate these distributions determined in different processes, including those at different scales. That is, for a TMDPDF $f_{i}^{\mathrm{TMD}}\big(x, \vec{b}_{T}, \mu_0, \zeta_0\big)$, defined for a parton of flavor $i$ with longitudinal momentum fraction $x$, transverse displacement $\vec{b}_T$ (the Fourier conjugate of the transverse momentum $\vec{q}_T$), virtuality scale $\mu_0$, and hadron momentum scale $\zeta_0$ which is related to the hard scale of the scattering process, it is critical to understand its evolution to scales $(\mu,\zeta)$:
\begin{align} \nonumber
    &f_{i}^{\mathrm{TMD}}\left(x, \vec{b}_{T}, \mu, \zeta\right)=f_{i}^{\mathrm{TMD}}\left(x, \vec{b}_{T}, \mu_{0},\zeta_{0}\right) \\ 
    &\ \times \exp \left[\int_{\mu_{0}}^{\mu} \frac{\mathrm{d} \mu^{\prime}}{\mu^{\prime}} \gamma_{\mu}^{i}\left(\mu^{\prime}, \zeta_{0}\right)\right] \exp \left[\frac{1}{2} \gamma_{\zeta}^{i}\left(\mu, b_{T}\right) \ln \frac{\zeta}{\zeta_{0}}\right],
\end{align}
where $b_T = |\vec{b}_T|$. The first exponential in this equation governs the $\mu$-evolution of the TMDPDF, which is perturbative for scales $\{\mu_0,\mu\}\gg \Lambda_\text{QCD}$. The evolution in $\zeta$ governed by the second exponential, however, is encoded in the Collins-Soper kernel\footnote{The Collins-Soper kernel is also often denoted by $K(b_T,\mu)$~\cite{Collins:1984kg}, and it is defined as $-{\cal D}(b_T,\mu)$ in Ref.~\cite{Scimemi:2017etj}.} $\gamma_\zeta^i(\mu,b_T)$, which is inherently non-perturbative for $q_T\sim b_T^{-1}\sim \Lambda_\text{QCD}$, even for $\mu \gg \Lambda_\text{QCD}$. Experimentally, the Collins-Soper kernel can be extracted by simultaneous global fits with the TMDPDF, and recent global analyses show some discrepancy in determinations of the kernel in the region $q_T\le 500$ MeV~\cite{Vladimirov:2020umg}. It would greatly improve systematic control if the Collins-Soper kernel could be independently determined from first-principles QCD calculations, and taken as input for global fits of experimental data.

Since TMDPDFs are defined in terms of light-cone correlation functions, they are challenging to calculate directly in the lattice formulation of QCD on a discrete Euclidean spacetime, which is the only known systematically improvable first-principles approach to nonperturbative QCD.  Nevertheless, efforts to calculate aspects of TMD physics from equal-time correlation functions in boosted hadron states have been made in Refs.~\cite{Musch:2010ka,Musch:2011er,Engelhardt:2015xja,Yoon:2016dyh,Yoon:2017qzo}, and the large-momentum effective theory (LaMET) framework~\cite{Ji:2013dva,Ji:2014gla} provides a promising pathway towards the determination of TMDPDFs by matching these matrix elements to the desired light-cone correlation functions at large hadron momentum~\cite{Ji:2014hxa,Ji:2018hvs,Ebert:2018gzl,Ebert:2019okf,Ebert:2019tvc,Ji:2019sxk,Ji:2019ewn,Vladimirov:2020ofp}.  In particular, it was recently shown in Refs.~\cite{Ebert:2018gzl,Ebert:2019okf} how this approach may be used to extract the Collins-Soper kernel nonperturbatively from computations of matrix elements of nonlocal quark bilinear operators with staple-shaped Wilson lines. 
Here, this approach is implemented numerically for the first time, in a proof-of-principle calculation in quenched QCD. The Collins-Soper kernel is extracted at a range of $q_T$ scales, including in the non-perturbative region.

Section~\ref{sec:th} outlines the procedure, developed in Refs.~\cite{Ebert:2018gzl,Ebert:2019okf}, for constraining the Collins-Soper kernel using lattice QCD and LaMET. Section~\ref{sec:LQCD} details the quenched lattice QCD calculation undertaken here, including discussion of the systematic uncertainties in the calculation, while Sec.~\ref{sec:conc} outlines the requirements for a fully-controlled calculation of the Collins-Soper kernel to be achieved by this method.

\section{Collins-Soper kernel from lattice QCD}
\label{sec:th}

In Refs.~\cite{Ebert:2018gzl,Ebert:2019okf} a method was proposed to determine the quark Collins-Soper kernel using lattice QCD and LaMET. Precisely, it was shown that $\gamma^q_\zeta(\mu, b_T)$ can be extracted from a ratio of nonsinglet quasi TMDPDFs $\tilde f_{\ns}^\TMD$ at different momenta, which are defined using equal-time correlation functions within hadron states at large momentum in the $z$-direction:
\begin{align}\label{eq:gamma_zeta1}
	\gamma^q_\zeta(\mu, b_T) & 
= \frac{1}{\ln(P^z_1/P^z_2)}\nn\\
&\times \ln \frac{C^\TMD_\ns (\mu,x P_2^z)\, \tilde f_{\ns}^\TMD(x, \bt, \mu, P_1^z)}	{C^\TMD_\ns (\mu,x P_1^z)\, \tilde f_{\ns}^\TMD(x, \bt, \mu, P_2^z)}\,,
\end{align}
up to power corrections which are discussed further below. In this expression, $P^z_i \gg \Lambda_\text{QCD}$ are the $z$-component of the hadron momenta and $C^\TMD_\ns$ is a perturbative matching coefficient that has been obtained at one-loop order~\cite{Ebert:2018gzl,Ebert:2019okf}. The quasi TMDPDF $\tilde f_{\ns}^\TMD$, defined below, approximates the physical TMDPDF involving light-like paths, as detailed in Ref.~\cite{Ebert:2019okf}, and complications involving matching in the soft sector~\cite{Ji:2018hvs,Ebert:2019okf,Ji:2019sxk,Ji:2019ewn} are eliminated in the ratio that gives the Collins-Soper kernel. Similar constructions have been used in calculations of ratios of $x$-moments of TMDPDFs from lattice QCD~\cite{Musch:2010ka,Musch:2011er,Engelhardt:2015xja,Yoon:2016dyh,Yoon:2017qzo}.

The unpolarized quasi TMDPDF is defined in terms of a quasi beam function $\tilde{B}^\Gamma_i$ and a quasi soft factor $\tilde{\Delta}_S$~\cite{Ji:2014hxa,Ji:2018hvs,Ebert:2018gzl,Ebert:2019okf}, both of which are calculable in lattice QCD:
\begin{align}\label{eq:quasiTMD}
&\tilde{f}_{i}^{\mathrm{TMD}}\big(x, \vec{b}_{T}, \mu, P^{z}\big)\equiv  \lim_{\substack{a\to 0 \\ \eta\to \infty}}
\int \frac{\mathrm{d} b^{z}}{2 \pi} e^{-\mathrm{i}b^{z}\left(x P^{z}\right)} \mathcal{Z}^{\MS}_{\gamma^4\Gamma}(\mu,b^z\!,a)\nonumber\\
&\qquad\qquad\times{P^z\over E_{\vec{P}}}\tilde{B}^{\Gamma}_{i}\big(b^{z},\vec{b}_{T},a,\eta,P^{z}\big) \tilde{\Delta}_{S}\left(b_{T},a,\eta\right)\,,
\end{align}
where $a$ denotes the lattice spacing, the subscript $i$ is the flavor index, and summation over Dirac structures is implied. This summation accounts for the operator mixings among different Dirac structures in lattice QCD calculations defined on a hypercubic space-time lattice~\cite{Constantinou:2019vyb,Shanahan:2019zcq,Green:2020xco}. Additional mixing with gluon operators, not shown in Eq.~\eqref{eq:quasiTMD}, cancels in the flavor nonsinglet combination used in Eq.~\eqref{eq:gamma_zeta1}, which is defined as $\tilde f^{\mathrm{TMD}}_\ns=\tilde f^{\mathrm{TMD}}_u-\tilde f^{\mathrm{TMD}}_d$. 
Both $\tilde{B}^\Gamma_i$ and $\tilde{\Delta}_S$ include logarithmic ($\sim \ln a$) and linear ($\sim 1/a$) ultraviolet divergences, with the latter proportional to the total lengths of the Wilson lines. Both functions also include contributions diverging linearly as $\sim \eta/b_T$ in the limit $\eta\to\infty$~\cite{Ebert:2019okf}. The $\eta/a$ and $b_T/a$ divergences, as well as $\eta/b_T$-dependence, cancel between $\tilde{B}^\Gamma_i$ and $\tilde{\Delta}_S$ in Eq.~\eqref{eq:quasiTMD}.
The factor $\mathcal{Z}^{\MS}_{\gamma^4\Gamma}(\mu,b^z,a)$ renormalizes the remaining linear ($\sim b^z/a$) and logarithmic divergences in the quasi TMDPDF and matches it to the quasi TMDPDF with Dirac structure $\gamma^4$ (where `4' indexes the temporal direction) in the $\MS$ scheme at scale $\mu$~\cite{Constantinou:2019vyb,Ebert:2019tvc,Shanahan:2019zcq}. An alternate choice is to consider the quasi TMDPDF with Dirac structure $\gamma^3$; both $\gamma^4$ and $\gamma^3$ can be boosted onto $\gamma^+$ and thus define quasi TMDPDFs which can be matched to the spin-independent TMDPDF in the infinite-momentum limit.

\begin{figure}[!t]
    \centering
    \includegraphics[width=0.7\columnwidth]{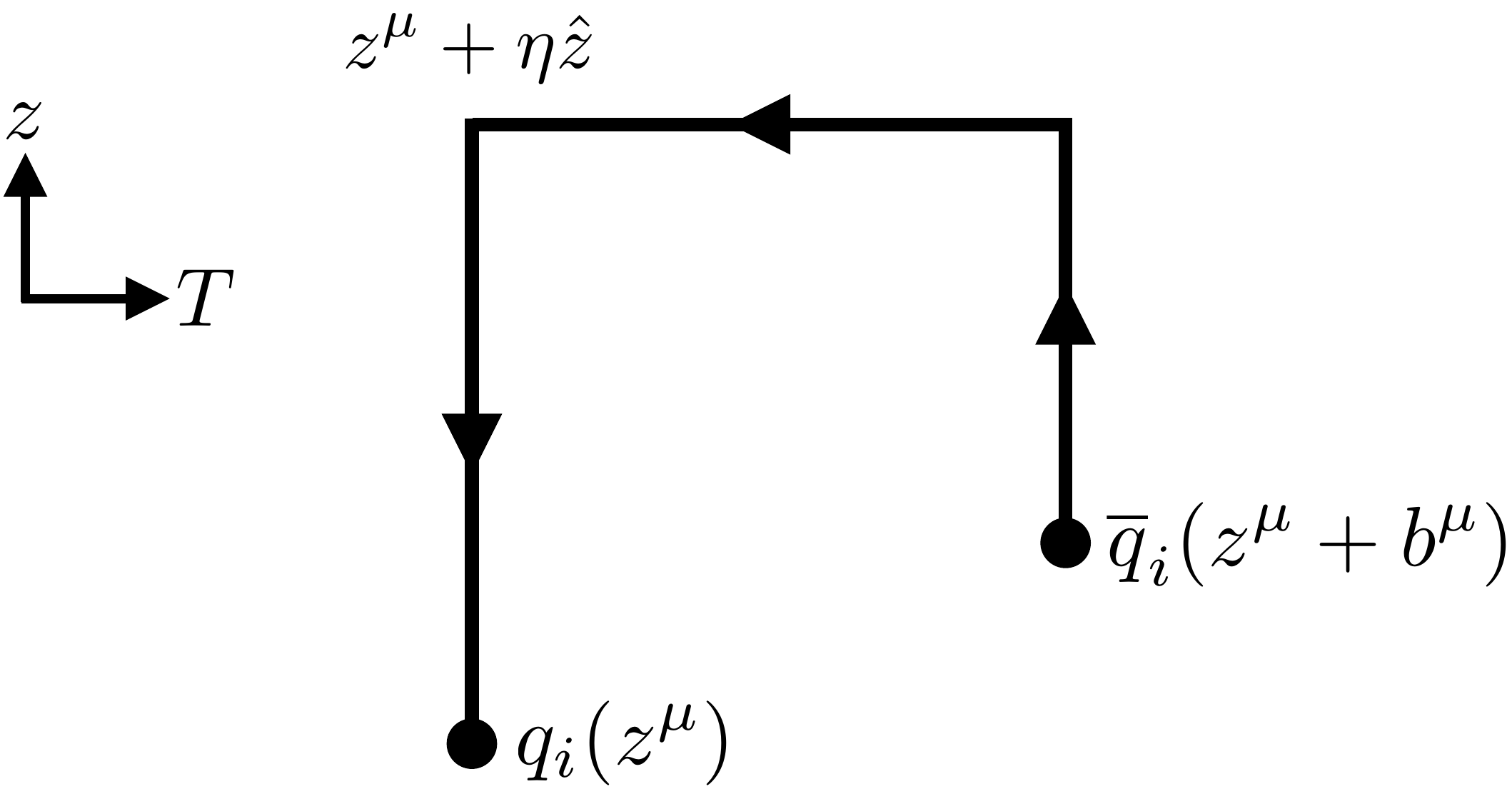}
    \caption{Illustration of the staple-shaped Wilson line structure of the non-local quark bilinear operators $\mathcal{O}^i_\Gamma(b^\mu,z^\mu,\eta)$ defining quasi beam functions, see Eq.~\eqref{eq:op}.}
    \label{fig:staple}
\end{figure}

{\it Quasi beam function:}
The quasi beam functions in \eq{quasiTMD} are defined as matrix elements of quark bilinear operators with staple-shaped Wilson lines: 
\begin{align} \label{eq:qbeam}
\tilde B^\Gamma_{i}(b^z, \bt,a,\eta,P^z)
=&
\Bigl\langle h(P^z) \big| \mathcal{O}_\Gamma^{i}(b^\mu,0,\eta) \big| h(P^z) \Bigr\rangle\,.
\end{align}
Here $h(P^z)$ denotes a boosted hadron state with four-momentum $P^{\mu}=(0,0,P^z,E^{(h)}_{\vec{P}})$, with $E^{(h)}_{\vec{P}}=\sqrt{\vec{P}^2+m_h^2}$ and where $m_h$ is the mass of the hadron $h$. States are normalized as  $\langle h(\vec{P}')|h(\vec{P})\rangle = 2E^{(h)}_{\vec{P}}(2\pi)^3\delta^{(3)}(\vec{P}-\vec{P}')$.
It is convenient to define a dimensionless `bare' nonsinglet beam function:
\begin{align}
    \nonumber
B^\text{bare}_\Gamma(b^z, \bt,a,\eta,P^z)  = & \frac{1}{2 E_{\vec{P}}}\left(\tilde B^\Gamma_{u}(b^z, \bt,a,\eta,P^z)\right. \\
\label{eq:qbeambare}& \left. \hspace{6mm}- \tilde B^\Gamma_{d}(b^z, \bt,a,\eta,P^z)\right).
\end{align}
The operator $\mathcal{O}_\Gamma^{i}(b^\mu,0,\eta)$ in Eq.~\eqref{eq:qbeam} is defined as a quark bilinear with a staple-shaped Wilson line, depicted in \fig{staple}: 
\begin{align}\nonumber
\mathcal{O}^{i}_\Gamma(b^\mu,z^\mu,\eta) &=\bar q_i(z^\mu + b^\mu) \frac{\Gamma}{2} W_{\hat z}(z^\mu + b^\mu ;\eta-b^z) \\\nonumber
&\times 
W^\dagger_{T}(z^\mu + \eta \hat{z}; b_T) W^\dagger_{\hat z}(z^\mu;\eta) q_i(z^\mu)\\\label{eq:op}
&\equiv \bar q_i(z^\mu + b^\mu) \frac{\Gamma}{2}\widetilde{W}(\eta;b^\mu;z^\mu)q_i(z^\mu)\,,
\end{align}
where $\widetilde{W}(\eta;b^\mu;z^\mu)$ is a spatial Wilson line of staple length $\eta$ in the $\vec{e}_{z}$ direction connecting endpoints separated by $b^\mu = (\bt,b^z,0)$. Here $T$ denotes a direction transverse to $\vec{e}_{z}$, and all spatial Wilson lines are defined as 
\begin{align} \label{eq:coll_Wilson_L}
W_{\hat \alpha}(x^\mu;\eta) &= P \exp\left[ \img g \int_0^{\eta} \df s \, \cA^\alpha(x^\mu + s \hat{\alpha}) \right]\,.
\end{align}

{\it Quasi soft factor:}
The quasi soft factor $\tilde{\Delta}_{S}(b_T,a,\eta)$ can be computed as the vacuum matrix element of a closed spatial Wilson loop, whose definition and properties are detailed in Refs.~\cite{Ji:2014hxa,Ji:2018hvs,Ebert:2018gzl,Ebert:2019okf}. This factor cancels in the ratios of quasi TMDPDFs which define the Collins-Soper evolution kernel by Eq.~\eqref{eq:gamma_zeta1}, and will thus not be discussed further here.

{\it Renormalization factor:}
The renormalization factor $\mathcal{Z}_{\gamma^4\Gamma}^{\MS}$ can be separated into two parts which renormalize the quasi beam function and soft factor respectively, denoted by $Z_{\cO_{\gamma^4\Gamma}}^{\MS}$ and $Z_{S}^{\MS}$:
\begin{align}\label{eq:separatedZ}
\mathcal{Z}_{\gamma^4\Gamma}^{\MS}(\mu,b^z\!,a) = Z_{\cO_{\gamma^4\Gamma}}^{\MS}(\mu,b^z\!,b_T,a,\eta) Z_{S}^{\MS}(\mu,b_T,a,\eta).
\end{align}
Both $Z_{\cO_{\gamma^4\Gamma}}^{\MS}$ and $Z_{S}^{\MS}$ include linear power divergences proportional to $\eta/a$ and $b_T/a$ that cancel between the two terms, such that the complete renormalization factor $\mathcal{Z}_{\gamma^4\Gamma}^{\MS}$ is independent of $\eta$ and $b_T$.
$Z_{\cO_{\gamma^4\Gamma}}^{\MS}$ can be computed nonperturbatively using the regularization independent momentum subtraction ($\RI$) scheme, with a perturbative matching to the $\MS$ scheme via a multiplicative factor $\mathcal{R}^{\MS}_{\cO_{\gamma^4\Gamma}}$ as described in Refs.~\cite{Constantinou:2019vyb,Ebert:2019tvc}. In this approach, $Z_{\cO_{\gamma^4\Gamma}}^{\MS}$ can be expressed as 
\begin{align}\nonumber
    Z_{\cO_{\gamma^4\Gamma}}^{\MS}(\mu,b^z\!,b_T,a,\eta) = & \mathcal{R}^{\MS}_{\cO_{\gamma^4\Gamma}}(\mu,p_R,b^z,\vec{b}_T,\eta)\\
    &\times Z^{\RI}_{\cO_{\gamma^4\Gamma}}(p_R,b^z\!,\vec{b}_T,a,\eta),
\end{align}
where $Z^{\RI}_{\cO_{\gamma^4\Gamma}}$ is the $\RI$ renormalization factor and $p_R$ denotes the matching scale introduced in the $\RI$ scheme.
At all orders in perturbation theory, the scheme conversion factor $\mathcal{R}^{\MS}_{\cO_{\gamma^4\Gamma}}$ cancels the dependence of $Z^{\RI}_{\cO_{\gamma^4\Gamma}}$ on $p_R$ and on the direction of $\vec{b}_T$ (up to discretization artefacts).

The authors have previously calculated $Z^{\RI}_{\cO_{\gamma^4\Gamma}}$, and thereby
$Z_{\cO_{\gamma^4\Gamma}}^{\MS}$, by this approach in a quenched lattice QCD study~\cite{Shanahan:2019zcq}; those results are used for the numerical study in this work. The renormalization factor $Z_{S}^{\MS}$ does not need to be evaluated for a computation of the Collins-Soper kernel, as detailed in the following subsection.

{\it Collins-Soper kernel:} In the ratio of quasi TMDPDFs which gives the Collins-Soper kernel in Eq.~\eqref{eq:gamma_zeta1}, $\tilde{\Delta}_{S}$ and its renormalization factor $Z_{S}^{\MS}$, which do not depend on $b^z$, cancel between the numerator and denominator. As a result, $\gamma^q_\zeta(\mu, b_T)$ can be expressed in terms of the quasi beam function and its renormalization only, at the cost of introducing power-law divergences in $\eta$ and $b_T$ separately in the numerator and denominator (divergences which were canceled by the quasi soft factor and its renormalization in the original expression for the kernel). Moreover, to ensure that the renormalization and matching between $\RI$ and $\MS$ is performed in the perturbative region, the scale $b_T$ must be taken to be much smaller than $\Lambda_{\rm QCD}^{-1}$, a condition which does not permit an extraction of the Collins-Soper kernel at $b_T$ values in the nonperturbative region. 
A perturbative renormalization matching scale $b_T=b_T^R\ll \Lambda_{\rm QCD}^{-1}$ in \eq{separatedZ} can, however, be defined by exploiting the $b_T$-independence of $\mathcal{Z}_{\gamma^4\Gamma}^{\MS}(\mu,b^z,a)$, as described in Ref.~\cite{Ebert:2019tvc}. 
In this approach, for the choice $\Gamma=\gamma^4$ in Eq.~\eqref{eq:quasiTMD}, the Collins-Soper kernel can be expressed as
\begin{align}\label{eq:finalCSexpression}
&\gamma^q_\zeta(\mu, b_T)  
= \frac{1}{\ln(P^z_1/P^z_2)}\ln\Biggr[ \frac{C^\TMD_\ns (\mu,x P_2^z)}{C^\TMD_\ns (\mu,x P_1^z)}\nn\\
&\! \times\! \frac{\int\! \df b^z e^{-ib^z\! xP_1^z}  P_1^z \lim_{\substack{a\to 0 \\ \eta\to \infty}}B^{\MS}_{\gamma^4}(\mu,b^z, \bt, a, \eta, P_1^z)}
    {\int\! \df b^z e^{-ib^z\! xP_2^z}\!  P_2^z \lim_{\substack{a\to 0 \\ \eta\to \infty}}B^{\MS}_{\gamma^4}(\mu,b^z, \bt, a, \eta,  P_2^z)}\Biggr]\,,
\end{align}
where a modified $\MS$-renormalized quasi beam function $B^{\MS}_{\Gamma}$ has been defined as
\begin{align}\nonumber
   B^{\MS}_{\gamma^4}(\mu,b^z, \bt, a, \eta, P^z)& =  Z_{\cO_{\gamma^4\Gamma}}^{\MS}(\mu,b^z, b_T^R,a,\eta) \\\label{eq:BMSbar}
    &\hspace{-25mm} \times \tilde{R}(b_T,b_T^R,a,\eta) B^{\text{bare}}_{\Gamma}(b^z, \bt, a, \eta, P^z).
\end{align}
Here, the additional factor $\tilde R$ has been introduced into the modified $\MS$-renormalized quasi beam function to compensate for the power-law divergences $\sim |b_T-b_T^R|/a$ which would otherwise affect both the numerator and denominator of Eq.~\eqref{eq:finalCSexpression}: 
\begin{align}\label{eq:R}
    &\tilde{R}(b_T,b_T^R,a,\eta) = \frac{{Z}_{\cO_{\gamma^4\gamma^4}}^{ \RI}(p_R=\tilde{p}_R,b^z=0,\vec{b}_T,a,\eta)}{{Z}_{\cO_{\gamma^4\gamma^4}}^{ \RI}(p_R=\tilde{p}'_R,b^z=0,\vec{b}_T^R,a,\eta)}\,.
 \end{align}
 In this definition, fixed choices of $\tilde{p}_R$, $\tilde{p}'_R$, and of the directions of $\vec{b}_T$ and $\vec{b}^R_T$, are taken. 
 Since the factor $\tilde R$ is independent of $b^z$, and thus cancels between the numerator and denominator of Eq.~\eqref{eq:finalCSexpression}, the specific choice of definition will not affect the determination of the Collins-Soper kernel\footnote{The definition of $\tilde R$ used here differs from that in Ref.~\cite{Ebert:2019tvc} by the omission of the quasi soft factor and by allowing $\tilde p'_R$ to be different from $\tilde p_R$.}. 
In the numerical study in this work, an average over $\bt$ and $\vec{b}^R_T$ orientations, and over several choices of $\tilde{p}_R$ and $\tilde{p}'_R$, is performed in the same manner detailed in Appendix~\ref{app:renbeams} in the numerator and denominator of $\tilde R$.\footnote{In the numerical study presented here, a set of ten momenta $p_R$ with $p_R^2$ ranging from $5.7$ to $28\text{ GeV}^2$, as described in Ref.~\cite{Shanahan:2019zcq}, are used to construct $\tilde R$.}

Several observations are pertinent to the computation of the Collins-Soper evolution kernel by Eq.~\eqref{eq:finalCSexpression}. First, since the kernel is independent of the external state~\cite{Ebert:2018gzl}, one may calculate the quasi beam functions in the state with the best signal-to-noise properties in a lattice QCD calculation, e.g., for the pion. In a quenched calculation, a heavier-than-physical valence quark mass can be chosen for the same reason. Moreover, since although the kernel is state-independent, the power-corrections to the kernel are not, and so variation of the choice of external state, and external state momenta, provides a test of systematic effects in a numerical calculation.
Second, the Collins-Soper kernel does not depend on the longitudinal momentum fraction $x$ or on the hadron momenta $P_i^z$, at $\mathcal{O}\left( {b_T}/{\eta},{1}/({b_T P^z})\right)$. Although the truncation in the $b^z$-space Fourier integral will induce oscillatory behavior in $x$-space, varying these parameters provides insight into these additional systematic uncertainties.

An alternative approach to extracting the Collins-Soper kernel by transforming the product of the matching coefficient and $\MS$ quasi beam function in \eq{finalCSexpression} into a convolution integral in $b^z$-space was advocated in Ref.~\cite{Ebert:2019tvc}.
Appendix~\ref{app:alt} provides an investigation of this approach and finds that it suffers from significant systematic uncertainties.

\section{Lattice QCD study}
\label{sec:LQCD}

\begin{table}[t]
	\begin{tabular}{ccccccc}\hline\hline
		Label & $\beta$ & $a$ [fm] & $L^3\times T$ & $\kappa$ & $n_\text{src} $ & $n_\text{cfg}$\\\hline
		$E_{32}$ & 6.3017 & 0.06 & $32^3\times 64$  & 0.1222 & 2 & 200 
		\\\hline
	\end{tabular}
	\caption{\label{tab:ensembles}The ensemble of quenched QCD gauge field configurations used in this work~\cite{Detmold:2018zgk,Endres:2015yca}. The lattice spacing $a$ is determined from an analysis of scale setting in Ref.~\cite{Asakawa:2015vta}, and the lattice geometry parameters $L$ and $T$ are specified in units of $a$. For operator structures with Dirac index $\Gamma=\gamma^4$, $n_\text{cfg}$ configurations are analyzed, with $n_\text{src}$ source locations chosen on each. For other operator Dirac structures $\Gamma\ne\gamma^4$, a subset with 25 configurations is analyzed, with 1 source location computed on each.
	}
\end{table}

The Collins-Soper evolution kernel is computed by Eq.~\eqref{eq:finalCSexpression} in a lattice QCD calculation using a single quenched ensemble, detailed in Table~\ref{tab:ensembles}.  The calculation is undertaken on gauge fields that have been subjected to Wilson flow to flow-time t = 1.0~\cite{Luscher:2010iy}, in order to increase the signal-to-noise ratio of the numerical results, and gauge-fixed to Landau gauge, in order to permit the use of gauge non-invariant quark wall sources.
Quasi beam functions are constructed for a pion external state using valence quark propagators that are computed with the tree-level $\mathcal{O}(a)$ improved Wilson clover fermion action~\cite{Sheikholeslami:1985ij} and a $\kappa$ value that corresponds to a heavy pion mass of 1.207(3)~GeV. This choice may be made without introducing systematic bias, since the Collins-Soper kernel is independent of state. Three external state momenta are studied, $\vec{P}=P^z\vec{e}_z$ with $P^z=n^z 2\pi/L$ for $n_z\in\{2,3,4\}$, corresponding to $P^z \in \{1.29,1.94,2.58\}$ GeV, allowing the kernel to be computed from three different momentum ratios. To improve the overlap of boosted pion interpolating operators onto their respective ground states and improve statistical precision, a combination of wall sources and momentum-smeared sinks~\cite{Bali:2016lva} are used to construct two-point and three-point correlation functions.

\begin{figure}[t!]
        \centering
      \includegraphics[width=0.46\textwidth]{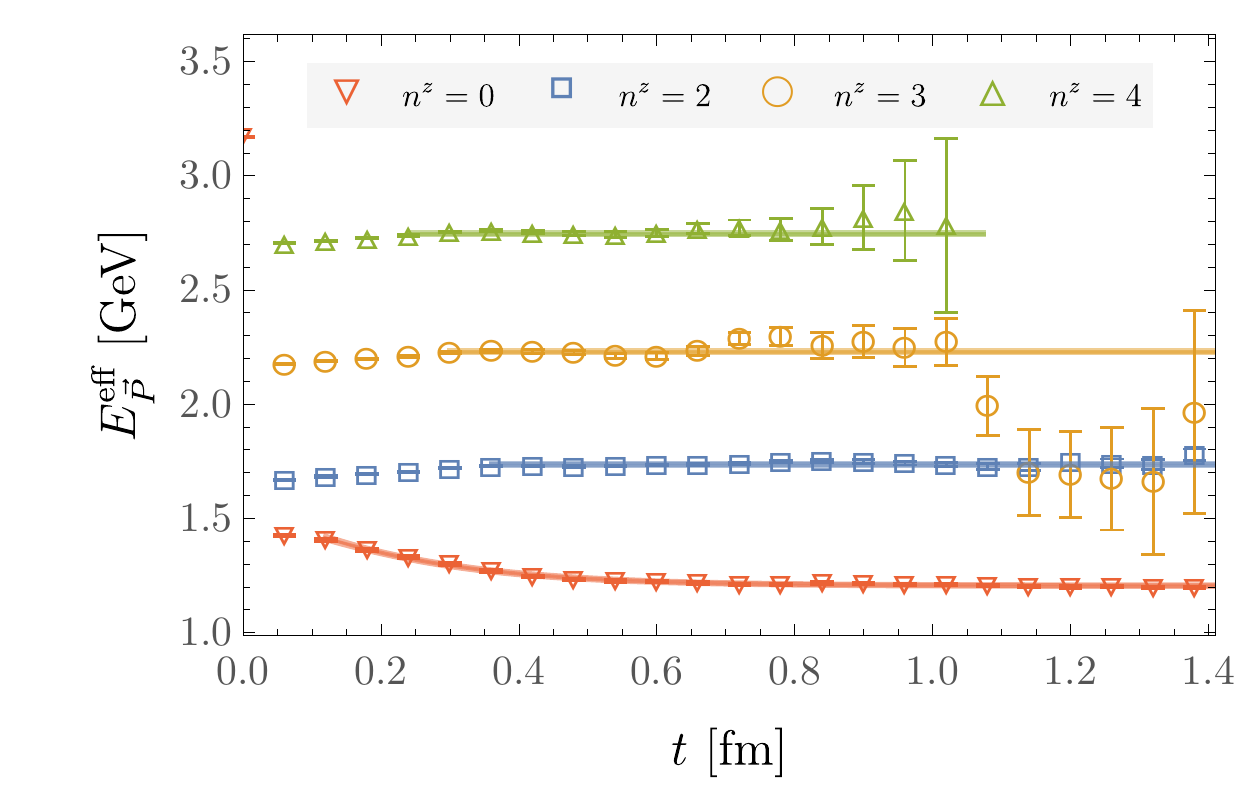}
      \caption{\label{fig:EMP}  Effective energy function defined by Eq.~\eqref{eq:Eeff} for pion states with momenta $|\vec{P}| = n^z 2\pi /L$. Shaded bands display the result of single-exponential fits to the two-point correlation functions for each non-zero momentum, and a two-exponential fit at zero momentum; the number of states in each fit is chosen to maximize an information criterion as described in the text, and the fit ranges shown correspond to the highest-weight fits in the weighted average over successful two-point function fits as discussed in Appendix~\ref{app:threetwofits}. }
\end{figure}

Bare quasi beam functions $B^{\text{bare}}_{\Gamma}(b^z, \bt, a, \eta, P^z)$ are extracted for non-local quark bilinear operators (Eq.~\eqref{eq:op}) with Wilson line staple geometries defined by staple extents $\eta$ ranging between 0.6 and 0.8 fm ($\eta/a\in\{10,12,14\}$), and with staple widths and asymmetries corresponding to $|b_T|$  and $b_z$ ranging from $-(\eta-a)$ to $(\eta-a)$. 
In order for the mixing contributions to Eq.~\eqref{eq:finalCSexpression} to be consistently included, bare quasi beam functions are computed for all Dirac operator structures $\Gamma$. As detailed in the caption of Table~\ref{tab:ensembles}, however, lower statistics are used for operators with Dirac structures $\Gamma\ne\gamma^4$, whose contributions to the Collins-Soper kernel are suppressed by the renormalization factors. Previously, the 16-dimensional vector of $\MS$ renormalization factors $Z_{\cO_{\gamma^4\Gamma'}}^{\MS}(\mu,b^z,\vec{b}_T,a,\eta)$ was computed for the same ensemble and operator parameters as studied here~\cite{Shanahan:2019zcq}, and those results are used in this work.

The two-point correlation function for the pion, projected to a given three-momentum $\vec{P}$, is defined as:
\begin{align}\nonumber\label{eq:twopt}
   C_\text{2pt}(t,\vec{P}) &= \sum_{\vec{x}}e^{i \vec{P}\cdot\vec{x}}\langle 0| \pi_{\vec{P},S} (\vec{x},t) \pi_{\vec{P},W}^\dagger (0)|0\rangle\\
    & \overset{t\gg 0}{\longrightarrow} \frac{Z_{\vec{P}}}{2 a E_{\vec{P}}}e^{-E_{\vec{P}}t}+\ldots,
\end{align}
where $Z_{\vec{P}}$ denotes the combination of overlap factors for the source and sink interpolation operators and the ellipsis in Eq.~\eqref{eq:twopt} denotes contributions from higher excitations, which are exponentially suppressed for large $t$ and discussed further in Appendix~\ref{app:threetwofits}. 
Wall-source interpolating operators $\pi_{\vec{P},W}(t) = \overline{u}(t,\vec{P}/2)\gamma_5d(t,\vec{P}/2)$  are used as sources for correlation functions, where momentum projected quark fields are defined by $q(t,\vec{P}) = \sum_{\vec{x}} e^{i\vec{P}\cdot\vec{x}} q(\vec{x},t)$ for $q=\{u,d\}$. 
Momentum-smeared interpolating operators $\pi_{\vec{P,S}}(\vec{x},t) = \overline{u}_{S(\vec{P}/2)}(\vec{x},t)\gamma_5d_{S(\vec{P}/2)}(\vec{x},t)$ are used as sinks, where $q_{S(\vec{P})}(\vec{x},t)$ are quasi local smeared quark fields obtained through iterative application of the Gaussian momentum-smearing operator defined in Ref.~\cite{Bali:2016lva}.
In particular, 50 steps of iterative momentum-smearing with smearing radius $\varepsilon = 0.25$, as defined in Ref.~\cite{Bali:2016lva}, are used to construct momentum-smeared sinks for each momentum corresponding to $n^z \in \{2,3,4\}$. 
An effective energy function that asymptotes to $E_{\vec{P}}$ can be defined from the two-point correlation function by
\begin{align}\label{eq:Eeff}\nonumber
   E^{\text{eff}}_{\vec{P}}(t) &= \frac{1}{a} \arccosh{\left({\frac{C_\text{2pt}(t+a,\vec{P})+C_\text{2pt}(t-a,\vec{P})}{2C_\text{2pt}(t,\vec{P})}}\right)}\\
    & \overset{t\gg 0}{\longrightarrow} E_{\vec{P}} + \ldots.
\end{align}
Two-point correlation functions for the three momenta which are considered here are displayed in Fig.~\ref{fig:EMP}.
The extracted energies are slightly smaller than those obtained with the continuum dispersion relation, with relative deviations from $E_{\vec{P}}=\sqrt{m_\pi^2+|\vec{P}|^2}$ ranging from $1.5(4)\%$ for $n^z = 2$ to $3.5(4)\%$ for $n^z = 4$.
These deviations are consistent with the expected size of lattice artifacts which are neglected in this exploratory work.

\subsection{Quasi beam functions}

\begin{figure}[!p]
    \subfigure[~Example of the computed bare quasi beam functions.]{\label{fig:baresample}
        \centering
        \includegraphics[width=0.46\textwidth]{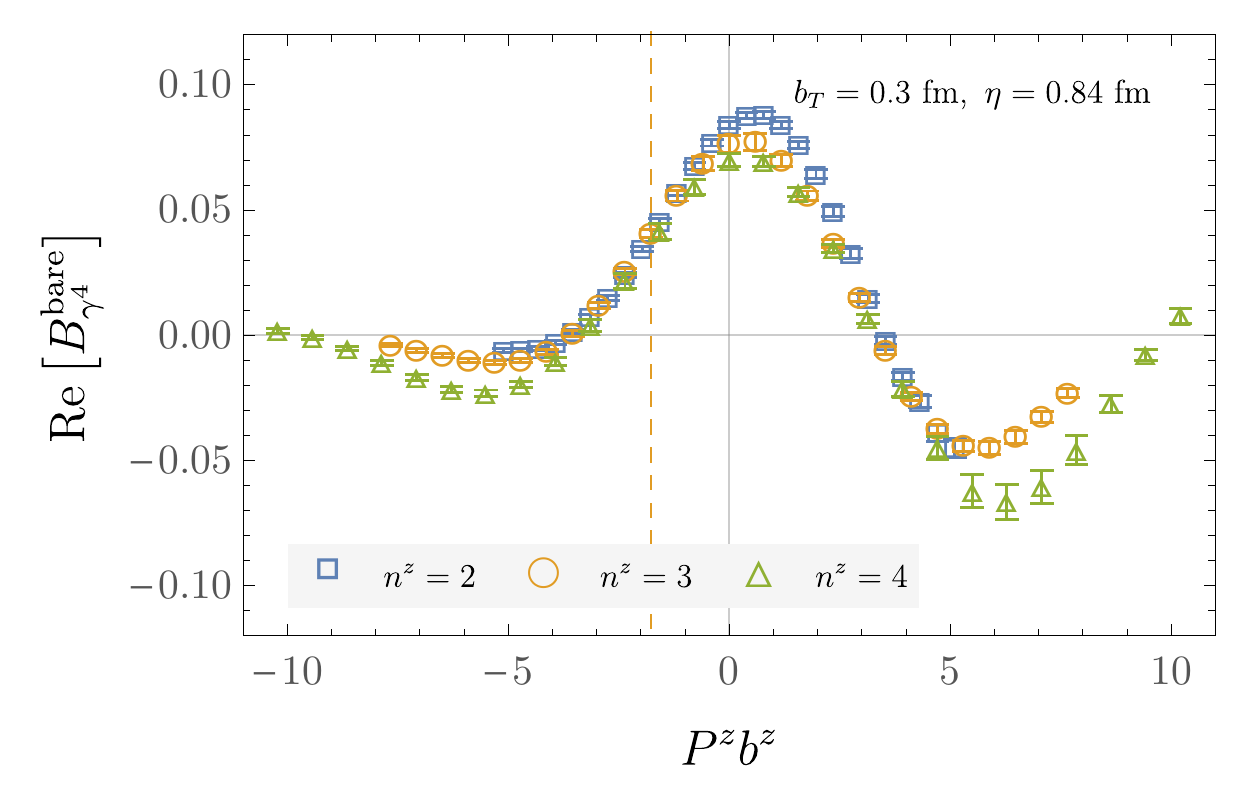}
        }\quad
    \subfigure[~Comparison of the values of bare quasi beam functions for different Dirac structures $\Gamma$, at the $b^z$ parameter indicated by the orange dotted vertical line in subfigure (a).]{\label{fig:baremixing}
        \centering
        \includegraphics[width=0.46\textwidth]{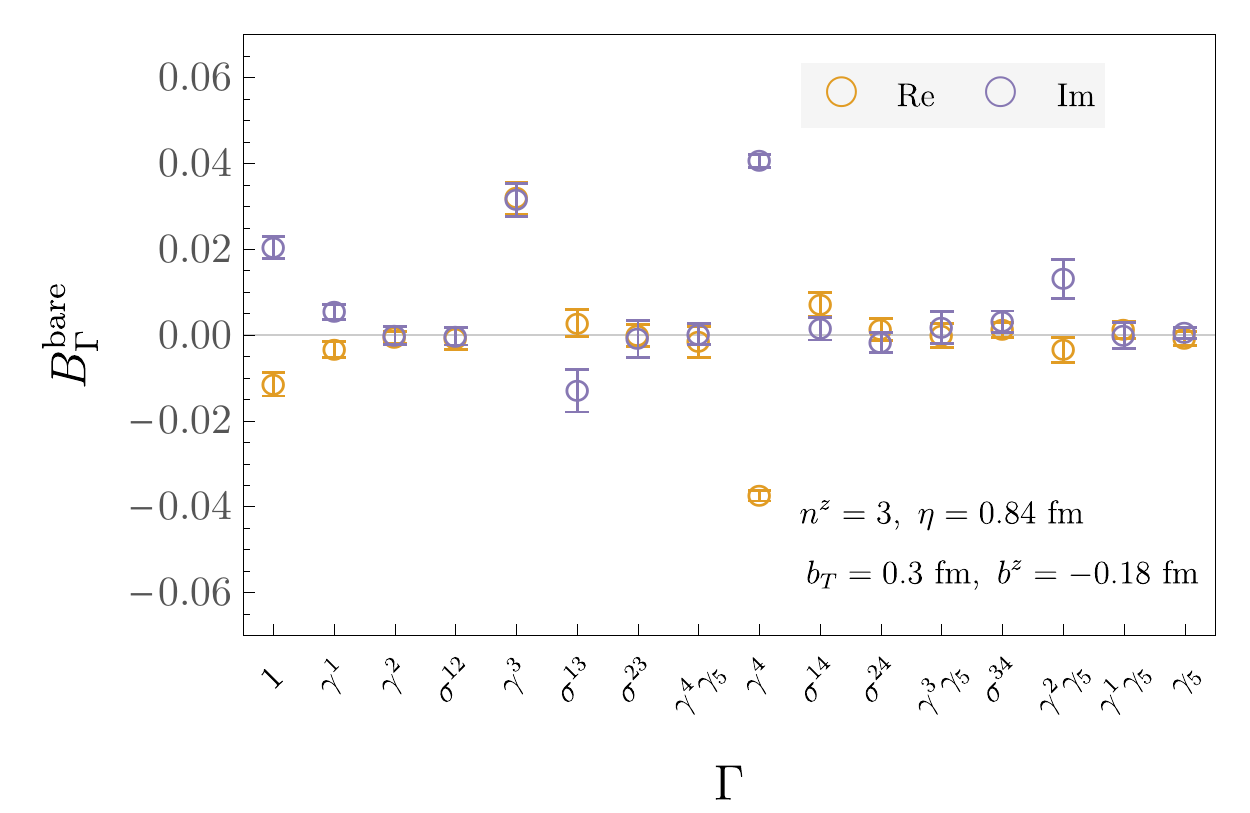}
        }\quad
    \subfigure[~Contribution to the renormalized quasi beam function $B_{\gamma^4}^{\MS}/\tilde{R}$ from each of the bare quasi beam functions shown in subfigure (b), as a fraction of the dominant contribution. The large relative uncertainties result from the lower statistics used to compute the off-diagonal renormalization factors and the bare beam functions with Dirac structures $\Gamma\ne\gamma^4$.]
        {\centering\label{fig:gammafracs}
        \includegraphics[width=0.46\textwidth]{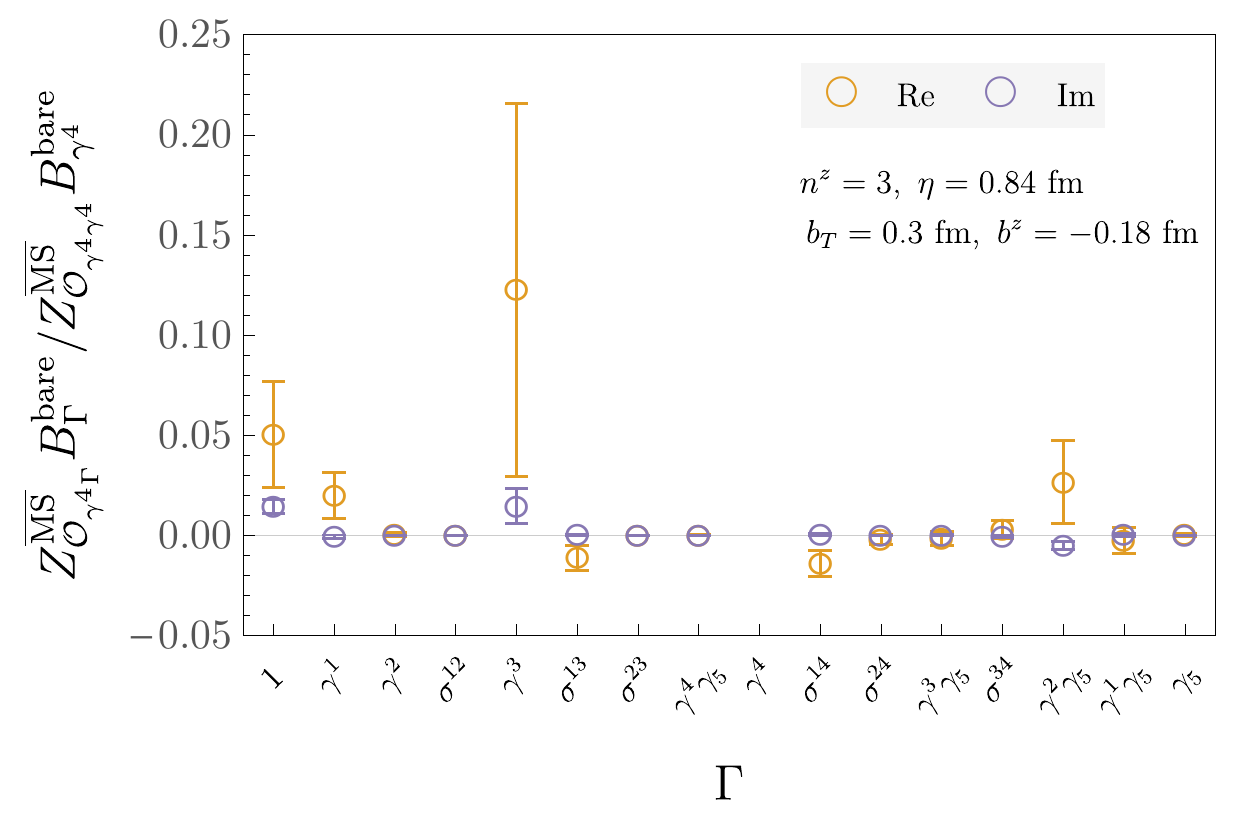}
        }\quad
        \caption{\label{fig:bare_MEs} Examples of the extracted bare quasi beam functions $B^\text{bare}_\Gamma(b^z, \bt = b_T \vec{e}_{x},a,\eta,P^z=n^z 2\pi /L)$, defined in Eq.~\eqref{eq:qbeambare}, for various parameter choices. 
        	Additional examples of the bare quasi beam functions with different parameter choices are displayed in Appendix~\ref{app:barebeams}.}
\end{figure}

Bare quasi beam functions can be computed from three-point correlation functions with insertions of the non-local quark bilinear operators $\mathcal{O}_\Gamma^i(b^\mu,z^\mu,\eta)$, defined in Eq.~\eqref{eq:op}. For the special case where pion momenta are taken only in the $z$-direction (i.e., consistent with the definition of quasi beam functions in Eq.~\eqref{eq:qbeam}), three-point correlation functions are defined as 
\begin{align}\nonumber
   &C_{\text{3pt}}^{\Gamma,i}(t,\tau,b^\mu,a,\eta,\vec{P}=P^z \vec{e}_z) \\\nonumber 
   &\ = \sum_{\vec{x},\vec{z}}e^{i\vec{P}\cdot\vec{x}}\langle 0 | \pi_{\vec{P},S} (\vec{x},t) \mathcal{O}^i_\Gamma(b^\mu,(\vec{z},\tau),\eta) \pi_{\vec{P},W}^\dagger (0)|0\rangle\\
    &\xrightarrow{t\gg \tau \gg 0}\frac{Z_{\vec{P}}}{4 a E^2_{\vec{P}}}e^{-E_{\vec{P}}t}\tilde B^\Gamma_{i}(b^z, \bt,a,\eta,P^z)+\ldots.
\end{align}
A ratio of three- and two-point correlation functions then enables the bare isovector quasi beam functions of Eq.~\eqref{eq:qbeambare} to be extracted:
\begin{align}\nonumber
    &\mathcal{R}_\Gamma(t,\tau,b^\mu,a,\eta,P^z) \\\nonumber
    &\qquad= \frac{C_{\text{3pt}}^{\Gamma,u}(t,\tau,b^\mu,a,\eta,P^z\vec{e}_z)-C_{\text{3pt}}^{\Gamma,d}(t,\tau,b^\mu,a,\eta,P^z\vec{e}_z)}{C_\text{2pt}(t,P^z\vec{e}_z)}\\\label{eq:Rcal}
    &\qquad\xrightarrow{t\gg \tau \gg 0}B^\text{bare}_\Gamma(b^z, \bt,a,\eta,P^z) + \ldots.
\end{align}

\begin{figure*}[!p]
    \subfigure[]{
        \centering
        \includegraphics[width=0.46\textwidth]{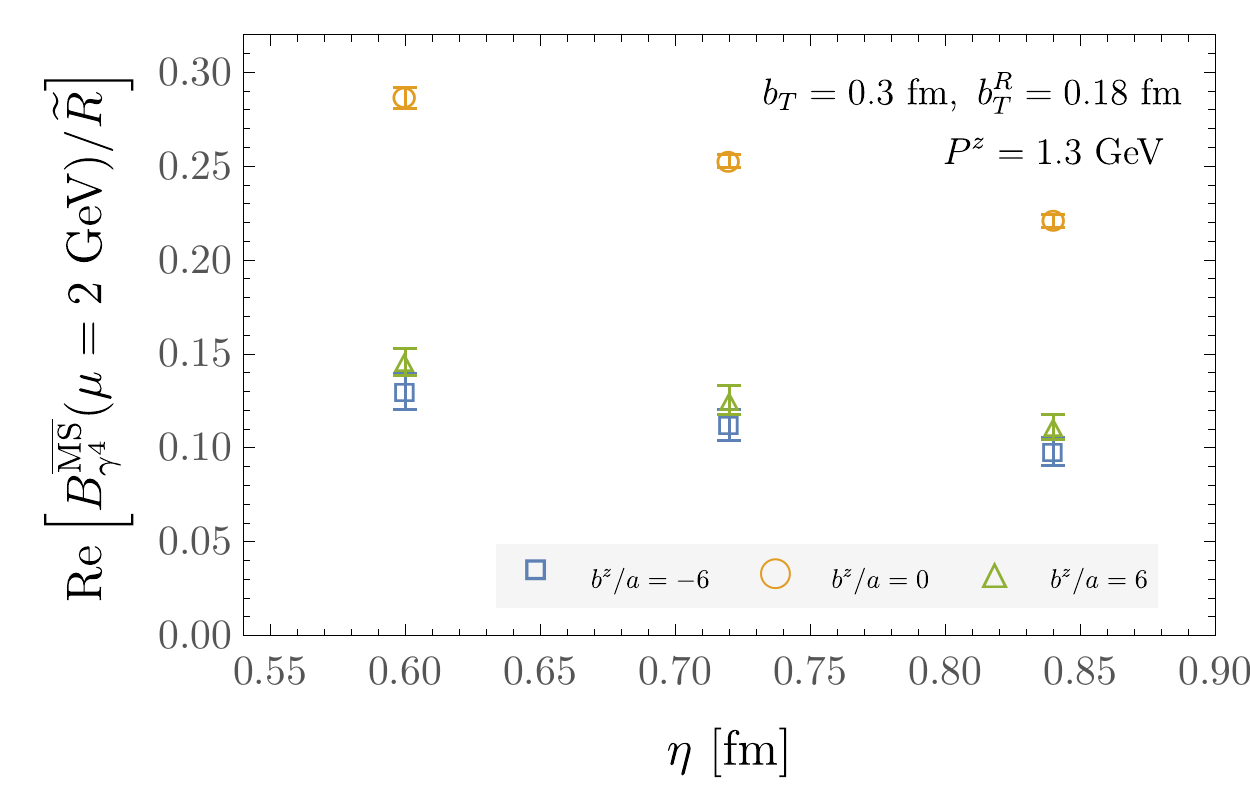}
        }\quad
    \subfigure[]{
        \centering
        \includegraphics[width=0.46\textwidth]{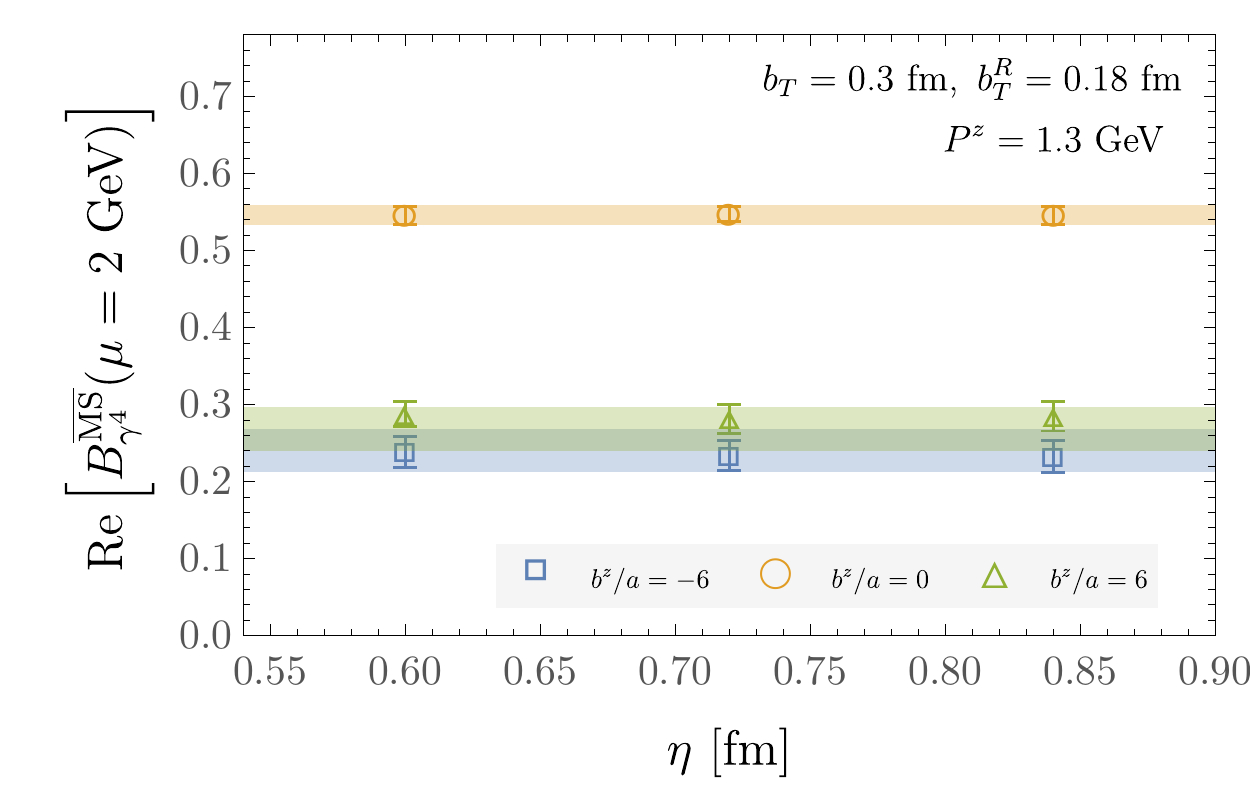}
        }\quad
    \subfigure[]{
        \centering
        \includegraphics[width=0.46\textwidth]{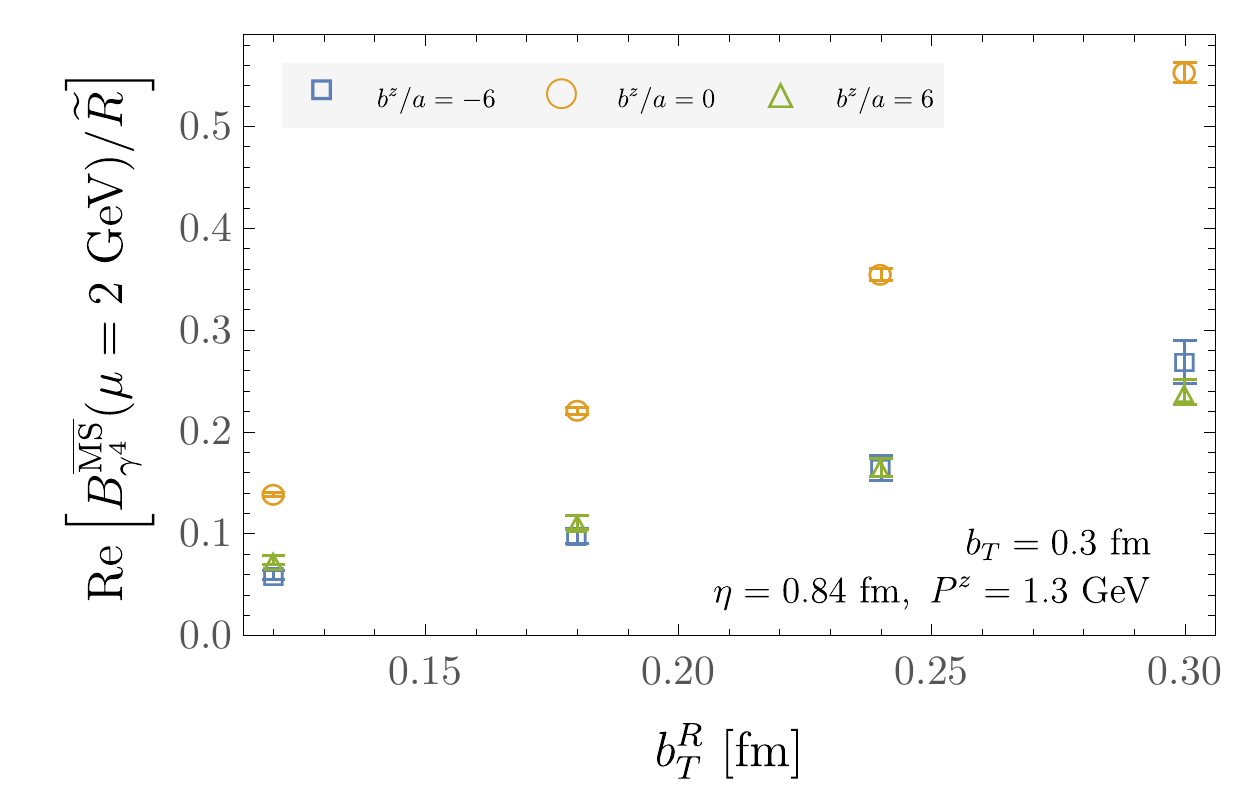}
        }\quad
    \subfigure[]{
        \centering
        \includegraphics[width=0.46\textwidth]{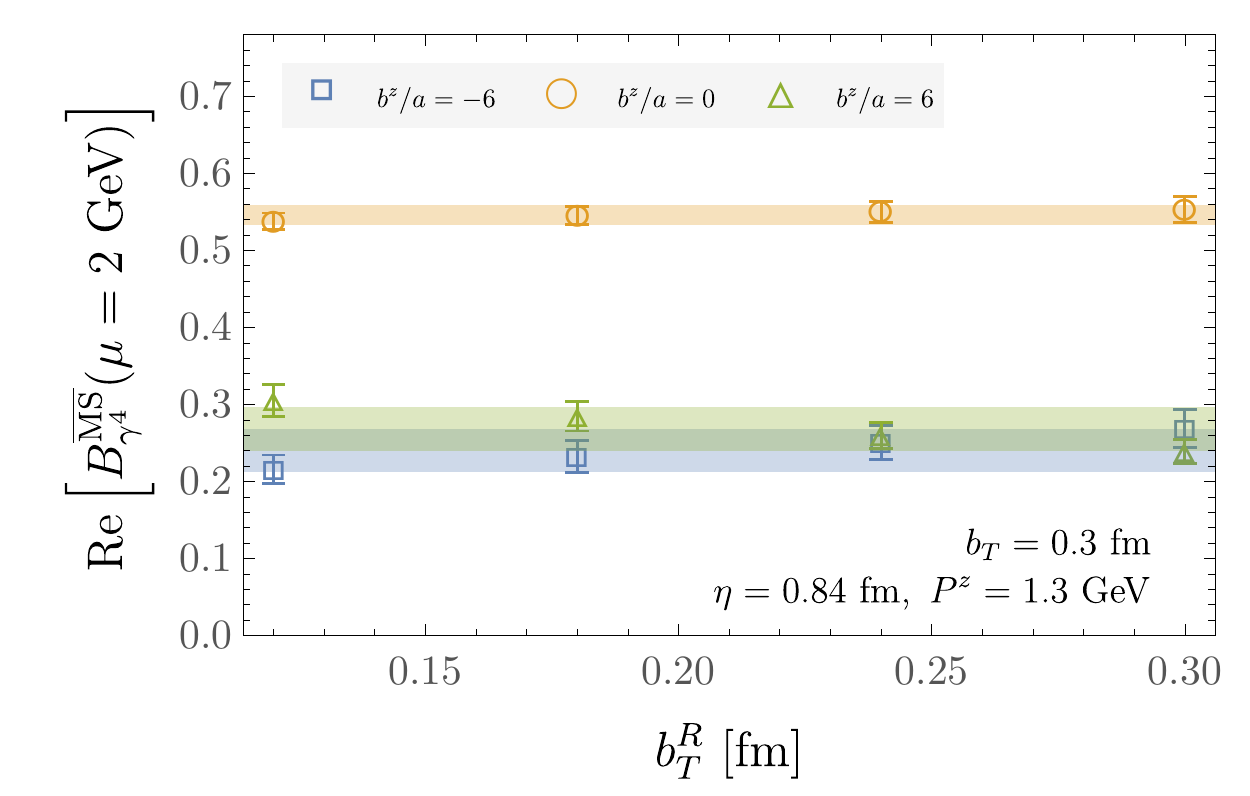}
        }\quad
        \caption{\label{fig:renorm_vs_bTR}  Renormalized quasi beam function $B^{\MS}_{\gamma^4}(\mu,b^z, \bt, a, \eta, b_T^R , P^z)$ in Eq.~\eqref{eq:BMSbar} (right column), and the same quantity divided by the factor $\tilde{R}(b_T,b_T^R ,a,\eta)$ in Eq.~\eqref{eq:R} (left column), similarly averaged, for various parameter choices. The horizontal shaded bands show the results of constant fits in $b_T^R$ and $\eta$ to the renormalized quasi beam function as a function of $b^z$ and $P^z$ (at the fixed $a$ of the calculation), as described in the text.  }
\end{figure*}

\begin{figure*}[!hp]
    \subfigure[]{
        \centering
        \includegraphics[width=0.46\textwidth]{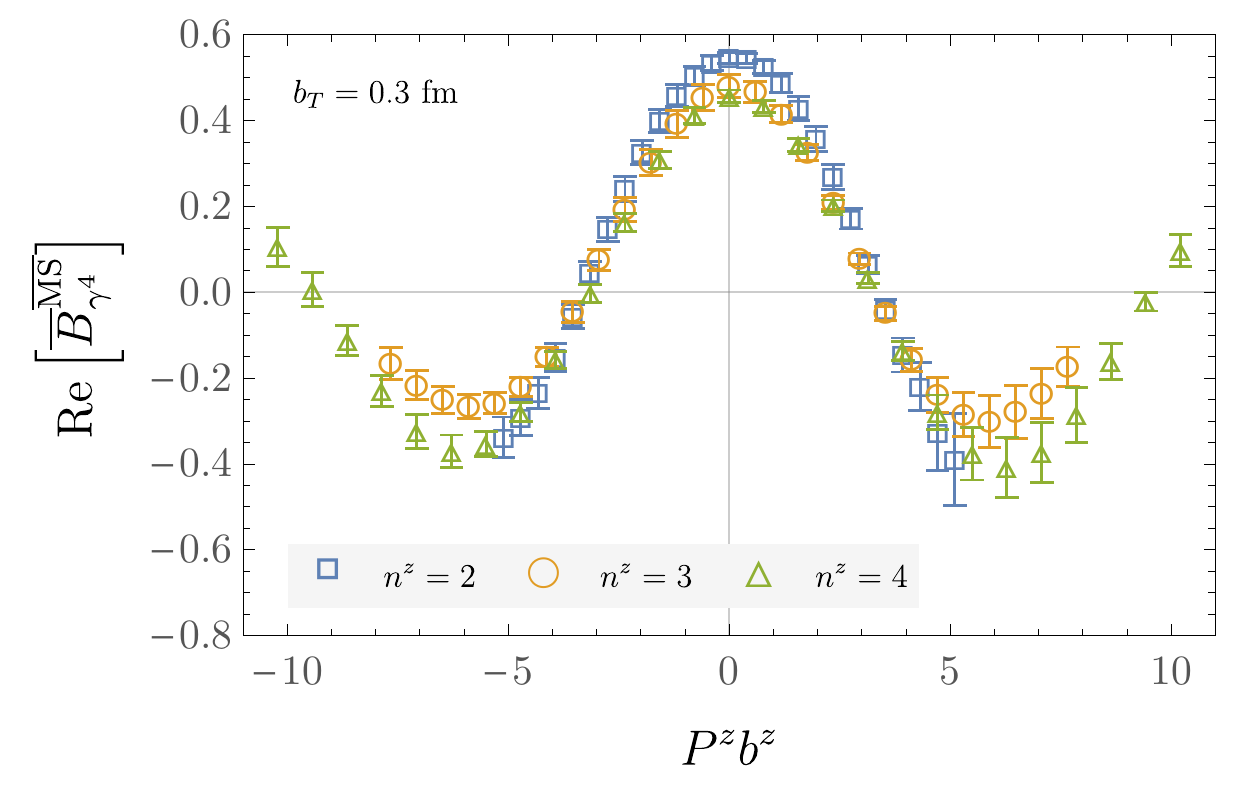}
        }\quad
        \subfigure[]{
        \centering
        \includegraphics[width=0.46\textwidth]{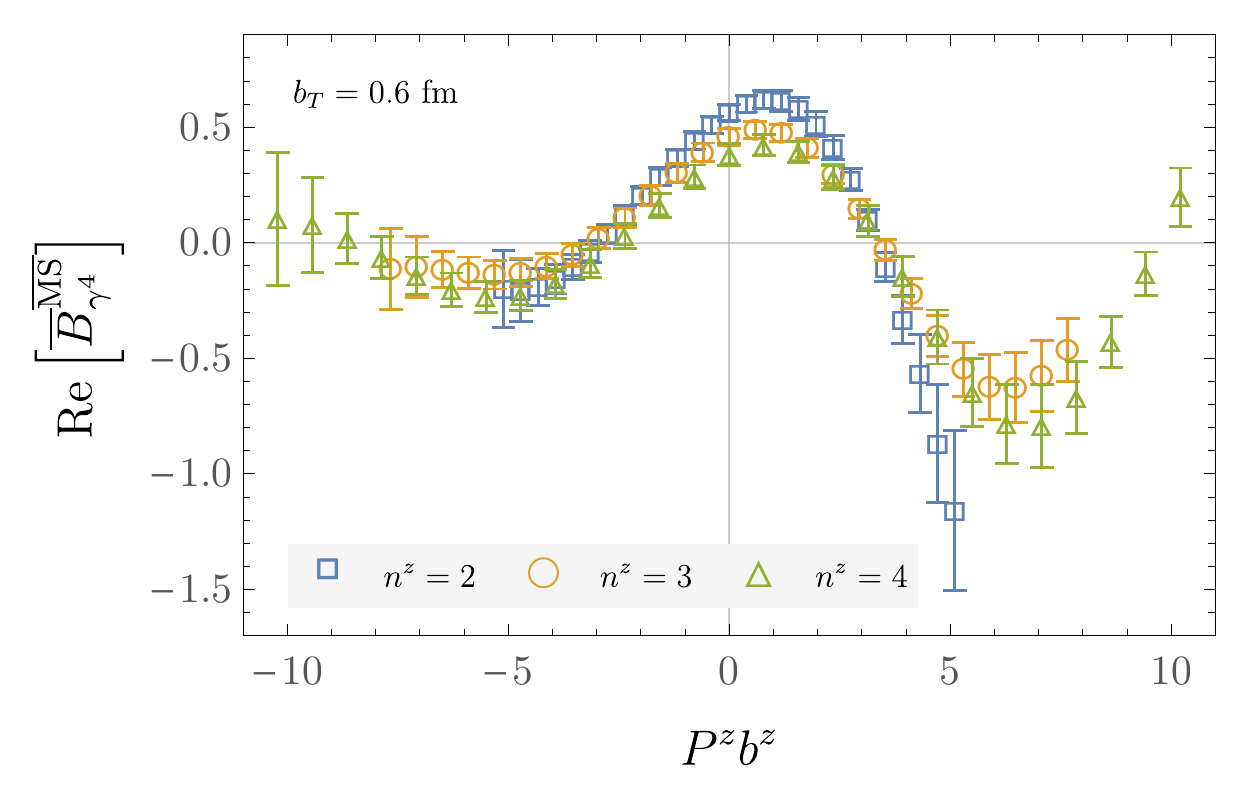}
        \label{bTReFit}
        }\quad
        \caption{\label{fig:renorm}  Averaged renormalized quasi beam function $\overline{B}^{\MS}_{\gamma^4}(\mu,b^z, b_T, a, P^z = n^z2\pi / L)$ at small (a) and large (b) $b_T$, after averaging over directions of $\vec{b}_T$, and weighted averaging over $b_T^R$ and $\eta$, as detailed in Appendix~\ref{app:renbeams}. Further examples of the averaged renormalized quasi beam functions at different choices of $b_T$ are also given in Appendix~\ref{app:renbeams}.}
\end{figure*}

Constraining the bare quasi beam functions $B^\text{bare}_\Gamma$ from ratios of two- and three- point functions $\mathcal{R}_\Gamma$ for all staple geometries (specified by $\{\eta,b^\mu\}$), all Dirac structures $\Gamma$, and all momenta $P^z$, considered in this work, requires fits for a very large number of operators (35,660) to be performed. These fits are automated using a fit procedure discussed in Appendix~\ref{app:threetwofits}. An example of the result of these fits, for $\Gamma=\gamma^4$, and specific choices of $b_T$ and $\eta$, is given in Fig.~\ref{fig:baresample}; a second example figure holding $b^z$ fixed, but showing all Dirac structures, is shown in Fig.~\ref{fig:baremixing}. Additional examples of the real and imaginary parts of the extracted bare quasi beam functions are displayed in Appendix~\ref{app:barebeams}.

The bare quasi beam functions obtained by Eq.~\eqref{eq:Rcal} are renormalized to the $\MS$ scheme by Eq.~\eqref{eq:BMSbar}, using renormalization factors $Z^{\MS}_{\mathcal{O}_{\gamma^4\Gamma}}$ which were computed for the same ensemble and operators as studied here in Ref.~\cite{Shanahan:2019zcq}. The fractional contributions to the renormalized quasi beam function from the bare quasi beam functions with different Dirac structures $\Gamma$ is shown in Fig.~\ref{fig:gammafracs}, for a particular choice of parameters. The size of these contributions is observed to grow with increasing $b_T$ and with increasing $(\eta - b^z)$; while the relative magnitudes of bare beam functions with different $\Gamma$ do not vary significantly with these parameters, the relative importance of the off-diagonal renormalization factors varies significantly as discussed in Ref.~\cite{Shanahan:2019zcq}.
Across the parameters studied, the combined contributions from mixing to the renormalized quasi beam function with Dirac structure $\Gamma=\gamma^4$ are at the $5\%$--$25\%$ level.
Calculations of the quasi beam functions, and the Collins-Soper evolution kernel, to better than this precision thus require bare quasi beam functions to be computed for several Dirac structures $\Gamma$. 

The functional dependence of the renormalized quasi beam function $B^{\MS}_{\gamma^4}(\mu,b^z, \bt, a, \eta, b_T^R, P^z)$ (defined in Eq.~\eqref{eq:BMSbar}) is shown in Fig.~\ref{fig:renorm_vs_bTR}. The factor $\tilde{R}(\vec{b}_T,b_T^R,a,\eta)$ (Eq.~\eqref{eq:R}) was included in the definition of the renormalized quasi beam function to cancel the dependence of the bare beam function on $\eta$ and on $b_T^R$. It is clear that over choices of $b_T^R$ within the perturbative region, this dependence is indeed removed to better than the statistical uncertainties of this study.
A weighted average of the renormalized quasi beam function over these parameters, as well as over different directions of $\vec{b}_T$, is thus taken as detailed in Appendix~\ref{app:renbeams} to define averaged quasi beam functions  $\overline{B}^{\MS}_{\gamma^4}(\mu,b^z, b_T, a, P^z)$. Examples of the $P^z b^z$-dependence of the resulting quasi beam functions are shown in Fig.~\ref{fig:renorm}, and additional examples are shown in Appendix~\ref{app:renbeams}. These are the key results used to extract the Collins-Soper kernel, as discussed in the next subsection.

\subsection{Collins-Soper kernel}\label{sec:CS}

\begin{figure}[!t]
            \subfigure[]{
        \centering
        \includegraphics[width=0.46\textwidth]{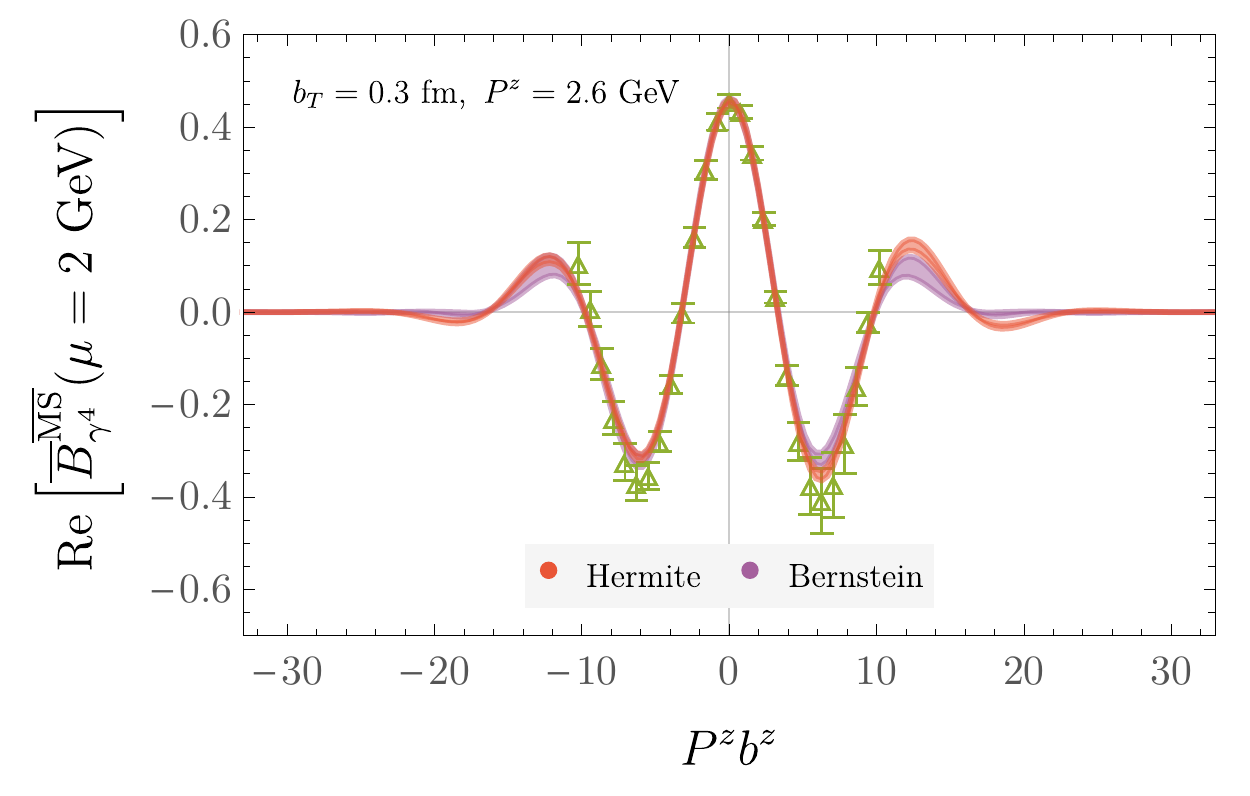}
        }\quad
        \subfigure[]{
        \centering
        \includegraphics[width=0.46\textwidth]{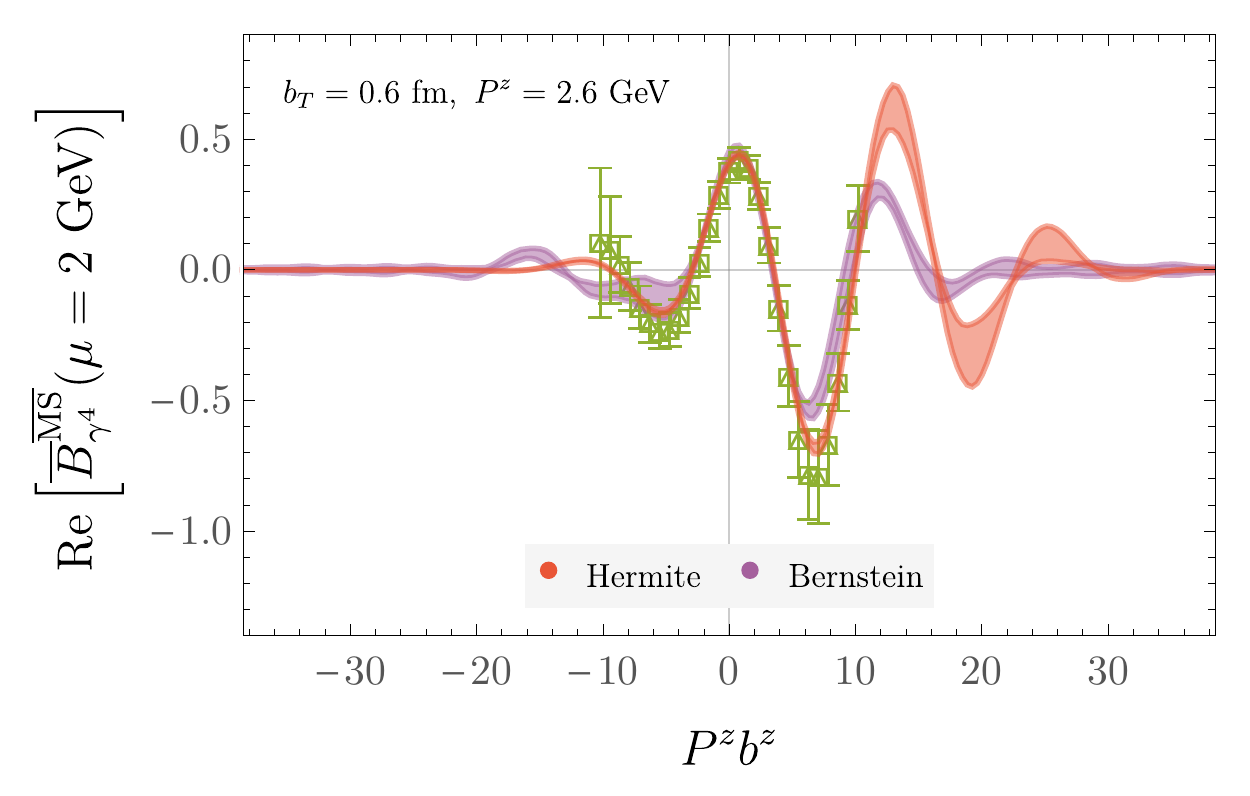}
        \label{bTReExtrap}
        }\quad
        \caption{\label{fig:renormextrap} Examples of fits to the averaged renormalized quasi beam functions $\overline{B}^{\MS}_{\gamma^4}(\mu,b^z, b_T, a, P^z)$ using  functional forms based on Hermite and Bernstein polynomials (Eqs.~(\ref{eq:herm}-\ref{eq:bern})). Further examples of fits at different choices of the $b_T$ and $P^z$ parameters are shown in Appendix~\ref{app:renbeams}.}
\end{figure}

\begin{figure*}[!t]
        \centering
        \includegraphics[width=0.65\textwidth]{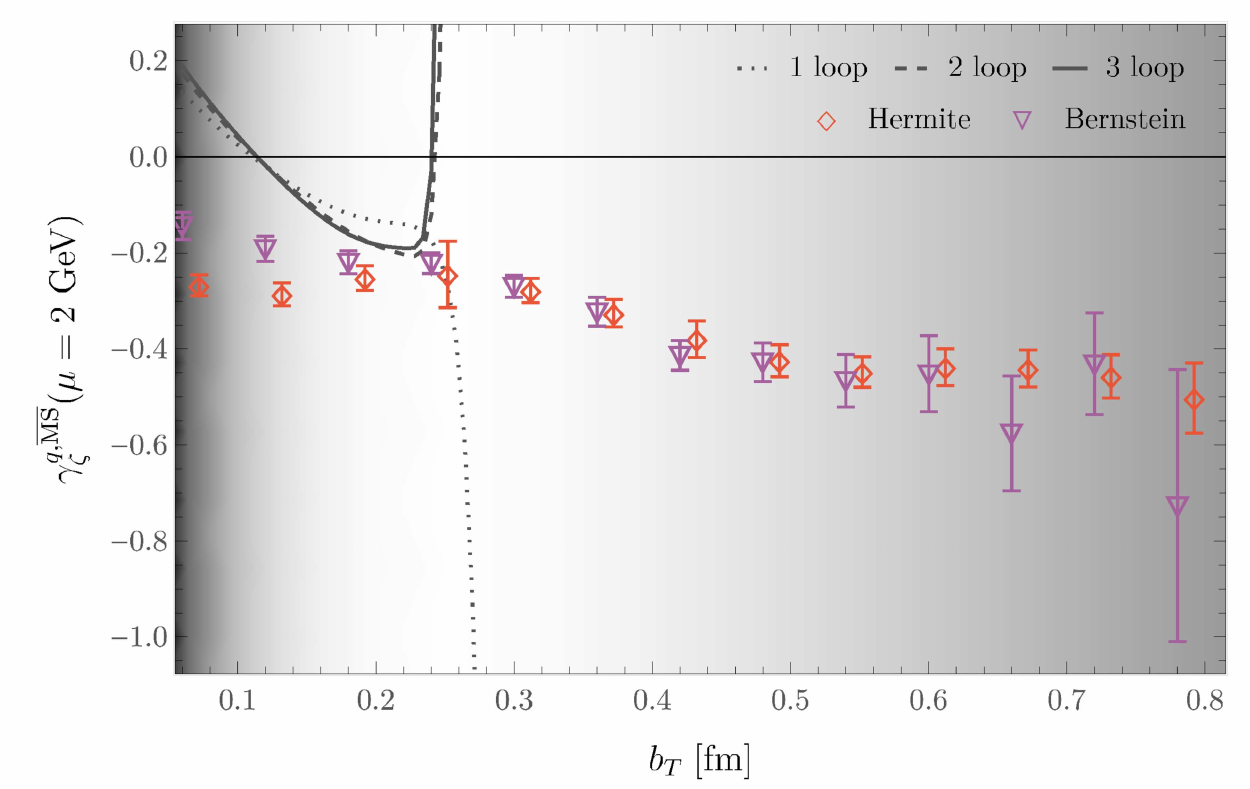}
        \label{CSvsbT}
        \caption{\label{fig:CS} Collins-Soper evolution kernel obtained using fits to the renormalized quasi beam functions based on Hermite and Bernstein polynomial bases (Eqs.~(\ref{eq:herm}-\ref{eq:bern})), computed as described in the text. The background shading density is proportional to $1/(b_TP^z) + b_T/\eta$, indicating regions of greater and lesser sensitivity to power corrections which are not included in the uncertainties presented. The black dotted, dashed and solid lines show perturbative results for the 0-flavor Collins-Soper kernel up to three-loop order~\cite{Li:2016ctv,Vladimirov:2016dll}. Perturbative results become singular at $b_T \sim 0.25$ fm because they reach the Landau pole associated with $\Lambda_{\rm QCD}^{\MS,N_f=0} = 639$ MeV. }
\end{figure*}

Computing the Collins-Soper evolution kernel by Eq.~\eqref{eq:finalCSexpression} requires taking the Fourier transform of the $\MS$-renormalized quasi beam function with respect to $b^z$. It is clear from the results shown in Fig.~\ref{fig:renorm}, however, that with the parameter ranges explored in this study the Fourier transform will suffer from significant truncation effects since the quasi beam function is not yet consistent with zero within uncertainties at the largest $b^z$ values that are used. 
For this reason, models are used to fit the $P^zb^z$-dependence of the lattice data for the quasi beam function before the Fourier transform is taken to evaluate the Collins-Soper kernel.
The results obtained by taking discrete Fourier transforms instead are shown in Appendix~\ref{app:DFT}, and a discussion of what will be required for future calculations to achieve robust results for the Collins-Soper kernel without this modeling step is presented in Sec.~\ref{sec:conc}.

Two models of the $P^zb^z$-dependence of the quasi beam functions are considered, based on Hermite and Bernstein polynomial bases. The models are constructed to yield $x$-independent Collins-Soper kernels, as would be expected in the absence of systematic artifacts, assuming the leading order value for the perturbative matching coefficient, i.e., $C^\text{TMD}_\text{ns}=1$. 
Including higher orders in the matching factor while guaranteeing an $x$-independent kernel would require more complicated functional forms to be fit to the quasi beam functions to compensate for the $x$-dependence of the matching. It is expected that the matching uncertainties are small relative to the systematic uncertainties inherent in introducing models for the $P^zb^z$-dependence of the quasi beam functions, and these effects are thus neglected in this work. While a model-independent result for the Collins-Soper kernel cannot be achieved from the data presented here, the comparison between results obtained using the two different models considered nevertheless provides some indication of the severity of the model-dependence, and the quality of fits to these functional forms not including power corrections also provides a measure of their importance.

\begin{figure*}
    \subfigure[]{
        \centering
        \includegraphics[width=0.46\textwidth]{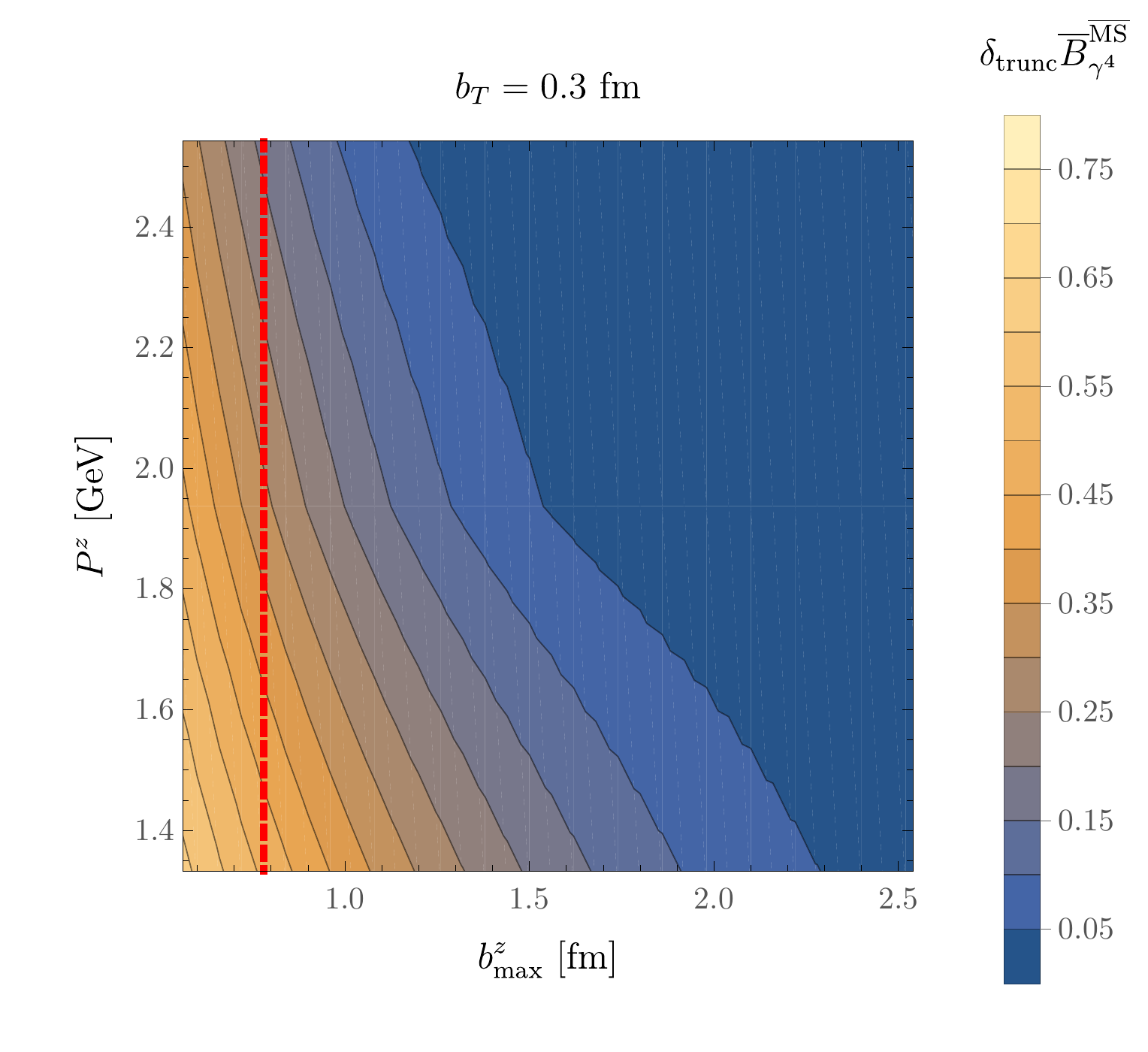}
        \label{TruncError}
        }\quad
    \subfigure[]{
        \centering
        \includegraphics[width=0.46\textwidth]{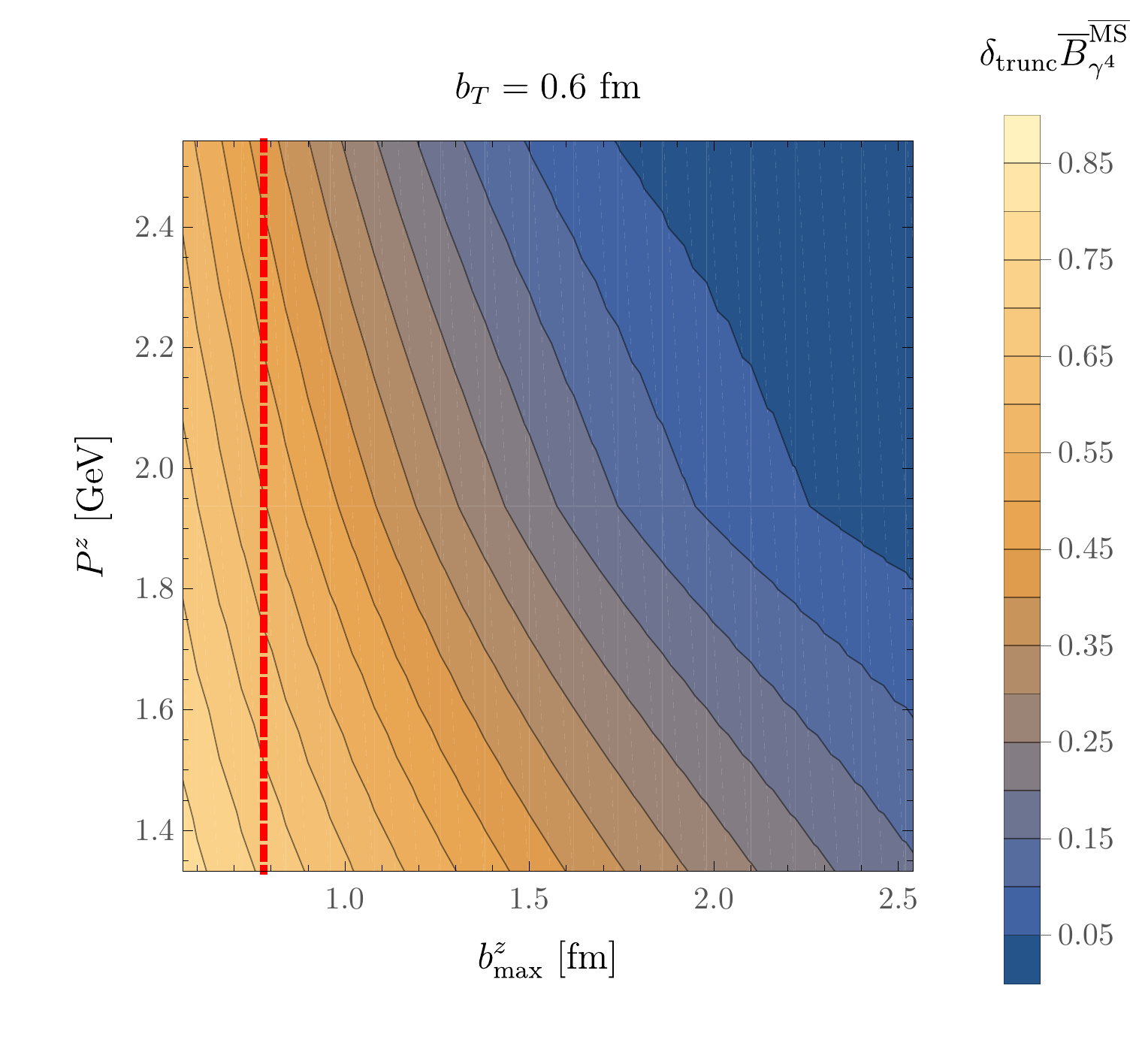}
        \label{TruncContour}
        }\quad
        \caption{\label{fig:truncation} Fractional truncation effects in the $\MS$-renormalized quasi beam functions, defined by Eq.~\eqref{eq:trunc}, evaluated at $x=0.5$ for two different $b_T$ values shown. The red vertical line denotes the maximum $b^z$ used in this study; the vertical axis range corresponds to the $P^z$ range of this study. }
\end{figure*}

The first functional form which is fit to the $\MS$-renormalized quasi beam function is 
\begin{align}\nonumber
    &\mathcal{F}_N^\text{Herm}(P^z,b^zP^z;\{a_k\},\gamma,\omega,\sigma) \\\label{eq:herm} &\quad=  \sum_{k=1}^N a_k \int_{-\infty}^{\infty}\hspace{-3mm}dx\, e^{i(b^zP^z) x}e^{-(x-\omega)^2/2\sigma} (P^z x)^\gamma \mathcal{H}_{k-1}(x),
\end{align}
where $\mathcal{H}_n(x)$ is the $n$-th Hermite polynomial. The fit parameter $\omega$ is taken to be complex, while the other free parameters are real.
Allowing $\text{Im}(\omega) \neq 0$ allows the Fourier transform of  $\mathcal{F}_N^\text{Herm}(P^z,b^zP^z;\{a_k\},\gamma,\omega,\sigma)$ with respect to $b^z P^z$ to be complex, and correspondingly enables $\mathcal{F}_N^\text{Herm}(P^z,b^zP^z;\{a_k\},\gamma,\omega,\sigma)$ to be an asymmetric function of $b^z P^z$. The real and imaginary parts of the quasi beam function are symmetric and antisymmetric functions of $b^z$ respectively in the $\eta\rightarrow \infty$ limit; however, the numerical results presented in this work show significant departures from these expectations, particularly for large $b_T$, as shown in Fig.~\ref{bTReFit}.
The observed asymmetry could arise from finite-volume effects: effective field theory calculations~\cite{Briceno:2018lfj} have demonstrated that finite-volume effects for pion matrix elements of non-local operators with separation $\ell$ generically take the form $e^{-m_\pi(L-\ell)}$. In this work, one therefore expects $b^z$-dependent finite-volume effects of the form $e^{-m_\pi(L - \eta + b^z)}$ as well as additional $b^z$ independent finite-volume effects. In addition, exponential dependence on $b^z$ could arise from an imperfect cancellation between power-law-divergent lattice artifacts in $B^\text{bare}_\Gamma(b^z, \bt,a,\eta,P^z)$ and $Z_{\cO_{\gamma^4\Gamma}}^{\MS}(\mu,b^z,\vec{b}_T,a,\eta){\tilde{R}(b_T,b_T^R,a,\eta)}$. Taking $\text{Im}(\omega) \neq 0$ allows the fit form in Eq.~\eqref{eq:herm} to include exponential dependence on $b^z$ and is found to significant improve the quality of fits to the numerical results with large $b_T \gtrsim 0.5$~fm.

The second model considered assumes that the Fourier transform of the quasi beam function has compact support on the interval $0 < x < 1$~\cite{Ji:2014hxa,Ji:2018hvs,Ebert:2019okf}, which is expected to become valid for large $P^z$, and takes the form
\begin{align}
    &\mathcal{F}_N^\text{Bern}(P^z,b^zP^z;\{a_n^r\},\gamma,A,B) \nonumber  \\
    & =  \sum_{r=0}^{N-1} a_r \int_0^1 dx\, e^{i (b^zP^z)x} \ x^A (1-x)^B(P^z x)^\gamma  \mathcal{B}_{r,N-1}(x)\,, \label{eq:bern}
\end{align}
where $\mathcal{B}_{r,N-1}$, for $r\in\{0,\ldots N-1\}$ are the $N$ Bernstein basis polynomials of degree $N-1$ normalized as in Ref.~\cite{PiegTill96}, and asymmetry in $b^z$ is accommodated by taking $\text{Im}(a_r)\neq 0$.

Using either functional form, $\mathcal{F}_N^{\text{Herm}}$ or $\mathcal{F}_N^{\text{Bern}}$, as a model for $\overline{B}^{\MS}_{\gamma^4}$, and evaluating Eq.~\eqref{eq:finalCSexpression} with the tree-level matching factor $C_{\text{ns}}^{\text{TMD}} = 1$, gives the result $\gamma_\zeta^{q,\MS} = \gamma$, where $\gamma$ is the model parameter appearing in Eqs.~\eqref{eq:herm}-\eqref{eq:bern}. That is, the resulting Collins-Soper kernel is independent of $x$ by construction. 
The full procedure by which each functional form is fit to the numerical results for the quasi beam function is described in Appendix~\ref{app:renbeams}, and examples of the resulting fits are shown both in Fig.~\ref{fig:renormextrap} and in Appendix~\ref{app:renbeams}. Briefly, the fits are undertaken simultaneously at all $P^z$ and $b^z$ values for a given $b_T$, and an information criterion is used to choose the model truncation $N$ for each fit. While both models fit the quasi beam function well within the range of $P^zb^z$ values constrained by the lattice data (with an average $\chi^2/N_\text{dof}$ over all fits of 0.9, tabulated in Appendix~\ref{app:renbeams}), it is clear from Fig.~\ref{fig:renormextrap} that they correspond to substantially different models outside this range.

The Collins-Soper kernel determined from each set of model fits is shown in Fig.~\ref{fig:CS}.
The results obtained using the two model forms, i.e., the Hermite polynomial model, in which the quasi beam function has support on $-\infty < x <\infty$, and the Bernstein polynomial model, with support on $0 < x < 1$, are consistent.
This encouragingly suggests that $\gamma_\zeta^{q,\MS}$ is well-constrained by the numerical results at the $P^z$ and $b^z$ values of this calculation, and that the model-dependence introduced in the Fourier transform is relatively mild. 
Perturbative results for the 0-flavor Collins-Soper kernel~\cite{Li:2016ctv,Vladimirov:2016dll} are also shown in Fig.~\ref{fig:CS} for comparison.\footnote{In this work, $\alpha_s^{\MS,N_f=0}$ is determined by evolving $\alpha_s^{\MS,N_f=5}(\mu = M_Z)$ from Ref.~\cite{Bethke:2009jm} to lower scales using the four-loop $\beta$ function~\cite{vanRitbergen:1997va}, integrating out bottom and charm quarks, and finally matching $\alpha_s^{\MS,N_f=0}(\mu)$ to $\alpha_s^{\MS,N_f=3}(\mu)$ at the scale $\mu = 2$ GeV where $Z_{\cO_{\Gamma\Gamma'}}^{\MS}(\mu,b^z,b_T,a,\eta)$ is calculated in Ref.~\cite{Shanahan:2019zcq}.
This procedure gives the result $\Lambda_{\rm QCD}^{\MS,N_f=0} = 639$ MeV, which determines $\alpha_s^{\MS,N_f = 0}(\mu)$ at all $\mu$; throughout this work $\alpha_s^{\MS,N_f = 0}(\mu=2\ {\rm GeV})=0.293951$.}
It is noteworthy that the lattice QCD results for $\gamma_\zeta^{q,\MS}$ obtained here are consistent with perturbative calculations of the 0-flavor Collins-Soper kernel~\cite{Li:2016ctv,Vladimirov:2016dll} in the region $b_T\sim 0.2$~fm. For $b_T < 0.2$~fm,  the results differ significantly from the perturbative calculation, which is likely due to power corrections to the lattice QCD results of the form $1/(b_TP^z)$, which have not been estimated here. 

Although the Collins-Soper kernel shown in Fig.~\ref{fig:CS} has been obtained in a 0-flavor calculation, it can also be compared qualitatively with the results of fits to experimental data, in which several different parametrizations of the nonperturbative behavior of the kernel at large $b_T$ have been used. 
In early fits to Drell-Yan data~\cite{Landry:1999an,Landry:2002ix,Konychev:2005iy}, the Collins-Soper kernel was parametrized as a quadratic function in $b_T$ in the nonperturbative region. It was later found in Refs.~\cite{Sun:2013dya,Sun:2013hua}, however, that these fits cannot describe SIDIS data. 
More recently, it has been argued that $\gamma_\zeta^{q,\MS}$ should approach a constant as $b_T \rightarrow \infty$; phenomenological fits to Drell-Yan data under this assumption suggest that this constant is $\sim -0.6$~\cite{Collins:2014jpa}. Finally, in a recent fit to both SIDIS and Drell-Yan data~\cite{Scimemi:2019cmh}, the kernel was parametrized to behave linearly at large $b_T$. The results of this numerical study are qualitatively consistent with constant or linear behavior of the kernel in the nonperturbative region.
Once lattice QCD results with controlled systematic uncertainties are available, it will be possible to test these and other phenomenological expectations for the large-$b_T$ behavior of the Collins-Soper kernel with QCD predictions, and begin incorporating lattice QCD constraints in phenomenological analyses.
The requirements for a fully controlled lattice QCD determination of the Collins-Soper kernel are discussed in the following section.

\section{Outlook}
\label{sec:conc}

This manuscript presents an exploratory calculation of the nonperturbative Collins-Soper kernel in quenched lattice QCD based on the method developed in Refs.~\cite{Ebert:2018gzl,Ebert:2019okf,Ebert:2019tvc}. 
In this approach, the kernel is computed from quasi beam functions defined from matrix elements of quark bilinear operators with staple-shaped Wilson lines in boosted hadron states. These beam functions are renormalized to the $\MS$ scheme via the $\RI$ prescription, and a ratio of Fourier-transformed quasi beam functions at different hadron boost momenta determines the Collins-Soper kernel. In this calculation, the kernel is extracted over a range of scales $b_T\in (0.1,0.8)~\text{fm}$. The final results presented here rely on modeling the $b^z$-space quasi beam functions to control truncation effects in the Fourier transform. Nevertheless, the results are robust under the variation of models considered here.

For a future controlled and model-independent determination of the Collins-Soper kernel by this method, several improvements will be essential. Critically, larger lattice volumes must be studied; the overwhelming systematic in this calculation arises from modeling to facilitate the Fourier transform, which is necessary because of truncation effects suffered due to the limited $b^z$ range over which quasi beam functions could be computed on the lattice volume used here. One measure of truncation effects is given by the model truncation error defined as 
   \begin{equation}\begin{split}\label{eq:trunc}
      \delta_{\text{trunc}}\overline{B}^{\text{MS}}  
      = \left|\frac{  \text{DFT}_{b^z_\text{max}} \left[\mathcal{F}_N^\text{Herm}\right](P^z,x) }{ \text{FT}_{\infty} \left[\mathcal{F}_N^\text{Herm}\right](P^z,x) } - 1\right| ,
   \end{split}
\end{equation} 
where the truncated discrete Fourier transform (DFT) and untruncated Fourier transform are defined as
\begin{subequations}\begin{align}\label{eq:FT}
   &\text{DFT}_{b^z_\text{max}} \left[\mathcal{F}_N^{\text{Herm}}(P^z,b^zP^z)\right](P^z,x) \\\nonumber
   &\hspace{30pt} = P^z \sum_{b^z = -b^z_\text{max}}^{b^z_\text{max}} e^{-i x P^z b^z} \mathcal{F}_N^{\text{Herm}}(b^z P^z), \\
   &\text{FT}_{\infty} \left[\mathcal{F}_N^{\text{Herm}}(b^zP^z)\right](P^z,x) \\\nonumber
   &\hspace{30pt} = \int_{-\infty}^{\infty} \frac{P^z db^z}{2\pi} e^{-i x P^z b^z} \mathcal{F}_N^{\text{Herm}}(b^z P^z).
\end{align}
\end{subequations}
Equation~\eqref{eq:trunc} is evaluated by applying both the DFT and untruncated Fourier transform to the best-fit model $\mathcal{F}_N^\text{Herm}(P^z,b^zP^z;\{a_k\},\gamma,\omega,\sigma)$ obtained by fitting the lattice QCD results at fixed $b_T$, as described in Sec.~\ref{sec:LQCD}. This provides an estimate, based on the model, of the effects of extending $b^z_\text{max}$ to larger values of $\eta$ than those used for the numerical calculations in this work.
Figure~\ref{fig:truncation} shows $\delta_{\text{trunc}}\overline{B}^{\text{MS}}(x)$ for $x=0.5$ and two values of $b_T$ over the range of $P^z$ values used in this work, and at values of $b^z_\text{max}$  both comparable to the value used here and considerably larger.
For the value of $b^z_\text{max}$ used here, truncation effects in the DFT results are $\cO(1)$ for large $b_T$ and prevent a reliable determination of the Collins-Soper kernel using a DFT approach.
Results with significantly larger $P^z b^z_\text{max}$, however, could be used to obtain a model independent prediction of the Collins-Soper kernel directly from a DFT of lattice QCD results (see Appendix~\ref{app:DFT}).
For example, based on the results of the present study, quasi beam function calculations with $(b^z_\text{max},P^z)~\sim (2.5~\text{fm},2.5~\text{GeV})$ are likely to suffer from truncation effects of less than 5\% for $b_T$ scales across the range of those studied in this work. 

In addition to truncation artifacts, extractions of the Collins-Soper kernel by the method pursued here suffer from power corrections at $\mathcal{O}\left( b_T/\eta,1/(b_TP^z)\right)$~\cite{Ebert:2019okf,Ji:2019ewn}.
These effects could not be resolved by model fits in this study, and as such, the coefficients of these power corrections could not be constrained. Nevertheless, larger physical lattice volumes, as needed to reduce truncation artifacts, will simultaneously enable ${\cal O}\left(b_T/\eta,1/(P^z\eta)\right)$ effects to be reduced by allowing larger staple extents $\eta$ to be investigated at fixed $b_T$. Larger boost momentum, again needed to control truncation effects, will also simultaneously enable control over power corrections of ${\cal O}(1/(b_TP^z))$. These artifacts make comparison of the Collins-Soper kernel extracted by the method pursued here with perturbative predictions, which are accurate in the region $b_T\ll \Lambda_{\rm QCD}^{-1}$, challenging. Precise comparisons in this region will be an important test of systematics in the lattice QCD approach. The infinite volume and continuum limits must also ultimately be taken for a fully controlled result.

Future studies would also gain significantly by exploiting the state-independence of the Collins-Soper kernel to obtain multiple constraints on the kernel from the same calculation and thus test systematic effects. An alternative, complementary, approach to extracting the Collins-Soper kernel from \eq{gamma_zeta1} was proposed in Ref.~\cite{Vladimirov:2020ofp}. This strategy uses the Mellin moments of the expressions so that one only needs to calculate the quasi beam function or its derivatives at $b^z=0$, which reduces the computational cost and has the advantage that renormalization factors cancel in the ratio. This approach also, however, requires a nontrivial integration over the TMDPDF that is extracted from experiments over a limited kinematic range. Comparison of results of the two approaches will also be valuable in future calculations.

Despite the significant challenges described above, the results presented here suggest that controlled first-principles calculations of the Collins-Soper kernel at nonperturbative scales as large as $b_T\sim 1$~\text{fm} are tractable with current methods. Refs.~\cite{Bacchetta:2017gcc,Bertone:2019nxa,Scimemi:2019cmh,Bacchetta:2019sam} indicate that such a calculation at 10\% precision at scales $b_T\in (0.2,1.0)~\text{fm}$ will be sufficient to differentiate different models of the Collins-Soper kernel and will thereby provide important input for fitting low-energy SIDIS data. Ultimately, larger values of $b_T$, e.g. $b_T \lesssim 2~\text{fm}$, will be important input for determinations of the TMDPDFs; this will be attainable with larger lattice volumes in the future.

\section*{Acknowledgments}

The authors thank Will Detmold, Markus Ebert, Michael Engelhardt, Andrew Pochinsky, and Iain Stewart for helpful discussions, and Michael Endres for providing the gauge field configurations used in this project.
Calculations were performed using the Qlua~\cite{qlua} and Chroma~\cite{Edwards:2004sx} software libraries.
This work is supported in part by the U.S.~Department of Energy, Office of Science, Office of Nuclear Physics, under grant Contracts No.~DE-SC0011090, No.~DE-SC0012704 and within the framework of the TMD Topical Collaboration. PES is additionally supported by the National Science Foundation under CAREER Award No.~1841699. MLW was additionally supported in part by an MIT Pappalardo fellowship. This research used resources of the National Energy Research Scientific Computing Center (NERSC), a U.S. Department of Energy Office of Science User Facility operated under Contract No. DE-AC02-05CH11231, and the Extreme Science and Engineering Discovery Environment (XSEDE), which is supported by National Science Foundation grant No.~ACI-1548562. This manuscript has been authored by Fermi Research Alliance, LLC under Contract No. DE-AC02-07CH11359 with the U.S. Department of Energy, Office of Science, Office of High Energy Physics.

\appendix

\section{Fits to three- and two-point functions}
\label{app:threetwofits}

As shown in Eq.~\eqref{eq:Rcal}, ratios of three-point and two-point correlation functions asymptote to the bare quasi beam function in the double limit $\{\tau,t-\tau\} \rightarrow \infty$.
Ratios computed at finite $t$ and $\tau$, however, will include contributions from matrix elements in excited states.
Two-point correlation functions have the spectral representation
\begin{equation}
   \begin{split}
      C_\text{2pt}(t,\vec{P}) &= \sum_n  \frac{Z^n_{\vec{P}}}{2aE^n_{\vec{P}}}   \left( e^{-E^n_{\vec{P}} t} + e^{-E^n_{\vec{P}} (T - t)} \right),
   \end{split}\label{eq:C2ptSpec}
\end{equation}
   where $n$ labels QCD energy eigenstates with the quantum numbers of a pion with momentum $\vec{P}$, $E^n_{\vec{P}}$ is the energy of state $n$, $\sqrt{Z^n_{\vec{P}}} = 2 E^n_{\vec{P}} \left<0 \right| \pi_{\vec{P}} \left| n \right>$ are overlap factors for the interpolating operator $\pi_{\vec{P}}$ onto state $n$, and thermal effects arising from the finite Euclidean time extent $T$ are included. 
Dependence of $Z^n_{\vec{P}}$ on the type of source/sink interpolating operators used (wall or momentum-smeared) is suppressed in Eq.~\eqref{eq:C2ptSpec} and throughout this section.
Fits of lattice QCD two-point correlation function results to Eq.~\eqref{eq:C2ptSpec} can be used to extract the ground-state energies $E^0_{\vec{P}}$ shown in Fig.~\ref{fig:EMP}, as well as excited-state energies and overlap factors.

Three-point functions have an analogous spectral representation
\begin{widetext}
\begin{equation}
   \begin{split}
      C_{\text{3pt}}^{\Gamma,i}(t,\tau,b^\mu,a,\eta,\vec{P})  &= \sum_{m,n} \frac{\sqrt{(Z^{m}_{\vec{P}})^* Z^{n}_{\vec{P}} } }{4 a E^m_{\vec{P}} E^n_{\vec{P}}} e^{-E^m_{\vec{P}}(t - \tau) } e^{-E^n_{\vec{P}}\tau  }  \mbraket{ m }{  \mathcal{O}^{i}_\Gamma(b^\mu,z^\mu,\eta)  }{ n } ,
   \end{split}\label{eq:C3ptSpec}
\end{equation}
where $n,m$ index energy eigenstates.
   Combing Eq.~\eqref{eq:C2ptSpec} and Eq.~\eqref{eq:C3ptSpec} and isolating the ground-state contributions yields a spectral representation for ratios of three- and two-point functions:
\begin{align}\nonumber
&\frac{ C_{\text{3pt}}^{\Gamma,i}(t,\tau,b^\mu,a,\eta,\vec{P})}
{C_\text{2pt}(t,a,\vec{P})} \times \left[  \sum_{n}\frac{Z_{\vec{P}}^n}{E^n_{\vec{P}}}\left(\frac{Z_{\vec{P}}^0}{E^0_{\vec{P}}} \right)^{-1}    \left( e^{-( E^n_{\vec{P}} - E^0_{\vec{P}}) t} + e^{-E^n_{\vec{P}} T} e^{(E^n_{\vec{P}} + E^0_{\vec{P}})t}  \right)     \right] \\ 
&\hspace{20pt} =  B^\text{bare}_\Gamma(b^z, \bt,a,\eta,P^z)  \left( 1 + \sum_{(n,m) \neq (0,0)} \frac{\sqrt{(Z^{m}_{\vec{P}})^* Z^{n}_{\vec{P}} } (E^0_{\vec{P}})^2}{ E^m_{\vec{P}} E^n_{\vec{P}}Z^0_{\vec{P}}} e^{-(E^m_{\vec{P}}-E^0_{\vec{P}})(t - \tau) } e^{-(E^n_{\vec{P}}-E^0_{\vec{P}})\tau  }  \frac{ \mbraket{ m }{  \mathcal{O}^{i}_\Gamma(b^\mu,z^\mu,\eta)  }{ n } }{ \mbraket{ 0}{  \mathcal{O}^{i}_\Gamma(b^\mu,z^\mu,\eta)  }{ 0 } } \right) .
\label{eq:CRatio}
\end{align}
\end{widetext}
After determining the spectral quantities appearing on the left-hand-side of Eq.~\eqref{eq:CRatio} from fits to lattice QCD results for $C_{\text{2pt}}$, where in practice the sum over states is truncated at $n = N_{\text{states}}$ as discussed below, the bare quasi beam functions and other parameters appearing on the right-hand-side of Eq.~\eqref{eq:CRatio} can be determined from fits to lattice QCD results for three-point to two-point function ratios.
Fitting directly to these ratios has the advantages that ground-state overlap factors cancel exactly between three- and two-point functions and that correlated ratios are determined more precisely than three-point functions alone. Including the additional factor on the left-hand-side of Eq.~\eqref{eq:CRatio}, which depends only on energies and overlaps obtained in two-point function fits, removes the need to model excited-state contamination in the two-point function during fits to the ratio (which would require fitting several additional parameters entering $\chi^2$-minimization nonlinearly) without spoiling these correlations. 

Three-point correlation functions are computed for six source/sink separations $t/a \in \{ 9,12,15,18,21,24 \}$ and all operator insertion points $0 < \tau < t$.
Signal-to-noise ratios of two-point and three-point correlation functions are proportional to $e^{-(E^0_{\vec{P}} - m_\pi)t}$, where $m_\pi$ is the pion mass, and for $n^z \geq 3$ the large-separation results are very noisy and so only results with $t/a \in \{9,12,15,18\}$ are used in fits.
Correlated $\chi^2$-minimization fits of two-point functions to Eq.~\eqref{eq:C2ptSpec}, followed subsequently by fits to Eq.~\eqref{eq:CRatio}, are performed in an automated manner as follows:
\begin{itemize}
   \item The minimum source/sink separation $t_\text{min}$ are varied over the range $2 \leq t_\text{min} \leq t_{\text{max}} - t_{\text{plateau}}$, where $t_\text{max}$ is chosen to be the largest $t$ for which the signal-to-noise ratio of $C_\text{2pt}(t,a,\vec{P})$ is greater than a fixed value (a threshold of 2 is used in the results presented here) and $t_{\text{plateau}}$ is a free parameter specifying the minimum number of points in a fit (results presented here use $t_{\text{plateau}} = 3$). The restriction $t_\text{min} \geq 2$ is set by the degree of nonlocality in the lattice action. For every possible choice of $t_{\text{min}}$ within this range, correlated $\chi^2$-minimization fits to Eq.~\eqref{eq:C2ptSpec} are performed using two-point function results with $t_\text{min} \leq t \leq t_\text{max}$.  The two-point function fitting procedure is identical to that described in Appendix B of Ref.~\cite{Beane:2020ycc}. To summarize, one-state fits are performed first, followed by two-state fits. If the Akaike information criterion (AIC)~\cite{1100705} is improved sufficiently by the addition of a second state to the fit (a threshold of $\Delta$AIC $< -2 N_\text{dof}$, where $N_\text{dof}$ is the number of degrees of freedom in the fit, is used in the final results), then a three-state fit is performed and the same criterion is used to judge whether the three-state fit is preferred. This procedure is repeated until adding additional states does not sufficiently improve the fit, in order to select the optimal number of states to include in the fit for each $t_\text{min}$. The best-fit parameters are determined using nonlinear optimization for the energies $E^n_{\vec{P}}$, with linear systems of equations solved to determine $Z^n_{\vec{P}}$ at each iteration. Covariance matrices are determined using optimal shrinkage~\cite{LEDOIT2004365} as described in Refs.~\cite{Rinaldi:2019thf,Beane:2020ycc} in order to reduce finite-statistic effects leading to poorly conditioned sample covariance matrices. Several checks on numerical $\chi^2$ optimization described in Ref.~\cite{Beane:2020ycc} are then performed to verify the reliability of the fit. If these checks are passed, an acceptable two-point function model has been found for this choice of $t_\text{min}$, and three- to two-point function ratios are subsequently analyzed using fits to Eq.~\eqref{eq:CRatio}.

   \item The minimum source/operator and source/sink separations  corresponding to $\tau \in [\tau_\text{min},\tau_\text{max}]$  are varied over the ranges $2 \leq \tau_{\text{min}} \leq (t_{\text{min}} - t_{\text{plateau}})/2$ and $2 \leq t - \tau_{\text{max}} \leq (t_{\text{min}} - t_{\text{plateau}})/2$.
      Three-point to two-point ratios using all available $t \in [t_\text{min}, t_\text{max}]$ and $\tau \in [\tau_\text{min}, \tau_\text{max}]$ are multiplied by the factor in brackets on the left-hand-side of Eq.~\eqref{eq:CRatio}. A correlated $\chi^2$-minimization fit is performed to extract the parameters on the right-hand-side of Eq.~\eqref{eq:CRatio} using the same methods applied to two-point functions. The excited-state matrix elements appearing in Eq.~\eqref{eq:CRatio} enter the $\chi^2$ function linearly, and their optimal values are determined by solving a linear system of equations at each step of iterative nonlinear optimization for the energies appearing in Eq.~\eqref{eq:CRatio} as done in variable projection methods~\cite{varpro0,varpro1}. Because the low-lying spectrum is imperfectly modeled by few-state fits, the energies extracted from fits to Eq.~\eqref{eq:CRatio} are not constrained to identically equal the energies extracted from fits to Eq.~\eqref{eq:C2ptSpec} (although the spectrum determined from Eq.~\eqref{eq:C2ptSpec} is used to provide initial conditions for nonlinear optimization). Optimal shrinkage is used to define the covariance matrix. The best-fit ground-state matrix element defines $(B^\text{bare}_\Gamma)^f$ for fit range choice $f$ defined by $t_\text{min}$, $\tau_\text{min}$ and $\tau_\text{max}$. Fits where two solvers disagree on $(B^\text{bare}_\Gamma)^f$ by more than a specified tolerance ($10^{-5}$ is used in final results) are discarded in analogy to the reliability checks applied to fits to two-point functions~\cite{Beane:2020ycc}.
   \item Confidence intervals for ground-state matrix elements and other fit parameters are determined using bootstrap resampling; see Refs.~\cite{davison_hinkley_1997,Young} for reviews. Fits to Eq.~\eqref{eq:CRatio} are repeated $N_\text{boot}$ times ($N_\text{boot} =200$ is used in final results) using ensembles constructed by randomly resampling with replacement from the two- and three-point functions in a correlated manner. Statistical uncertainties on fit parameters are obtained from empirical confidence intervals of bootstrap fit results as detailed in Ref.~\cite{Beane:2020ycc}. The $68\%$ confidence interval defines $(\delta B^\text{bare}_\Gamma)^f$. Further reliability checks are applied: the median of the bootstrap distribution is verified to be within a specified tolerance of $(B^\text{bare}_\Gamma)^f$ (2$\sigma$ is used in final results), and uncorrelated fit results are verified to be within a specified tolerance of $(B^\text{bare}_\Gamma)^f$ (5$\sigma$ is used in the final results). Fits passing all reliability checks define an ensemble of $f=\{1,\ldots, N_\text{success}\}$ successful fit range choices.
   \item A weighted average of ground-state matrix element results from all successful fits is used to determine the final results. Each fit result $(B^\text{bare}_\Gamma)^f$ provides an unbiased estimate of the bare quasi beam function (in the infinite-statistics limit), and the relative weights between successful fits are arbitrary in the large statistics limit. Given a finite statistical ensemble, it is advantageous to define a weighted average that penalizes fits with worse goodness-of-fit and fits with larger uncertainties. The weighted average procedure used in Refs.~\cite{Rinaldi:2019thf,Beane:2020ycc} is followed: averages are defined by
\begin{subequations}
   \begin{align}
      B^\text{bare}_\Gamma &\equiv \sum_{f=1}^{N_{\text{success}}} w^f (B^\text{bare}_\Gamma)^f,\\
      (\delta_\text{stat} B^\text{bare}_\Gamma)^2 &\equiv \sum_{f=1}^{N_{\text{success}}} w^f  \left( (\delta B^\text{bare}_\Gamma)^f \right) ^2,\\
      (\delta_\text{sys} B^\text{bare}_\Gamma)^2 &\equiv \sum_{f=1}^{N_{\text{success}}} w^f \left( (B^\text{bare}_\Gamma)^f - B^\text{bare}_\Gamma \right)^2,\\
      (\delta B^\text{bare}_\Gamma)^2 &\equiv \sqrt{(\delta_{\text{stat}} B^\text{bare}_\Gamma)^2 + (\delta_{\text{sys}} B^\text{bare}_\Gamma)^2 },\label{eq:unc}
   \end{align}\label{eq:weightedave}
\end{subequations}
where the weights $w^f$ are defined as
\begin{subequations}
   \begin{align}
      w^f &\equiv \frac{\tilde{w}^f}{\sum_{f=1}^{N_{\text{success}}} \tilde{w}^f }, \\
      \widetilde{w}^f &\equiv \frac{p_f \left( (\delta B^\text{bare}_\Gamma)^f \right)^{-2} }{ \sum_{f^\prime = 1}^{N_{\text{success}}} p_{f^\prime} \left(  (\delta B^\text{bare}_\Gamma)^f \right)^{-2}  }.
 \end{align}\label{eq:weights}
\end{subequations}
      Here, $p_f = \Gamma(N_\text{dof}/2, \chi_f^2/2)/\Gamma(N_\text{dof}/2)$  where $\Gamma$ is the gamma function (not to be confused with the Dirac spinor index elsewhere in this work) in order to penalize fits with large $\chi^2/N_\text{dof}$ and large $(\delta B^\text{bare}_\Gamma)^f$. See Refs.~\cite{Rinaldi:2019thf,Beane:2020ycc} for further discussion of this procedure.
\end{itemize}

\begin{figure*}[!p]
    \centering
        \includegraphics[width=0.46\textwidth]{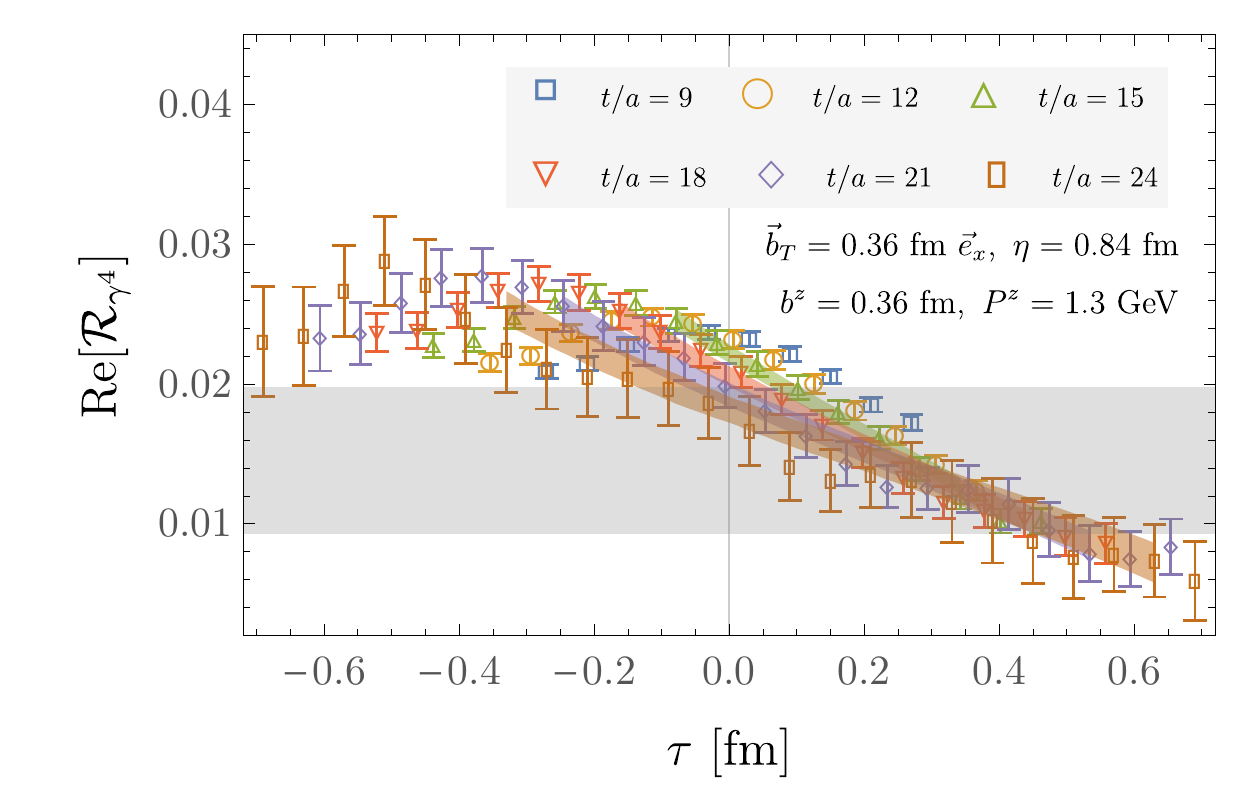} \hspace{20pt}
        \includegraphics[width=0.46\textwidth]{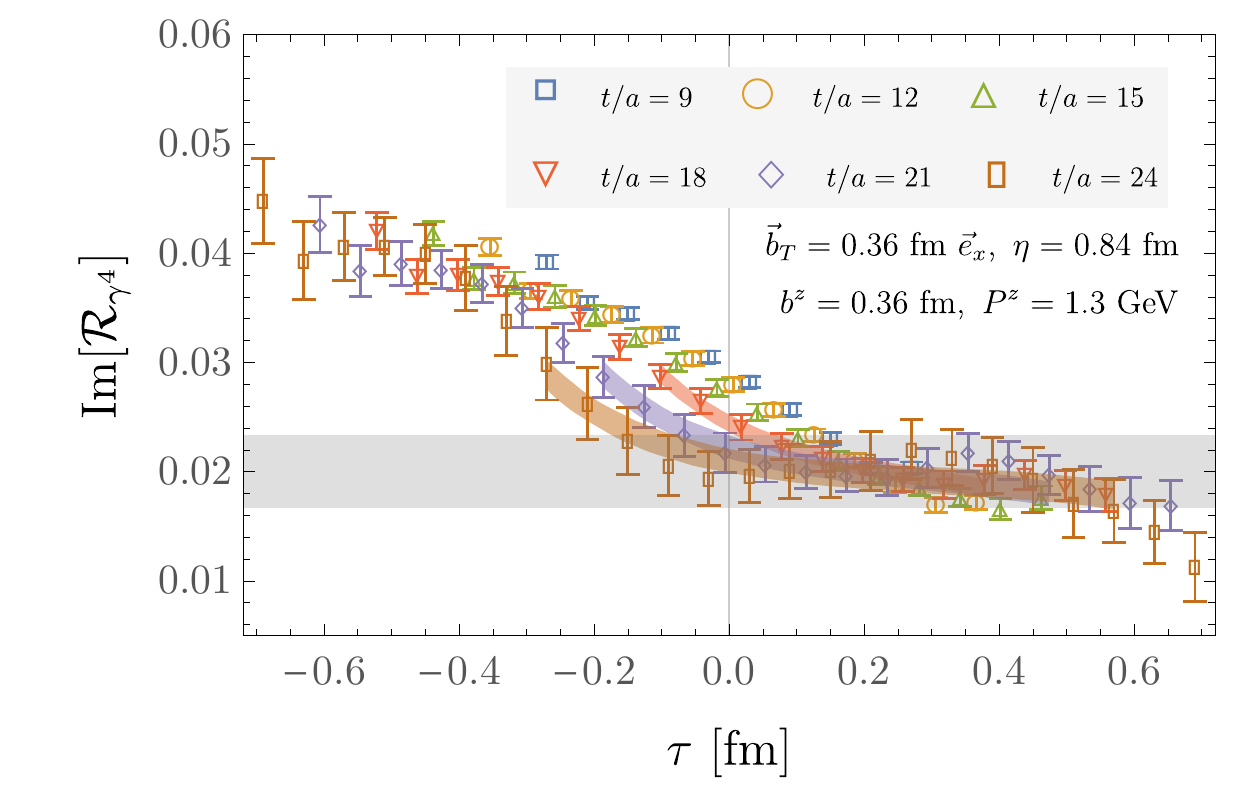}
        \includegraphics[width=0.46\textwidth]{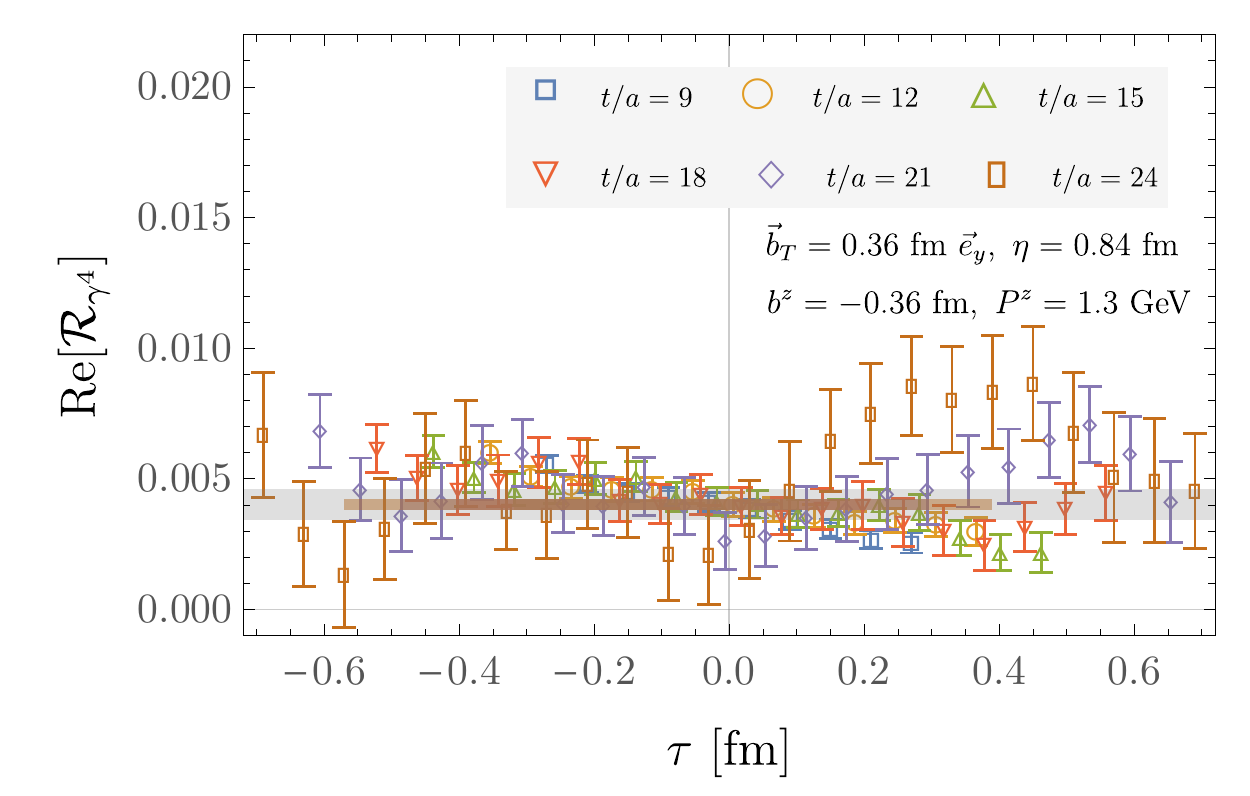} \hspace{20pt}
        \includegraphics[width=0.46\textwidth]{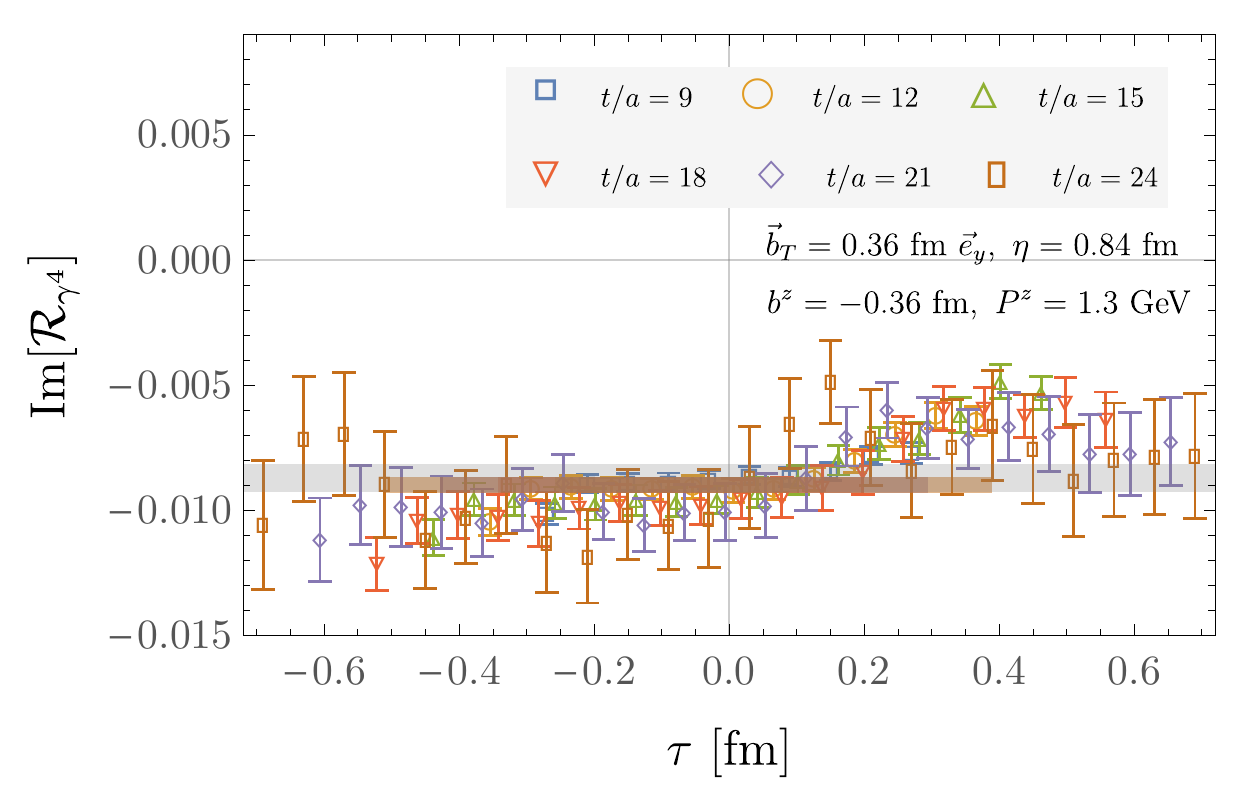}
        \includegraphics[width=0.46\textwidth]{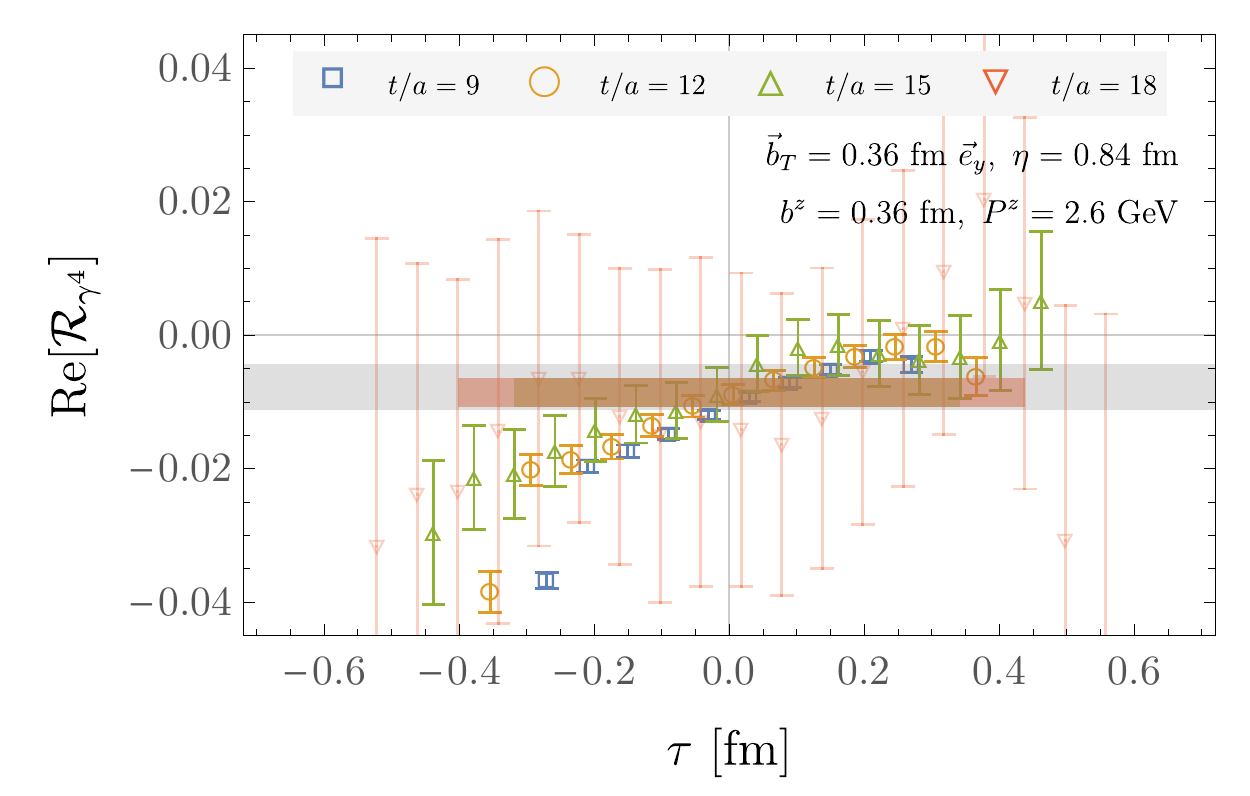} \hspace{20pt}
        \includegraphics[width=0.46\textwidth]{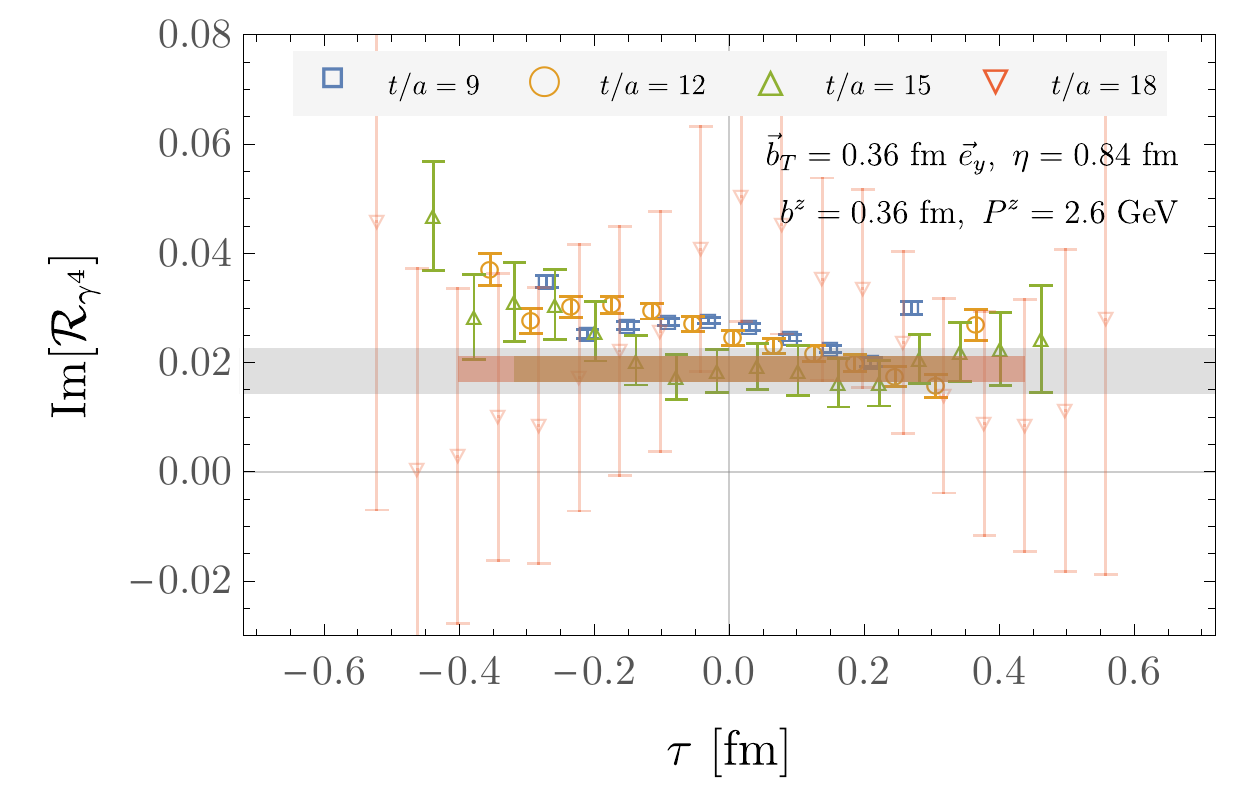}
        \includegraphics[width=0.46\textwidth]{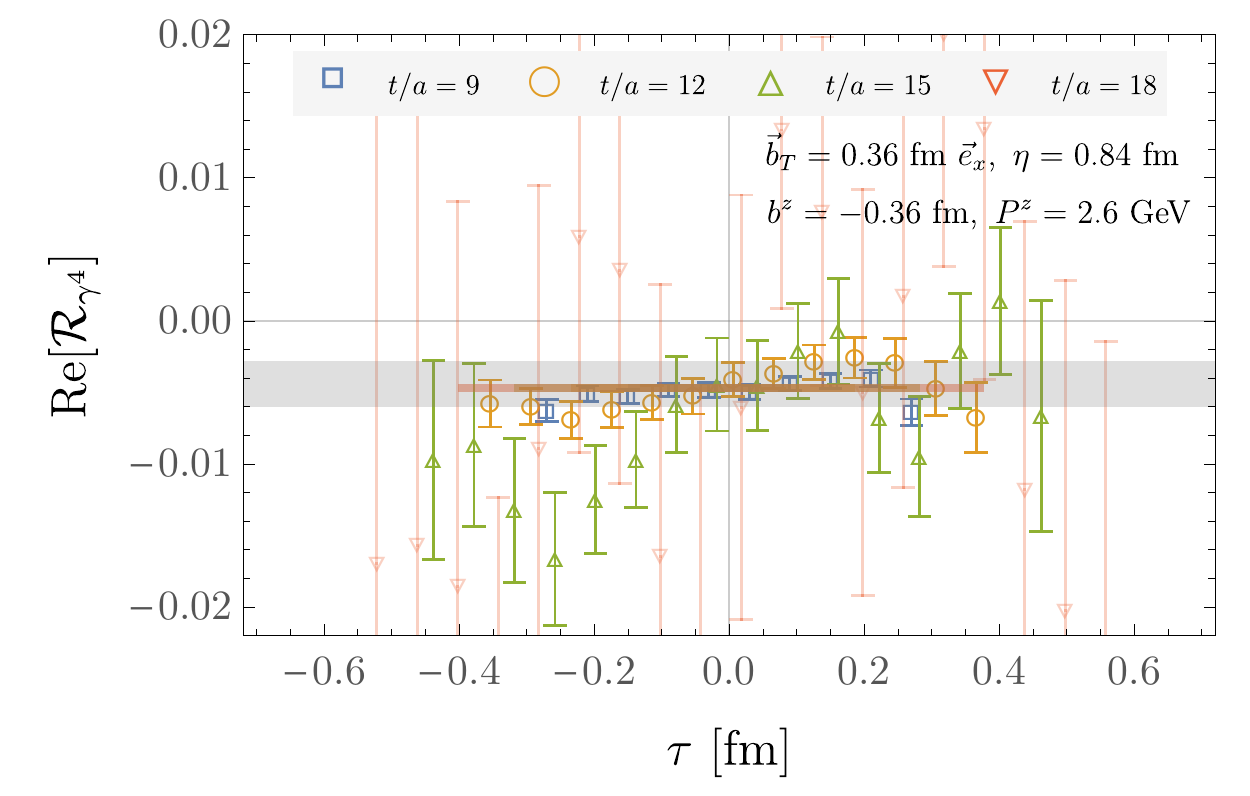} \hspace{20pt}
        \includegraphics[width=0.46\textwidth]{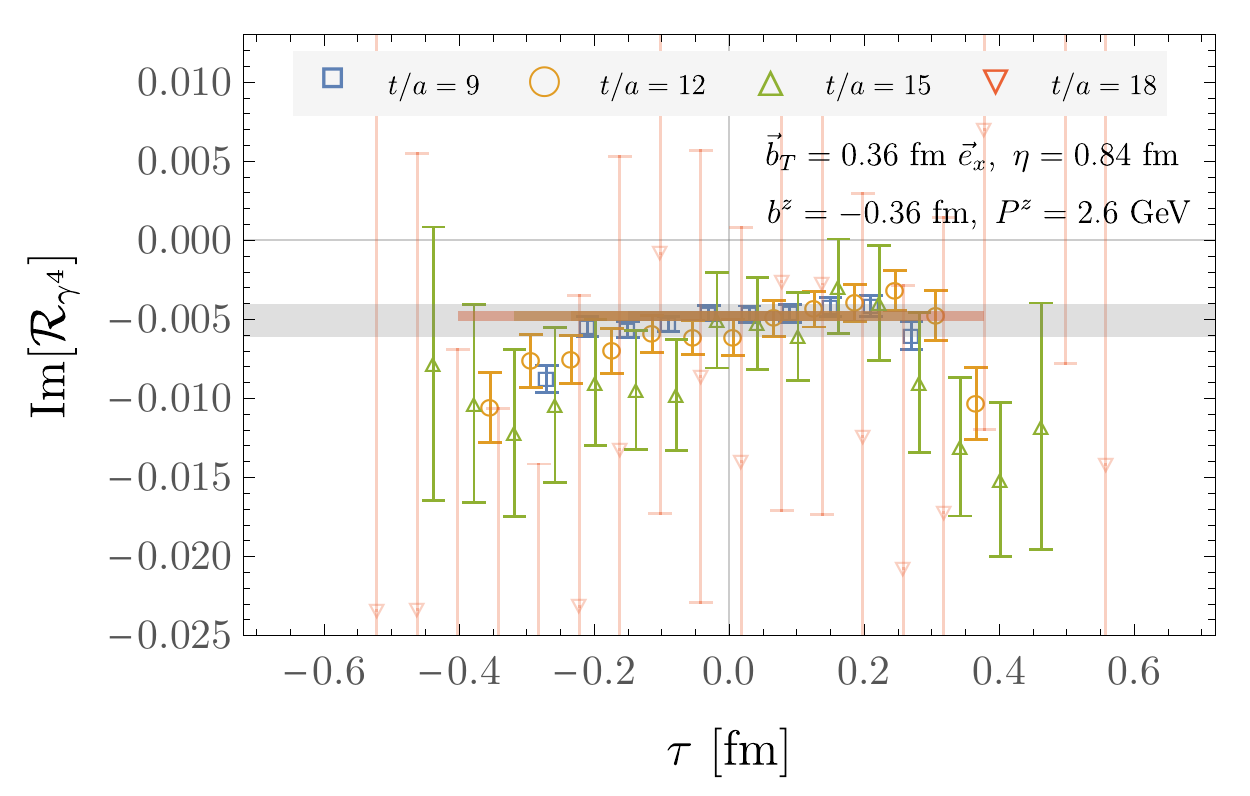}
    \caption{\label{fig:ratio} Examples of fits to the ratio of three- and two-point functions $\mathcal{R}_\Gamma(t,\tau,b^\mu,a,\eta,P^z)$ (Eq.~\eqref{eq:Rcal}), obtained as described in the text. Shaded bands of the same colors as the points show $68\%$ bootstrap confidence intervals of the $\tau$ and $t$-dependent fits from the fit range (specifically the choice of $t_\text{min}$, $\tau_\text{min}$, and $\tau_\text{max}$) that had the highest weight in the weighted average of successful fits. Gray horizontal bands show the total uncertainty on the bare quasi beam functions extracted from the fits, including the statistical uncertainty and the systematic uncertainty from variation of the results between different fit range choices. }
\end{figure*}

The resulting average values and uncertainties computed as in Eq.~\eqref{eq:unc} define the bare quasi beam functions $B^\text{bare}_\Gamma(b^z, \bt,a,\eta,P^z)$ and $\delta B^\text{bare}_\Gamma(b^z, \bt,a,\eta,P^z)$ used in this work.
A representative sample of three- and two-point function ratio fit results are shown in Fig.~\ref{fig:ratio} for the smallest and largest momenta used in this work and for both small and large Wilson-line extents among the set studied here.

\begin{figure*}[!p]
    \centering
        \includegraphics[width=0.46\textwidth]{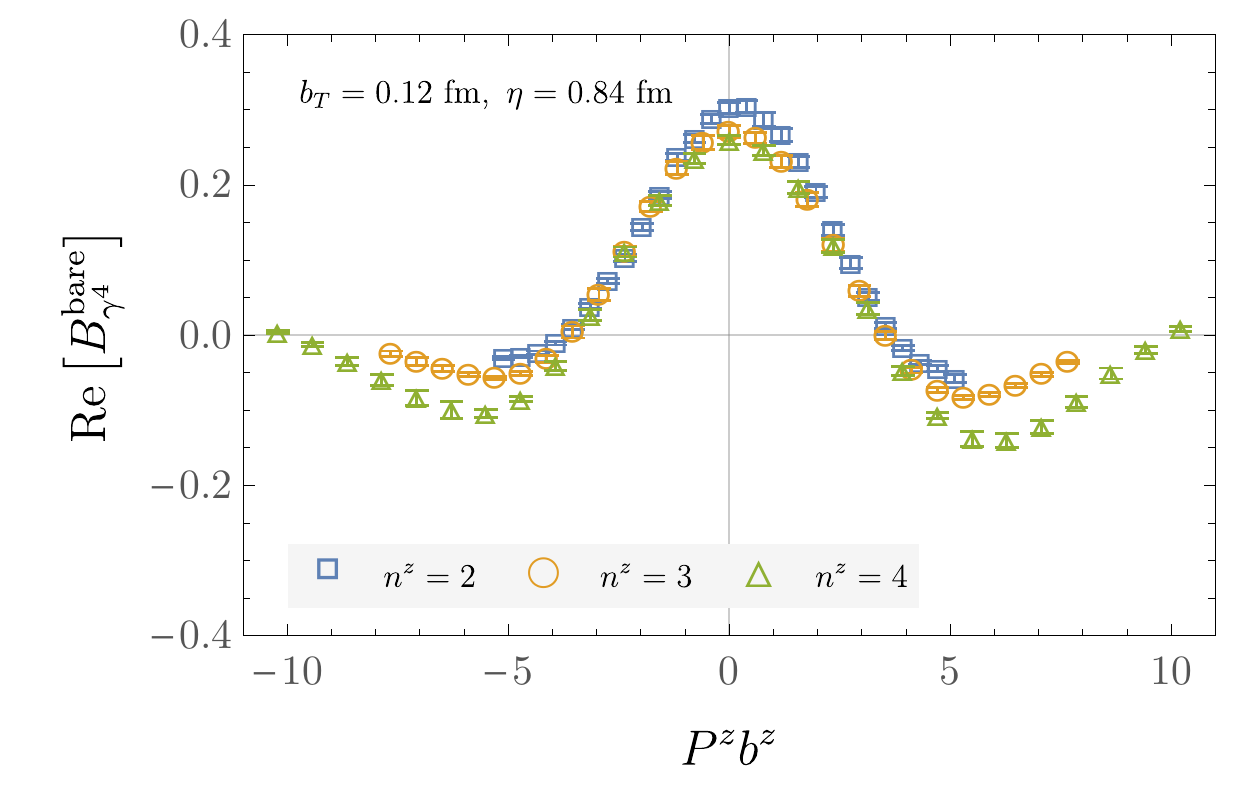} \hspace{20pt}
        \includegraphics[width=0.46\textwidth]{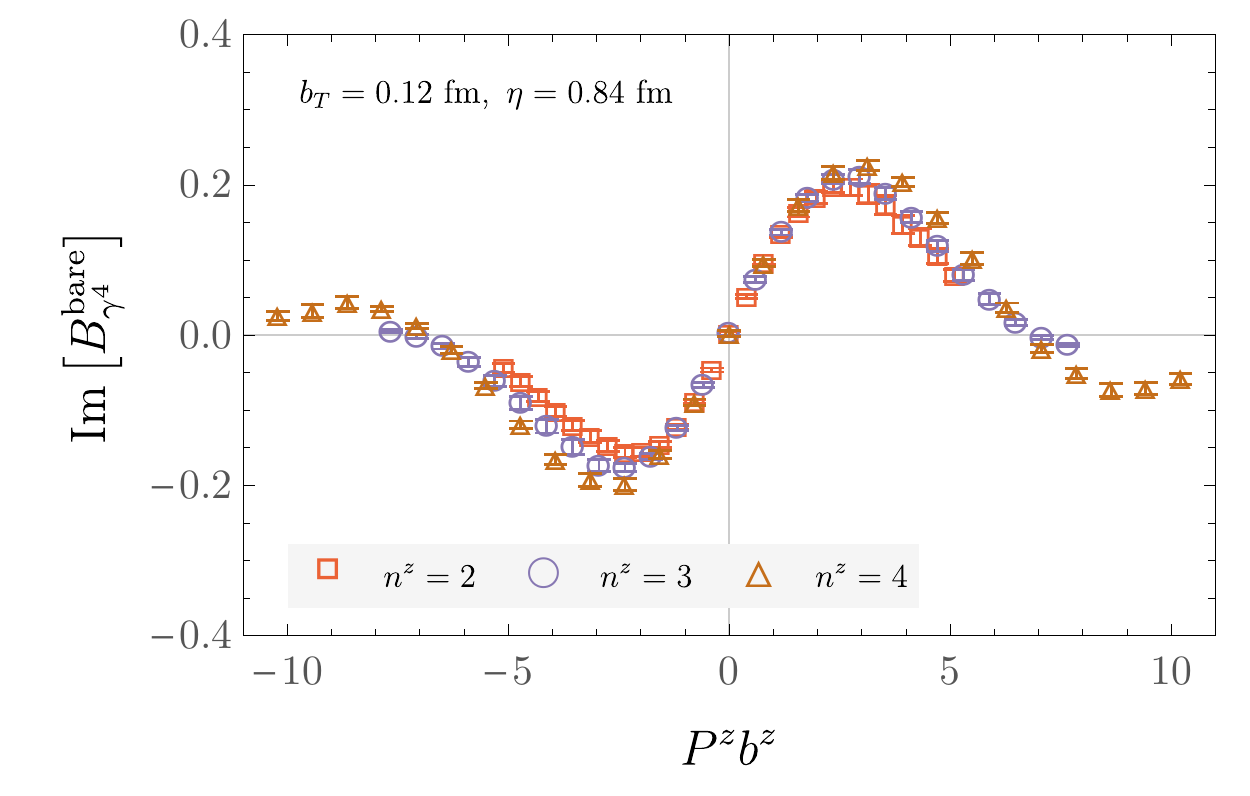}
        \includegraphics[width=0.46\textwidth]{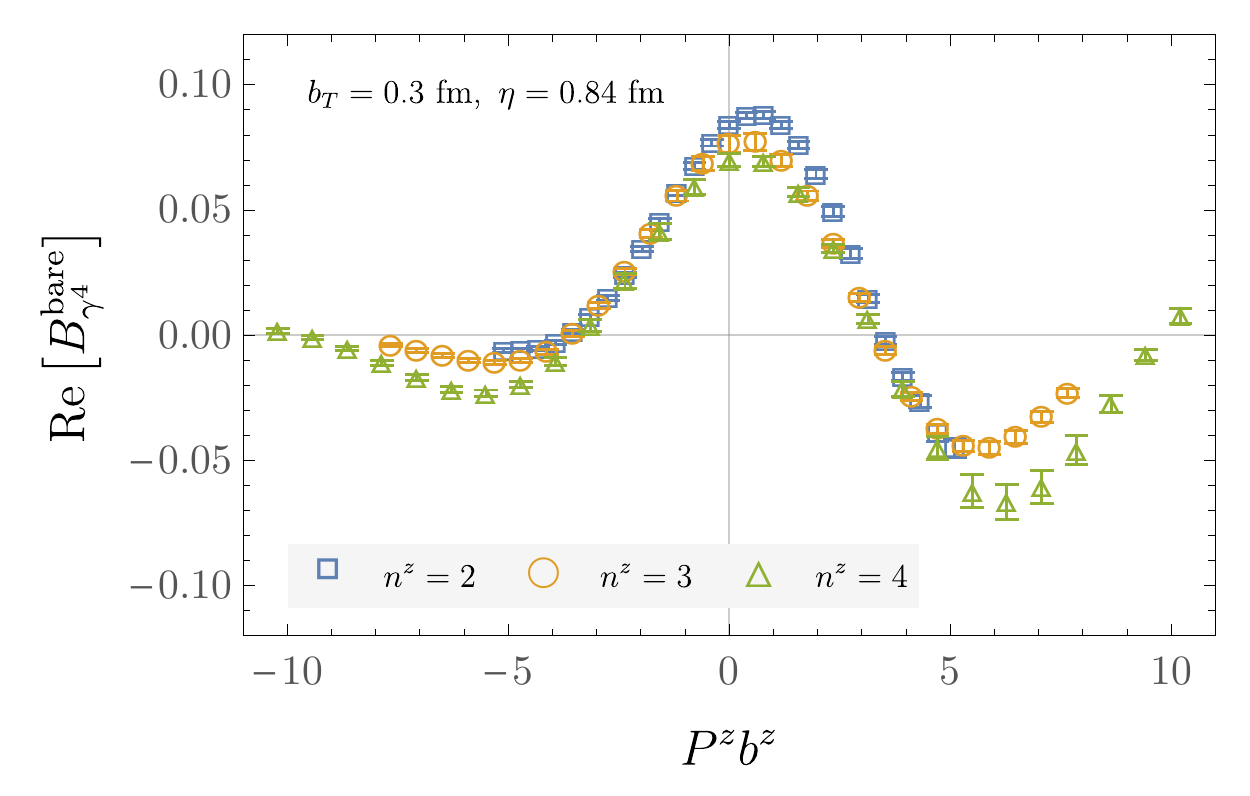} \hspace{20pt}
        \includegraphics[width=0.46\textwidth]{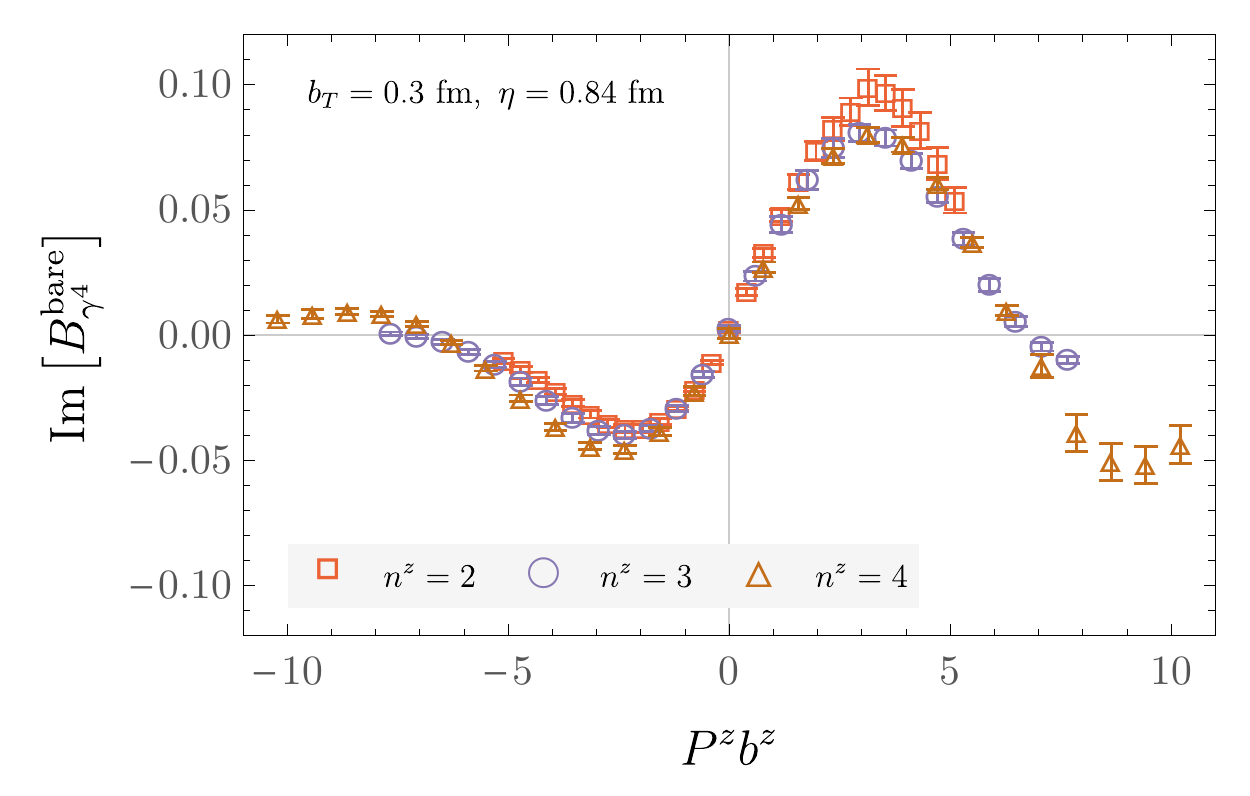}
        \includegraphics[width=0.46\textwidth]{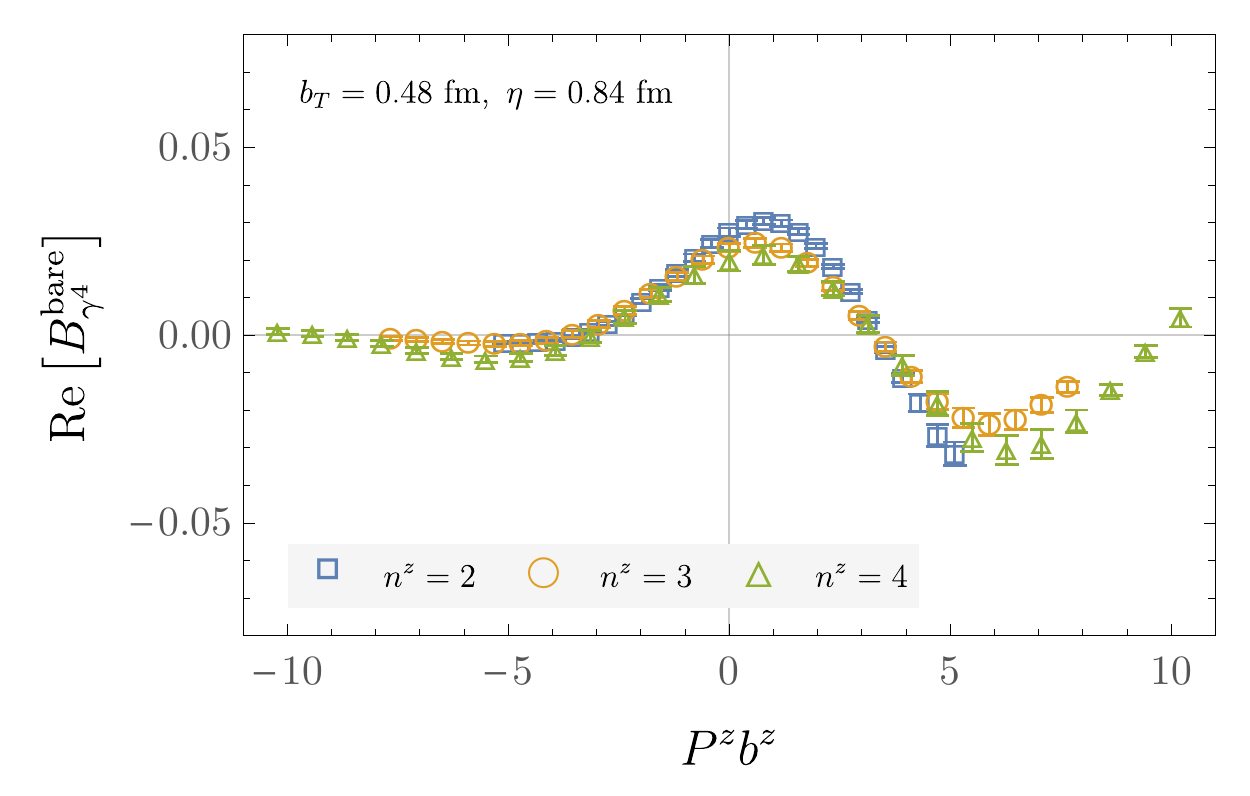} \hspace{20pt}
        \includegraphics[width=0.46\textwidth]{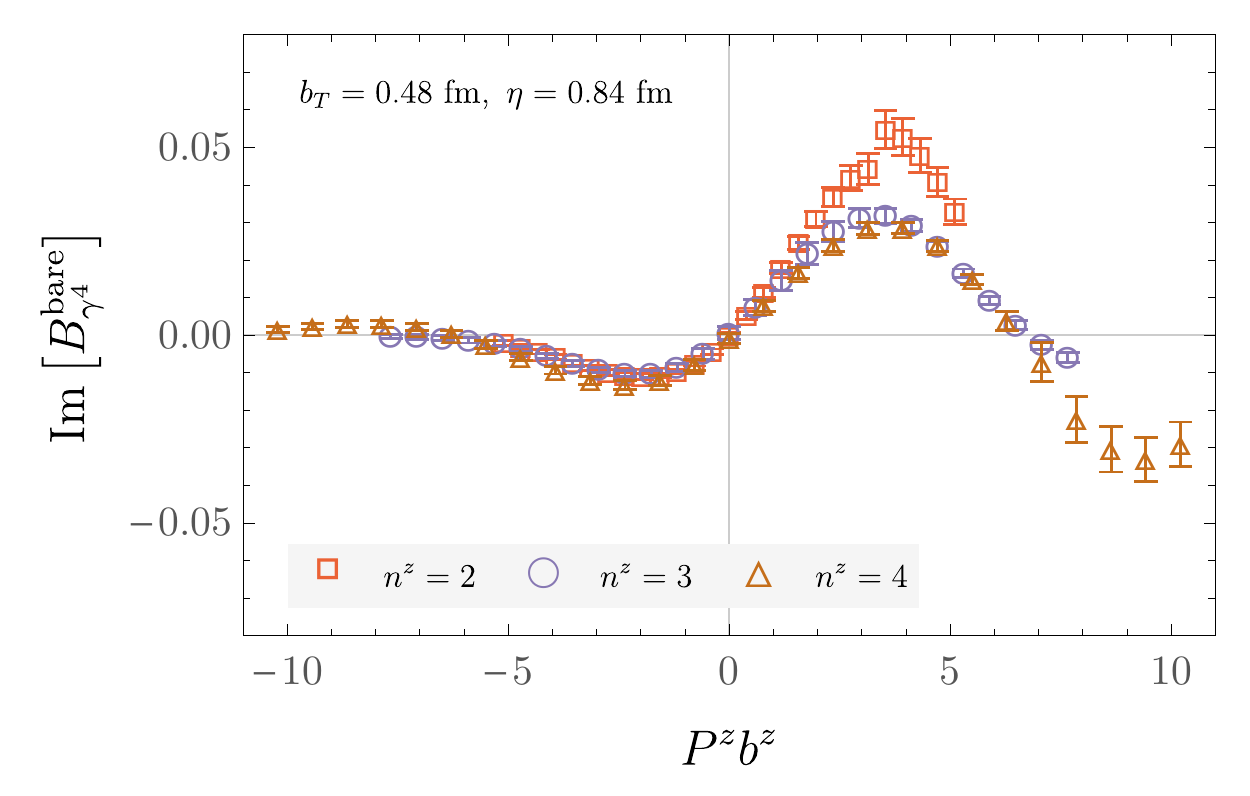}
        \includegraphics[width=0.46\textwidth]{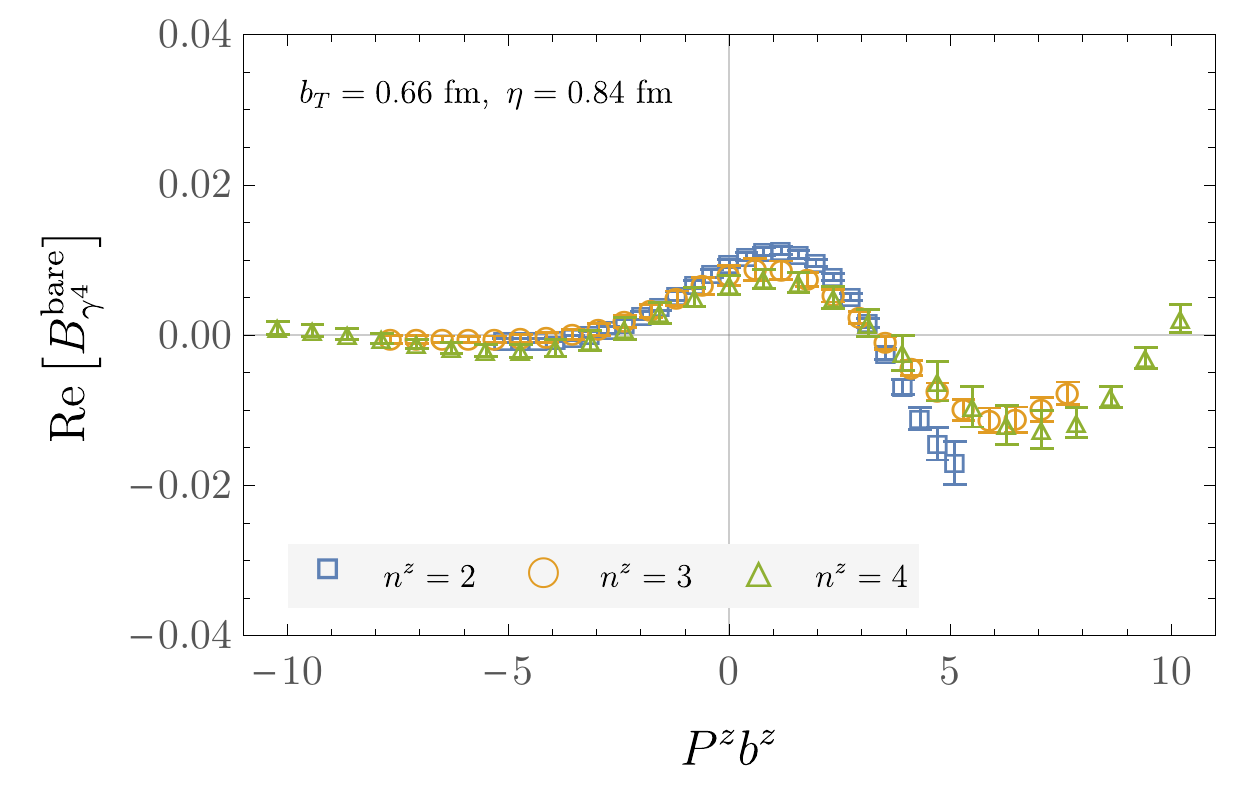} \hspace{20pt}
        \includegraphics[width=0.46\textwidth]{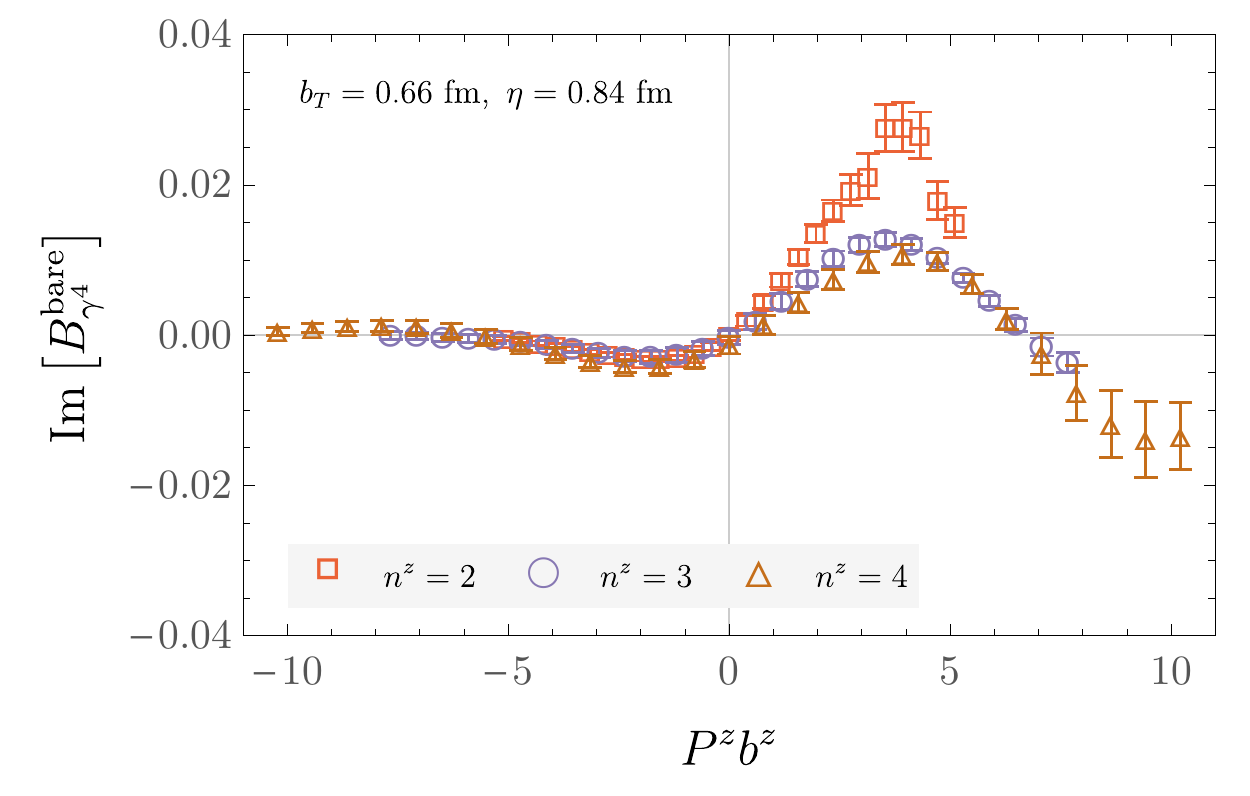}
    \caption{\label{fig:beam} Bare quasi beam functions $B^\text{bare}_\Gamma(b^z, \bt=b_T \vec{e}_{x},a,\eta,P^z=n^z 2\pi /L)$, defined in Eq.~\eqref{eq:qbeambare}, for various parameter choices. }
\end{figure*}

\section{Bare quasi beam functions}
\label{app:barebeams}

Additional examples of the real and imaginary parts of the extracted bare quasi beam functions $B^\text{bare}_\Gamma(b^z, \bt,a,\eta,P^z)$, defined in Eq.~\eqref{eq:qbeambare}, and determined as described in Sec.~\ref{sec:LQCD}, are shown for various parameter choices in Fig.~\ref{fig:beam}. A general trend can be observed that at increasing $b_T$ both the real and imaginary parts of the functions become more asymmetric in $P^zb^z$.
This asymmetry arises primarily from linear divergences in the bare quasi beam function that are canceled when the renormalization factors  $Z_{\mathcal{O}_\gamma^4\Gamma}^{\MS}$ are included, as can be seen by comparing Fig.~\ref{fig:beam} and Fig.~\ref{fig:MSbeam}. 
As discussed in Sec.~\ref{sec:CS}, residual $b^z$ asymmetries in $B^{\MS}_{\gamma^4}$ visible in Fig.~\ref{fig:MSbeam} could arise from  finite-volume effects coupled with imperfect cancellation of exponential $b^z$ dependence between the bare quasi beam functions and the renormalization factors.

\section{Renormalized beam functions}
\label{app:renbeams}

The renormalized quasi beam function is computed by combining the bare quasi beam function determined from three- to two-point function ratios as described in Appendix~\ref{app:threetwofits} with the renormalization factors computed in Ref.~\cite{Shanahan:2019zcq}, as shown in Eq.~\eqref{eq:BMSbar}.
The uncertainty on the renormalized quasi beam function is obtained by combining the total uncertainties of the bare quasi beam function and the renormalization factors in quadrature. 
Results are computed using two different staple orientations corresponding to $\bt = b_T \vec{e}_{x}$ and $\bt = b_T \vec{e}_{y}$.
Interchanging these orientations $\vec{e}_x \leftrightarrow \vec{e}_y$ is an exact symmetry of $B^{\MS}_{\gamma^4}$ but is not a symmetry of $B^{\text{bare}}_\Gamma$ for some $\Gamma$, and thus results with both orientations can be averaged only after renormalization.
A weighted average~\cite{Aoki:2019cca} is used to combine the $B^{\MS}_{\gamma^4}$ results with $\bt = b_T \vec{e}_{x}$ and $\bt = b_T \vec{e}_{y}$ by

\begin{figure*}
    \centering
        \includegraphics[width=0.46\textwidth]{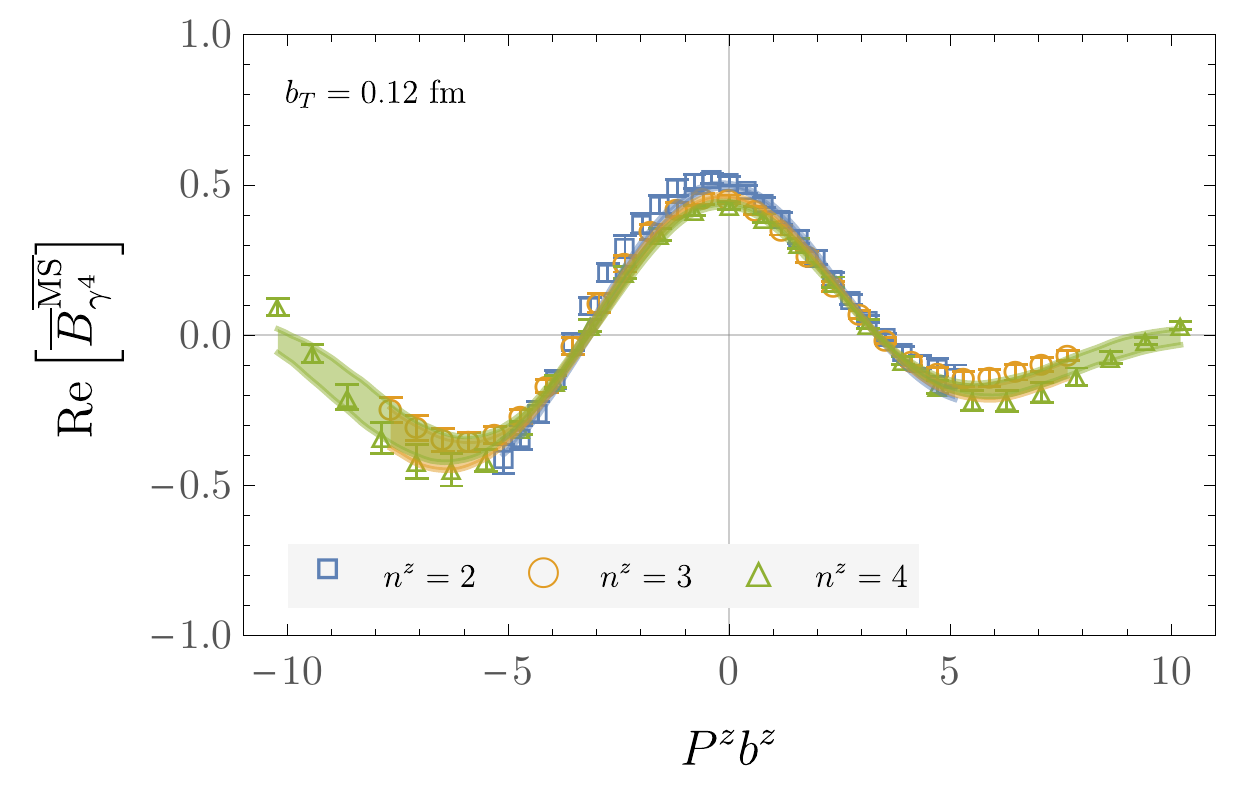} \hspace{20pt}
        \includegraphics[width=0.46\textwidth]{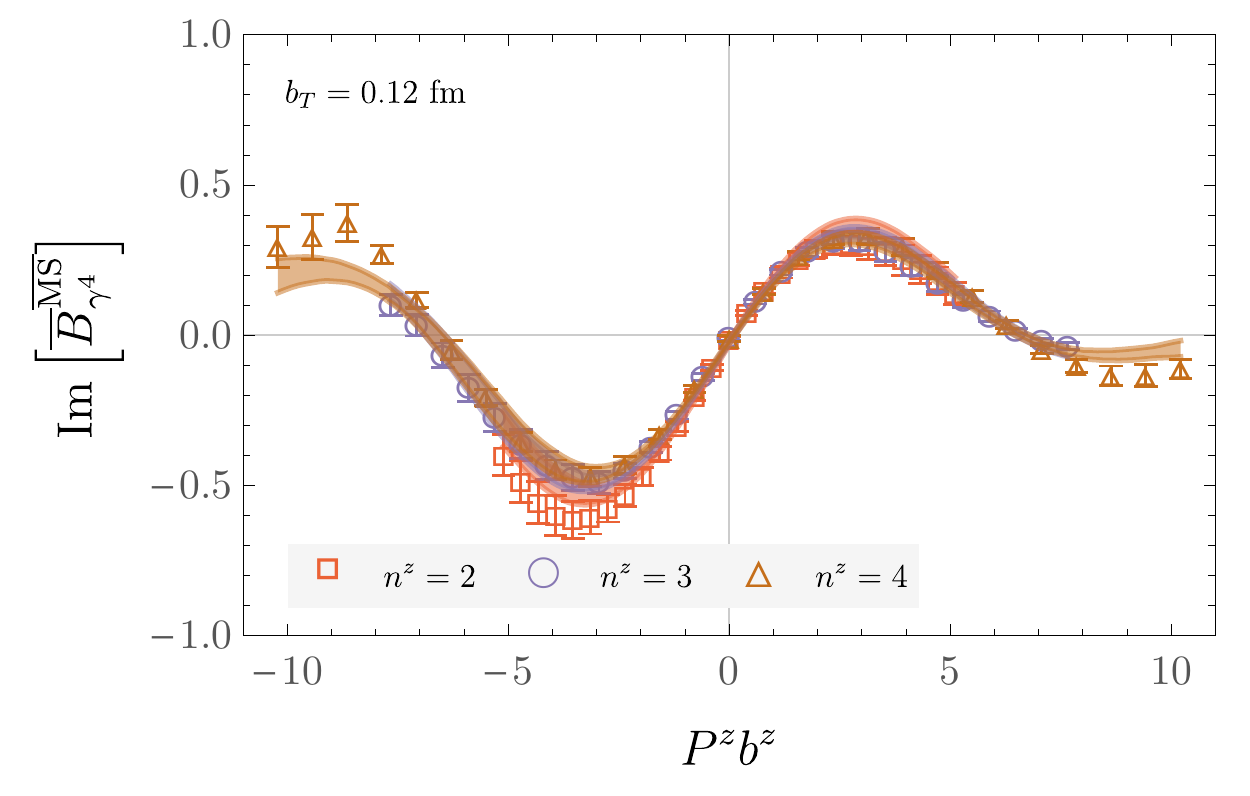}
        \includegraphics[width=0.46\textwidth]{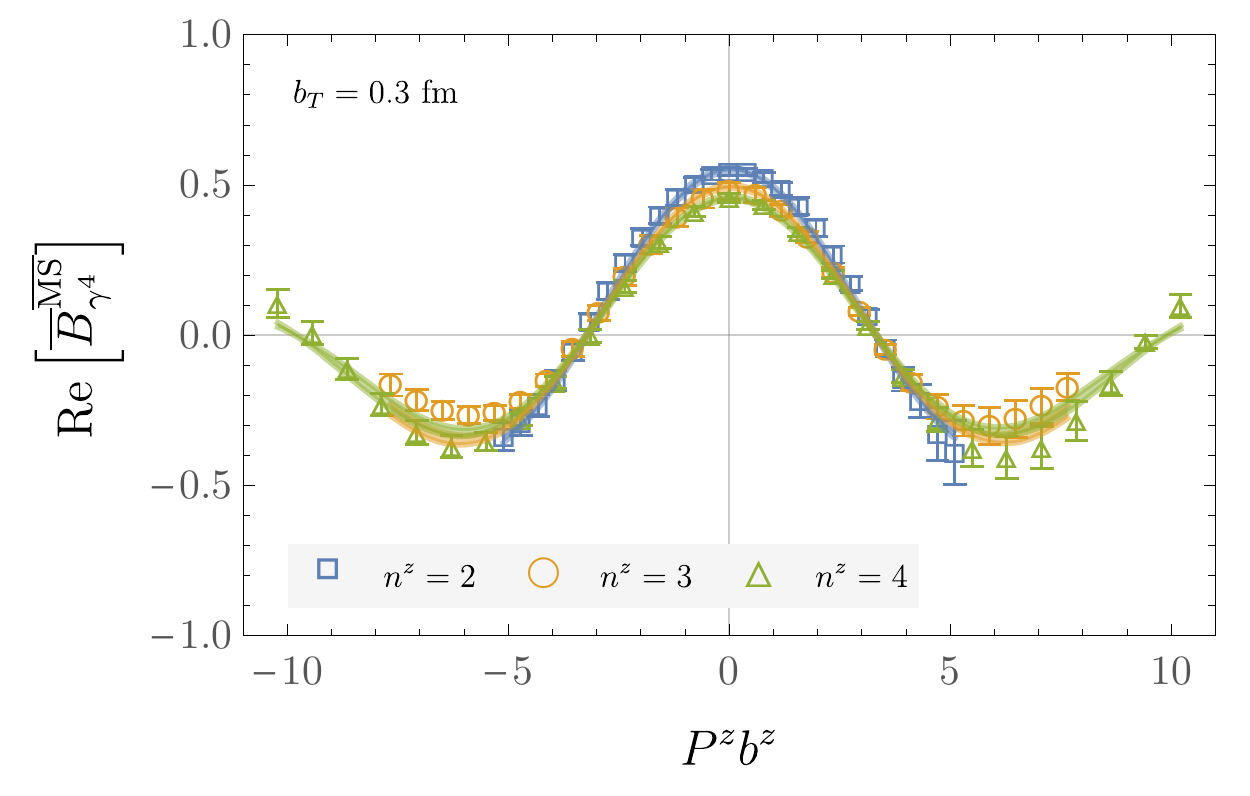} \hspace{20pt}
        \includegraphics[width=0.46\textwidth]{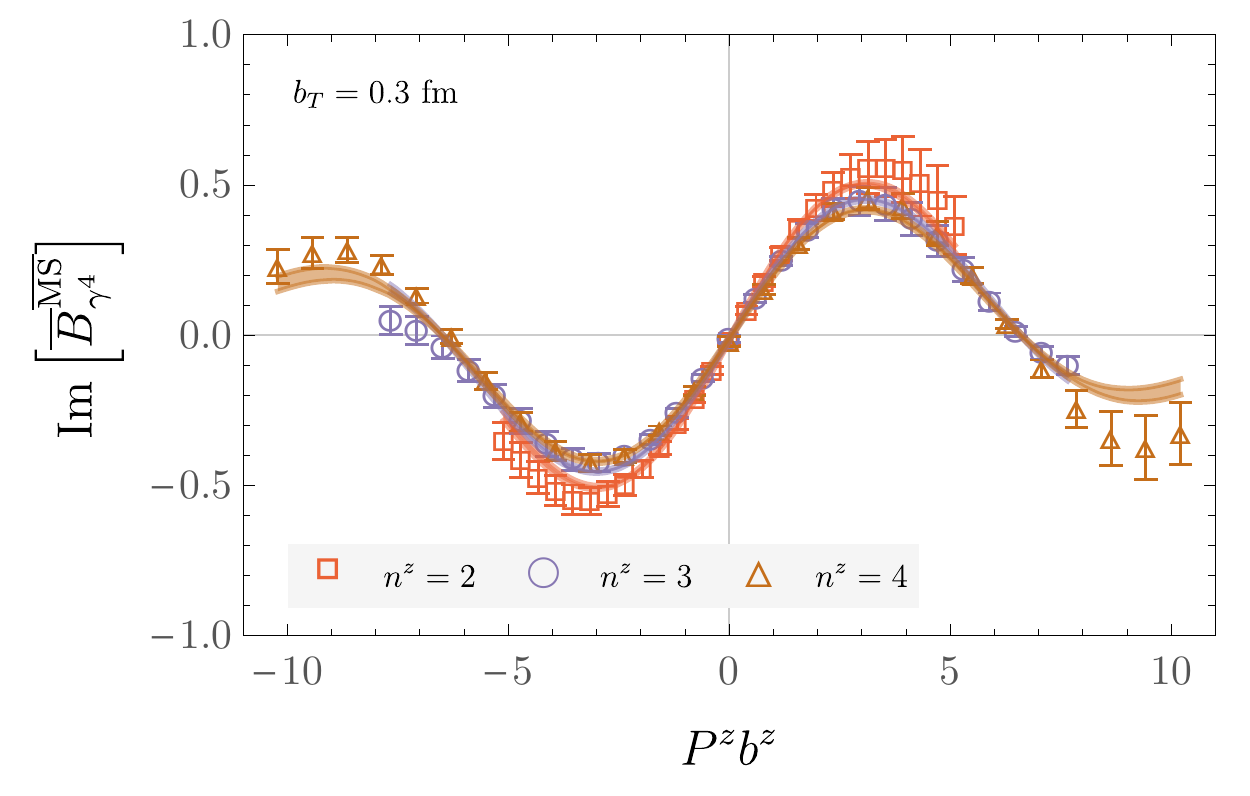}
        \includegraphics[width=0.46\textwidth]{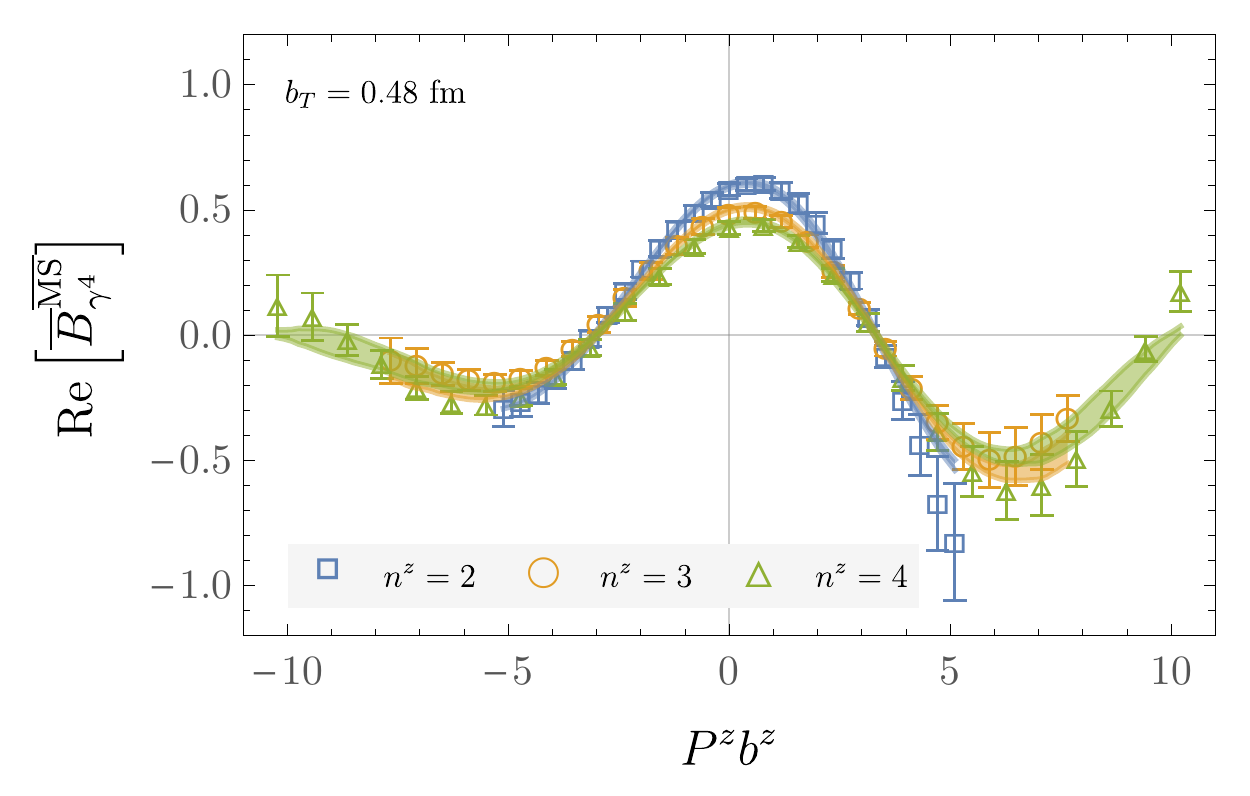} \hspace{20pt}
        \includegraphics[width=0.46\textwidth]{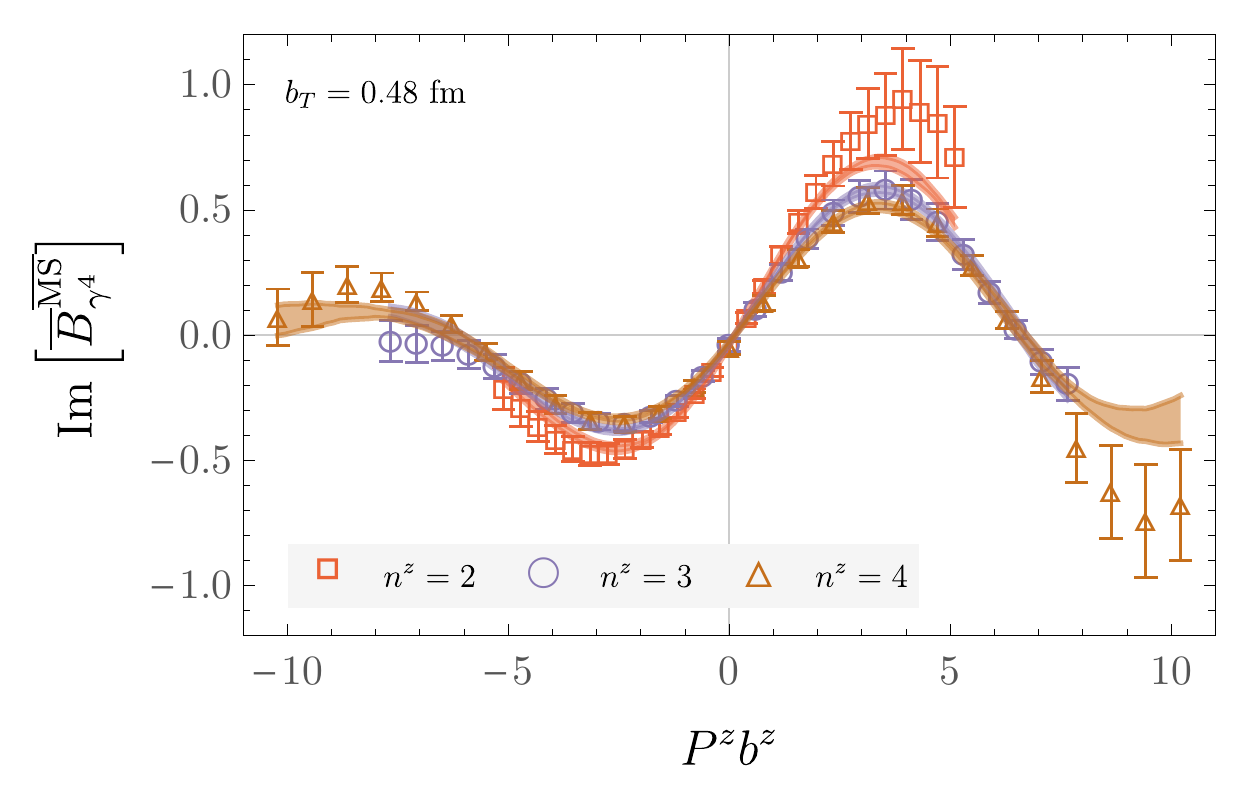}
        \includegraphics[width=0.46\textwidth]{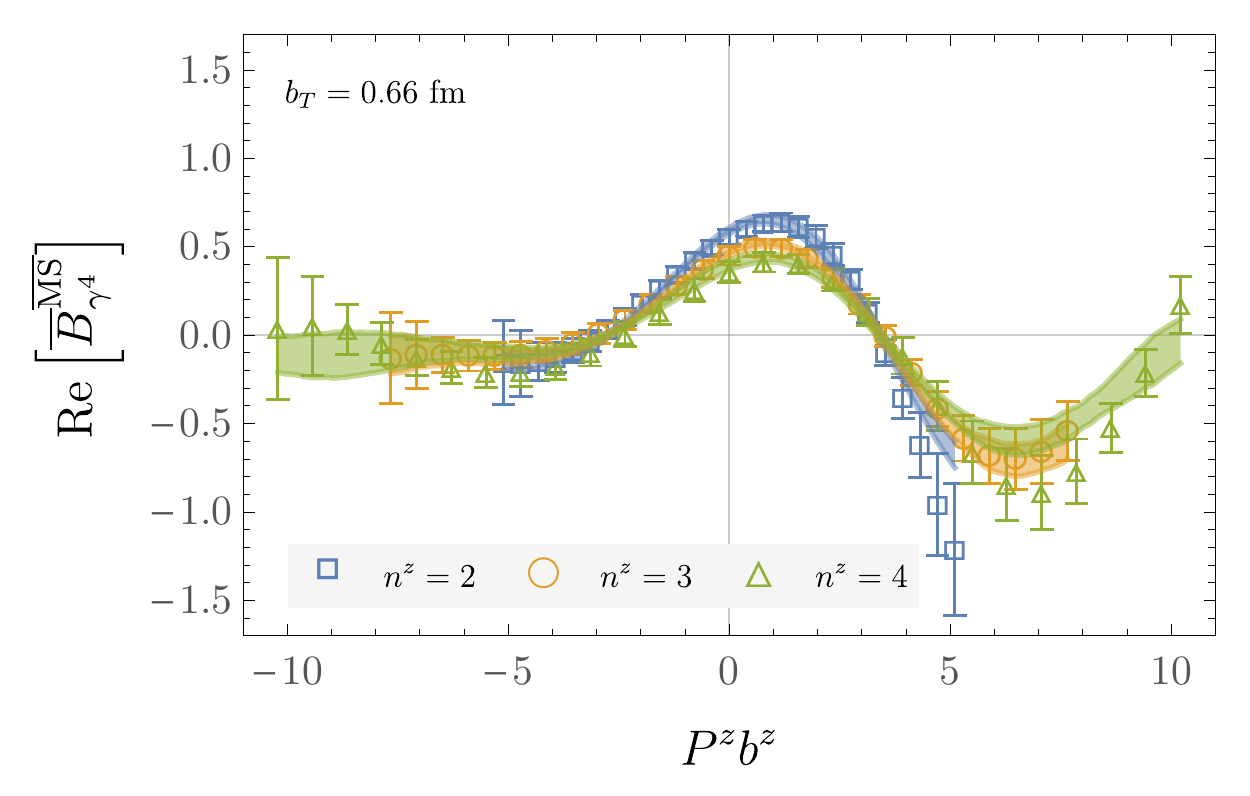} \hspace{20pt}
        \includegraphics[width=0.46\textwidth]{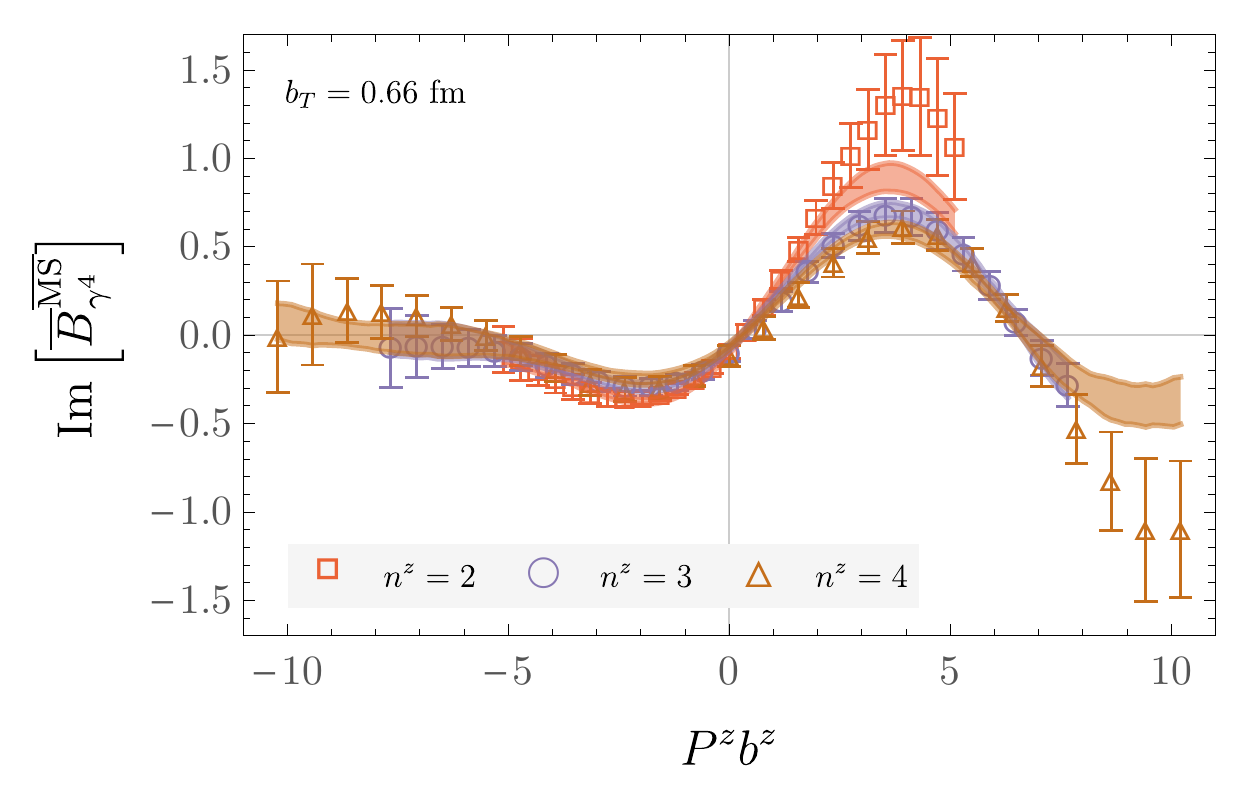}
        \caption{\label{fig:MSbeam} Averaged renormalized quasi beam function $\overline{B}^{\MS}_{\gamma^4}(b^z, b_T, a, P^z = n^z2\pi / L)$, for various parameter choices. The bands show fits to a functional form based on Bernstein polynomials (Eq.~\eqref{eq:bern}), as described in the text.  }
\end{figure*}

\begin{widetext}
\begin{equation}
   \begin{split}
      B^{\prime\; \MS}_{\gamma^4}(\mu,b^z, b_T, a, \eta, b_T^R, P^z) &\equiv \sum_{k=1}^2 w_{k} \; B^{\MS}_{\gamma^4}(\mu,b^z, b_T \vec{e}_k, a, \eta, b_T^R , P^z) \,, \\
      \delta_{\text{stat}} B^{\prime\; \MS}_{\gamma^4}(\mu,b^z, b_T, a, \eta, b_T^R, P^z)^2 &\equiv \sum_{k=1}^2 w_{k} \; \delta B^{\MS}_{\gamma^4}(\mu,b^z, b_T \vec{e}_k, a, \eta, b_T^R , P^z)^2 \,, \\
      \delta_{\text{sys}} B^{\prime\; \MS}_{\gamma^4}(\mu,b^z, b_T, a, \eta, b_T^R, P^z)^2 &\equiv \sum_{k=1}^2 w_{k} \; \left[ B^{\prime\; \MS}_{\gamma^4}(\mu,b^z, b_T, a, \eta, b_T^R, P^z) -  B^{\MS}_{\gamma^4}(\mu,b^z, b_T \vec{e}_k, a, \eta, b_T^R , P^z) \right]^2 \,, \\
      \delta B^{\prime\; \MS}_{\gamma^4}(\mu,b^z, b_T, a, \eta, b_T^R, P^z)^2 &\equiv \delta_{\text{stat}} B^{\prime\; \MS}_{\gamma^4}(\mu,b^z, b_T, a, \eta, b_T^R, P^z)^2 + \delta_{\text{sys}} B^{\prime\; \MS}_{\gamma^4}(\mu,b^z, b_T, a, \eta, \vec{b}_T^R, P^z)^2,
   \end{split}\label{eq:Bprime}
\end{equation}
where the weights are chosen to sum to unity and to be proportional to the inverse variance of each result:
\begin{equation}
   w_k \equiv \frac{\tilde{w}_k}{\sum_{k=1}^2 \tilde{w}_k}, \hspace{5mm}
      \tilde{w}_k \equiv \frac{1}{\delta B^{\MS}_{\gamma^4}(\mu,b^z, b_T \vec{e}_k, a, \eta, \vec{b}_T^R , P^z)^2}.
\label{eq:kweights}
\end{equation}

As shown for one example in Fig.~\ref{fig:renorm_vs_bTR}, the renormalized quasi beam functions do not depend on $\eta$ or $b_T^R$ within uncertainties. The formal extrapolation to $\eta \rightarrow \infty$, and an average over possible choices of $b_T^R$ in the window $a \ll b_T^R \ll \Lambda_{QCD}^{-1}$, are thus implemented with an analogous weighted average:
\begin{equation}
   \begin{split}
      \overline{B}^{\MS}_{\gamma^4}(\mu, b^z, b_T, a, P^z) &\equiv \sum_{\eta/a\in\{10,12,14\}} \sum_{(b_T^R/a)=2}^5   w_{\eta,b_T^R} \; B^{\prime\; \MS}_{\gamma^4}(\mu,b^z, b_T, a, \eta, b_T^R , P^z) \,, \\
      \delta_{\text{stat}} \overline{B}^{\MS}_{\gamma^4}(\mu,b^z, b_T, a, P^z)^2 &\equiv \sum_{\eta/a\in\{10,12,14\}} \sum_{(b_T^R/a)=2}^5   w_{\eta,b_T^R}  \; \delta B^{\prime\; \MS}_{\gamma^4}(\mu,b^z, b_T , a, \eta, b_T^R , P^z)^2 \,, \\
      \delta_{\text{sys}} \overline{B}^{\MS}_{\gamma^4}(\mu,b^z, b_T, a, P^z)^2 &\equiv \sum_{\eta/a\in\{10,12,14\}} \sum_{(b_T^R/a)=2}^5  w_{\eta,b_T^R} \left[ \overline{B}^{\MS}_{\gamma^4}(\mu,b^z, b_T, a, \eta, b_T^R, P^z) -  B^{\prime\; \MS}_{\gamma^4}(\mu,b^z, b_T , a, \eta, b_T^R , P^z) \right]^2 \,, \\
      \delta \overline{B}^{\MS}_{\gamma^4}(\mu,b^z, b_T, a, P^z)^2 &\equiv \delta_{\text{stat}} \overline{B}^{\MS}_{\gamma^4}(\mu,b^z, b_T, a, P^z)^2 + \delta_{\text{sys}} \overline{B}^{\MS}_{\gamma^4}(\mu,b^z, b_T, a, P^z)^2,
   \end{split}\label{eq:BBar}
\end{equation}
where the weights are
\begin{equation}
   w_{\eta,b_T^R} \equiv \frac{\tilde{w}_{\eta,b_T^R}}{\sum_{\eta/a\in\{10,12,14\}} \sum_{(b_T^R/a)=2}^5 \tilde{w}_{\eta,b_T^R}}, \hspace{5mm}
      \tilde{w}_{\eta,b_T^R} \equiv \frac{1}{\delta B^{\prime\; \MS}_{\gamma^4}(\mu,b^z, b_T , a, \eta, b_T^R , P^z)^2}.
\label{eq:etabTRweights}
\end{equation}
The resulting renormalized quasi beam functions are shown in Fig.~\ref{fig:MSbeam} along with fits to the Hermite and Bernstein functional forms shown in Eqs.~(\ref{eq:herm}-\ref{eq:bern}).
\end{widetext}

Joint fits to Eqs.~(\ref{eq:herm}-\ref{eq:bern}) for the real and imaginary parts of $\overline{B}^{\MS}_{\gamma^4}(b^z, b_T, a, P^z)$ for fixed $b_T$, all $b^z \in [-\eta+a,\eta-a]$, and all $P^z$ corresponding to $n^z\in \{2,3,4\}$, are performed using uncorrelated $\chi^2$-minimization\footnote{ Uncorrelated fits are performed because correlations between quasi beam functions with different staple geometries are not accounted for in the total statistical plus systematic uncertainties of each $\{\Gamma,b^z,\bt,\eta\}$, which are determined using weighted averages over multiple fit range choices as described in Appendix~\ref{app:threetwofits}. Although nonzero correlations exist for different matrix elements computed using the same gauge field configurations, the systematic uncertainties arising from using uncorrelated $\chi^2$-minimization are expected to be small compared to those inherent in modeling the $b^z$ dependence of the quasi beam function. } 
In order to determine the appropriate polynomial to use in Eqs.~(\ref{eq:herm}-\ref{eq:bern}), the AIC is employed.
Fits to Eq.~\eqref{eq:herm} with Hermite polynomials of degree $N=\{0,1,2\}$ are performed, and minimum-AIC fits with $b_T/a \in [1,13]$ are obtained with minimum $\chi^2/N_\text{dof}$ as given in Table~\ref{tab:chi2Herm}.
Fits to Eq.~\eqref{eq:bern} with Bernstein polynomials of degree $N=\{0,1,2,3\}$ are similarly performed, and minimum-AIC fits are obtained with minimum $\chi^2/N_\text{dof}$ as given in Table~\ref{tab:chi2Bern}.
Both fit forms describe the numerical quasi beam function results, although the Hermite form achieves a lower $\chi^2/N_\text{dof}$ in particular for small $b_T$.

\begin{table*}[t]
	\begin{tabular}{|c||c|c|c|c|c|c|c|c|c|c|c|c|c|}\hline
		$b_T$ & 1 & 2 & 3 & 4 & 5 & 6 & 7 & 8 & 9 & 10 & 11 & 12 & 13 \\\hline
		$N$              & 2   & 2   & 0   & 2   & 0   & 0   & 0   & 0   & 0   & 0   & 0   & 0   & 0 \\\hline
		$\chi^2/N_\text{dof}$ & 1.3 & 1.1 & 1.5 & 1.2 & 1.3 & 1.2 & 1.0 & 0.7 & 0.5 & 0.3 & 0.2 & 0.1 & 0.1 \\\hline
      $\gamma$ & -0.27(2) & -0.29(2) & -0.25(3) &  -0.24(7)&  -0.28(3) & -0.33(3) & -0.38(4) & -0.42(3) & -0.45(3) & -0.44(4) & -0.44(4) & -0.46(4) & -0.50(7) \\\hline
	\end{tabular}
	\caption{\label{tab:chi2Herm}Order $N$ and minimum $\chi^2/N_\text{dof}$ obtained in fits of the quasi beam functions to the Hermite polynomial based model form of Eq.~\eqref{eq:herm}, performed as described in the text. \\}
	\begin{tabular}{|c||c|c|c|c|c|c|c|c|c|c|c|c|c|}\hline
		$b_T$ & 1 & 2 & 3 & 4 & 5 & 6 & 7 & 8 & 9 & 10 & 11 & 12 & 13 \\\hline
		$N$              & 1 & 3 & 3 & 2 & 0 & 3 & 2 & 2 & 2 & 1 & 3 & 1 & 2 \\\hline
		$\chi^2/N_\text{dof}$ & 2.0 & 2.0 & 1.3 & 1.1 & 1.0 & 1.1 & 0.9 & 0.8 & 0.7 & 0.8 & 0.6 & 0.8 & 0.6 \\\hline
      $\gamma$ & -0.14(3)\! & -0.19(3)\! & -0.22(2)\! & -0.22(2)\! & -0.27(2)\! & -0.32(3)\! & -0.41(3)\! & -0.43(4)\! & -0.47(5)\! & -0.45(8)\! & -0.58(12)\! & -0.43(11)\! & -0.73(28)\! \\\hline
	\end{tabular}
	\caption{\label{tab:chi2Bern}As in Table~\ref{tab:chi2Herm}, for the Bernstein polynomial based model form of Eq.~\eqref{eq:bern}.}
\end{table*}

Substituting either of the fit forms in Eqs.~(\ref{eq:herm}-\ref{eq:bern}) into Eq.~\eqref{eq:finalCSexpression} and analytically performing the Fourier transforms gives the result $\gamma^{q,\MS}_\zeta = \gamma$ where $\gamma$ is the fit parameter appearing in Eqs.~(\ref{eq:herm}-\ref{eq:bern}) (neglecting one-loop matching effects as discussed in Sec.~\ref{sec:CS}).
Best-fit values for $\gamma$ from these fits therefore provide determinations of $\gamma^{q,\MS}_\zeta$.
The statistical uncertainties of these determinations are calculated by bootstrap resampling renormalized quasi beam function results from a Gaussian distribution with mean $B^{\MS}_{\gamma^4}$ and width $\delta B^{\MS}_{\gamma^4}$, refitting each resampled ensemble, and taking $68\%$ empirical confidence intervals of $\gamma$ in the resulting fits.
This procedure provides the results for $\gamma_\zeta^{q,\MS}$ for the Hermite and Bernstein fit forms shown in Fig.~\ref{fig:CS}.

\section{CS kernel from DFT}
\label{app:DFT}

\begin{figure}[!t]
    \subfigure[~The points show the DFT of the averaged renormalized quasi beam function $\overline{B}^{\MS}_{\gamma^4}(b^z, b_T, a, P^z = n^z2\pi / L)$. The bands show the results of an untruncated Fourier transform applied to the Bernstein polynomials fit to the data (Eq.~\eqref{eq:bern}).]{
        \centering
        \includegraphics[width=0.46\textwidth]{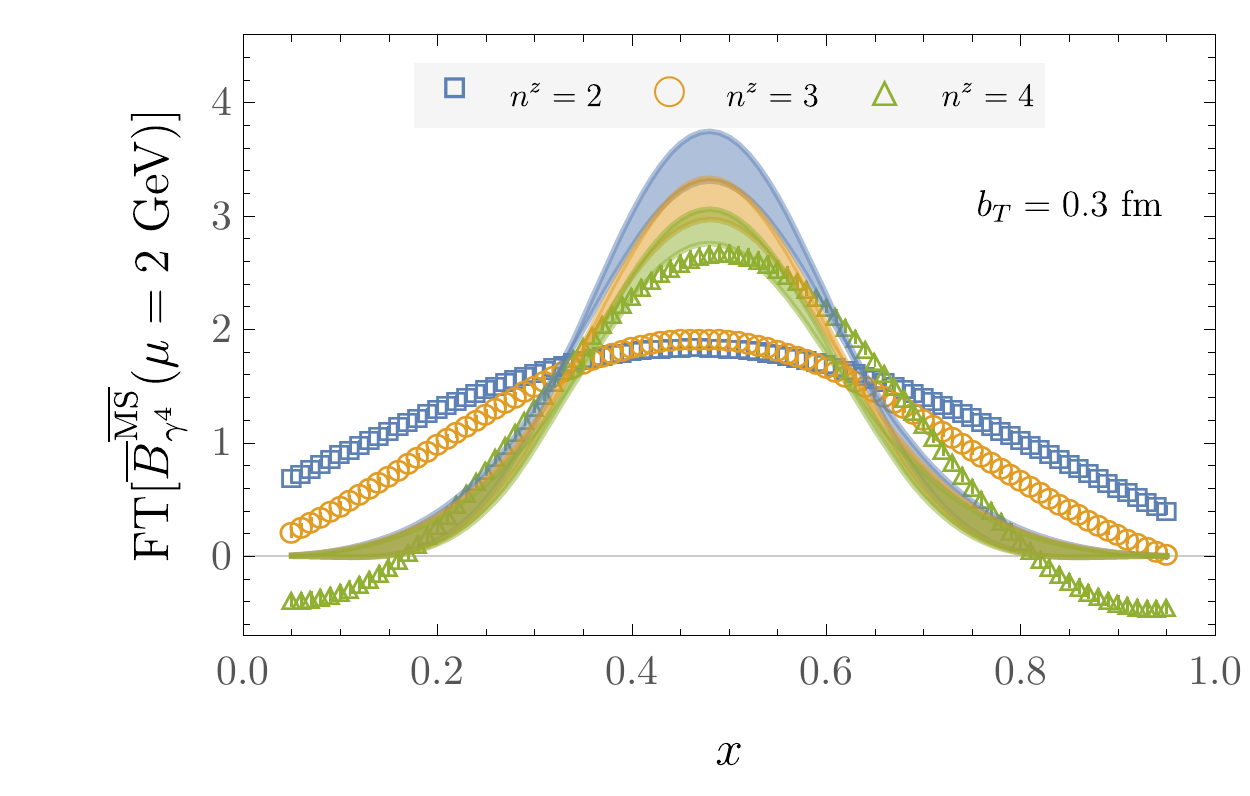}
        \label{DFT_Beam}
        }\quad
    \subfigure[~Collins-Soper kernel extracted based on the DFT of the quasi beam function shown in (a). The solid grey line shows the result obtained using the Bernstein polynomial model fit.]{
        \centering
        \includegraphics[width=0.46\textwidth]{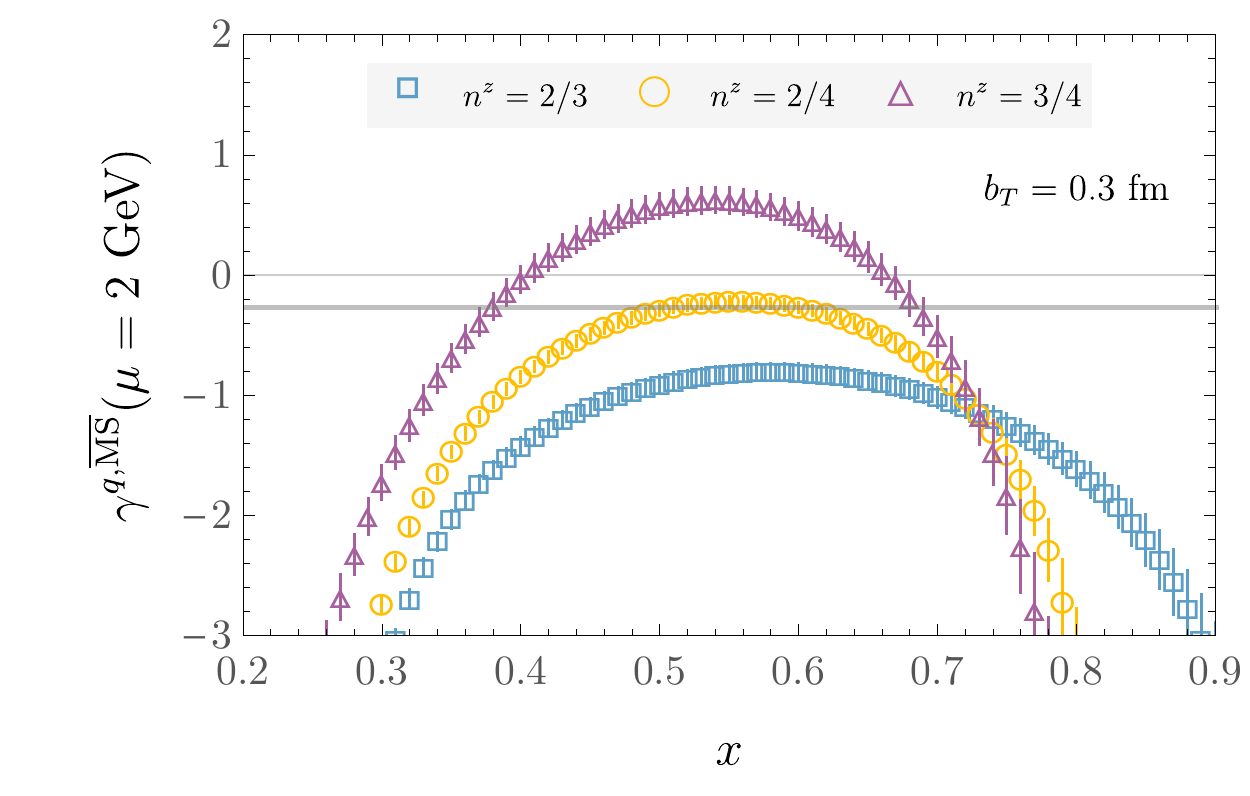}
        \label{DFT_CS}
        }\quad
        \caption{\label{fig:DFT} Fourier transformed quasi beam functions and a DFT calculation of the Collins-Soper kernel. }
\end{figure}

In this appendix, a strategy to extract the Collins-Soper kernel from the quasi beam function via the DFT method, as proposed in Refs.~\cite{Ebert:2018gzl,Ebert:2019tvc,Ebert:2019okf}, is discussed. 
Naively taking a DFT of the quasi beam function obtained in this study at different momenta (Fig.~\ref{fig:DFT}), and extracting the Collins-Soper evolution kernel using~\eq{finalCSexpression}, yields the results shown in Fig.~\ref{DFT_CS}. Clearly, the results from ratios formed using three different momentum pairs have significant $x$-dependence, and are different from each other by over $3\sigma$ at the peak values; convergence is not apparent. For this reason,  untruncated Fourier transforms using models of the large $P^z b^z$ behavior of the quasi beam function are used as described in the main text, rather than the DFT approach. 

\begin{figure}[!t]
	\includegraphics[width=0.9\linewidth]{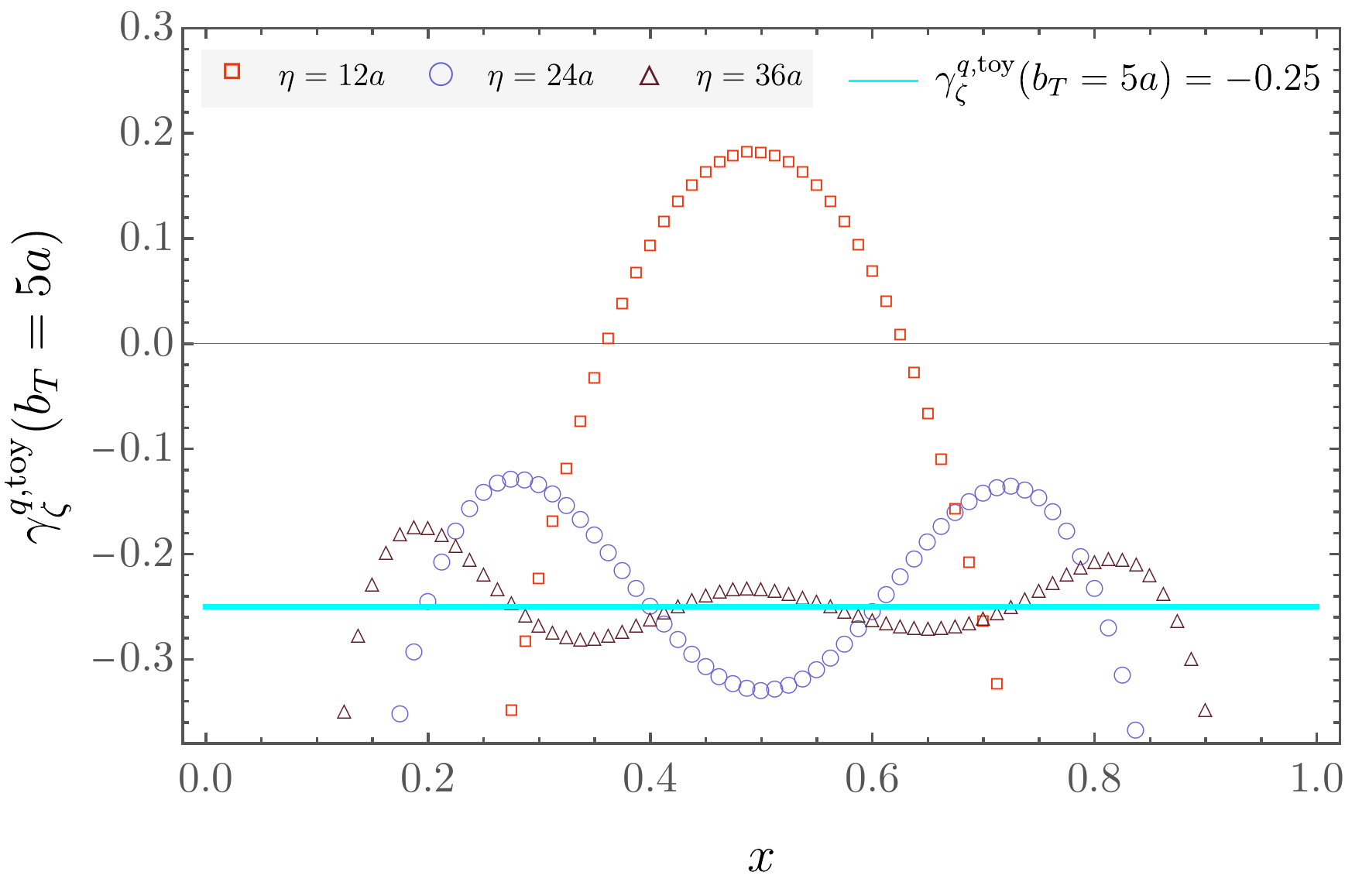}
	\caption{Extraction of the Collins-Soper kernel from the toy model of Eq.~\eqref{eq:toy} with the DFT method at $\{P_1^z,P_2^z\}=\{3,4\}\times(2\pi/32)$. The exact Collins-Soper kernel is $\gamma^{q,\text{toy}}_\zeta(b_T)=-0.01b_T^2/a^2$ with the $\mu$-dependence suppressed.}
	\label{fig:cstoydft}
\end{figure}

\begin{figure}[!t]
	\includegraphics[width=0.9\linewidth]{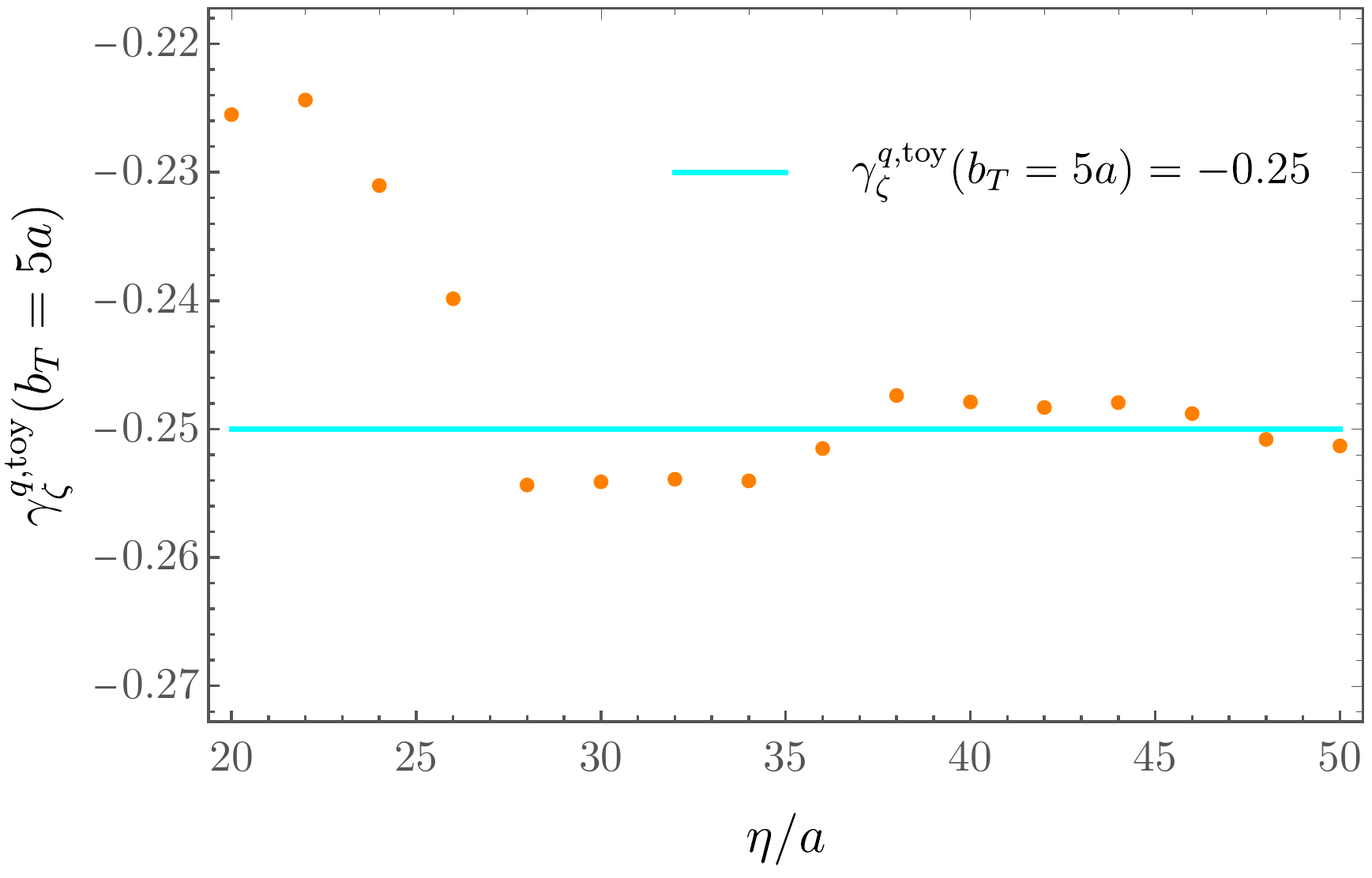}
	\caption{Collins-Soper kernel from the toy model of Eq.~\eqref{eq:toy} with the DFT method at $\{P_1^z,P_2^z\}=\{3,4\}\times(2\pi/32)$ and different $\eta$ values. The Collins-Soper kernel is determined by taking the average of the central peak and trough values near $x=0.5$, and the result converges to the original value with increasing $\eta$.}
	\label{fig:cstoydfteta}
\end{figure}

The instability in Fig.~\ref{DFT_CS} can be understood as a consequence of the limited range of $b^z$ and $\eta$ considered in this study. From \eq{finalCSexpression}, the ratio of quasi beam functions should stabilize in a certain $x$-region for a given momentum pair $\{P_1^z, P_2^z\}$. 
Given a limited range of $b^z$ in the Fourier transform, this stabilization can be expected to be robust for values of $x\sim 0.5$ only. 
One might naively choose the peak at around $x\sim 0.5$ as the central value, and take some variation around the peak to define the systematic uncertainties.
However, since $|b^z|\le(\eta-a)$ in this calculation, the Fourier transform will introduce an oscillatory term to the quasi beam function with frequency $P^z(\eta-a)$. The same issue has been encountered in lattice QCD calculations of collinear PDFs~\cite{Constantinou:2017sej,Chen:2017mzz,Izubuchi:2019lyk,Joo:2019jct}, and certain model assumptions have been suggested to avoid this challenge~\cite{Chen:2017lnm,Lin:2017ani}.
For $P^z=6\pi/L$, $\eta=12a$, as in this calculation, the period of this oscillation is $T = 2\pi/(P^z(\eta-a))\sim 1.0$, and the oscillatory behavior is not apparent within the region $0<x<1$, as shown in Fig.~\ref{DFT_Beam}. The oscillatory behavior will persist in ratios of quasi beam functions at different momenta $P^z_1$, $P^z_2$, as an interference between oscillations with frequencies $P^z_1(\eta-a)$ and $P^z_2(\eta-a)$. As a result, the Collins-Soper kernel extracted from the peak can be shifted significantly, which adds an important systematic error to numerical calculations via this approach.  

In future calculations with increased ranges of $\eta$ or $P^z$, such that there are more rapid oscillations of the DFTs of quasi beam functions within the range $0<x<1$, this approach may nevertheless be used robustly. For example, if the frequency $P^z(\eta-a)$ were doubled, then the Collins-Soper kernel would oscillate around the true value for at least two cycles, which would allow it to be determined by taking an average of the central local maximum and minimum within the oscillating region. To illustrate this point, a toy model for the quasi beam function $\tilde B_\ns$ in $x$-space is constructed:
\begin{align}\label{eq:toy}
&\tilde{B}_{\ns}(x,b_T,\mu,P^z) = 10^4C_{\rm ns}(\mu,xP^z) x^2(1-x)^2\nn\\
&\qquad\times\exp \left[-0.005 (b_T/a)^2 \ln\left(xP^za\right)^2-0.2(b_T/a)^2\right]\,,
\end{align}
where $x\in[0,1]$, and $P^z$ and $b_T$ are in lattice units. The $\overline{\rm MS}$ scale is set to $\mu=2.0$ GeV, and the lattice spacing and size are $a=0.06$ fm and $L=32a$.

To study the oscillatory behavior in this toy model, the inverse FT of the quasi beam function $\tilde{B}_{\ns}(x,b_T,\mu,P^z)$ is taken first. Then, a DFT of the truncated quasi beam functions back to $x$-space is performed for $\eta=\{12a, 24a, 36a\}$, and the Collins-Soper kernel is computed using \eq{finalCSexpression}. The results are shown in \fig{cstoydft}. It is apparent that the kernel suffers from oscillations due to the truncated DFT, and for $\eta=12a$ the shape of the curve is qualitatively similar to those in Fig.~\ref{DFT_CS}. Moreover, the peaks or local maximums around $x=0.5$ are significantly shifted from the true value of the Collins-Soper kernel for all $\eta$ choices. Nevertheless, for $\eta=24a$ and $36a$, taking the average of the central peak and trough values provides a close approximation to the Collins-Soper kernel, as shown in \fig{cstoydfteta}. With more rapid oscillations, this averaging method will lead to more accurate results. Future calculations with larger lattices sizes and higher pion momenta will thus likely enable reliable determination of the Collins-Soper kernel with the DFT method, although the toy model results shown in \fig{cstoydfteta} suggest that very large $\eta$ values may be required to achieve percent-level precision.

\section{Alternate approach in position space}
\label{app:alt}

\begin{figure}[!t]
	\centering
	\includegraphics[width=0.9\linewidth]{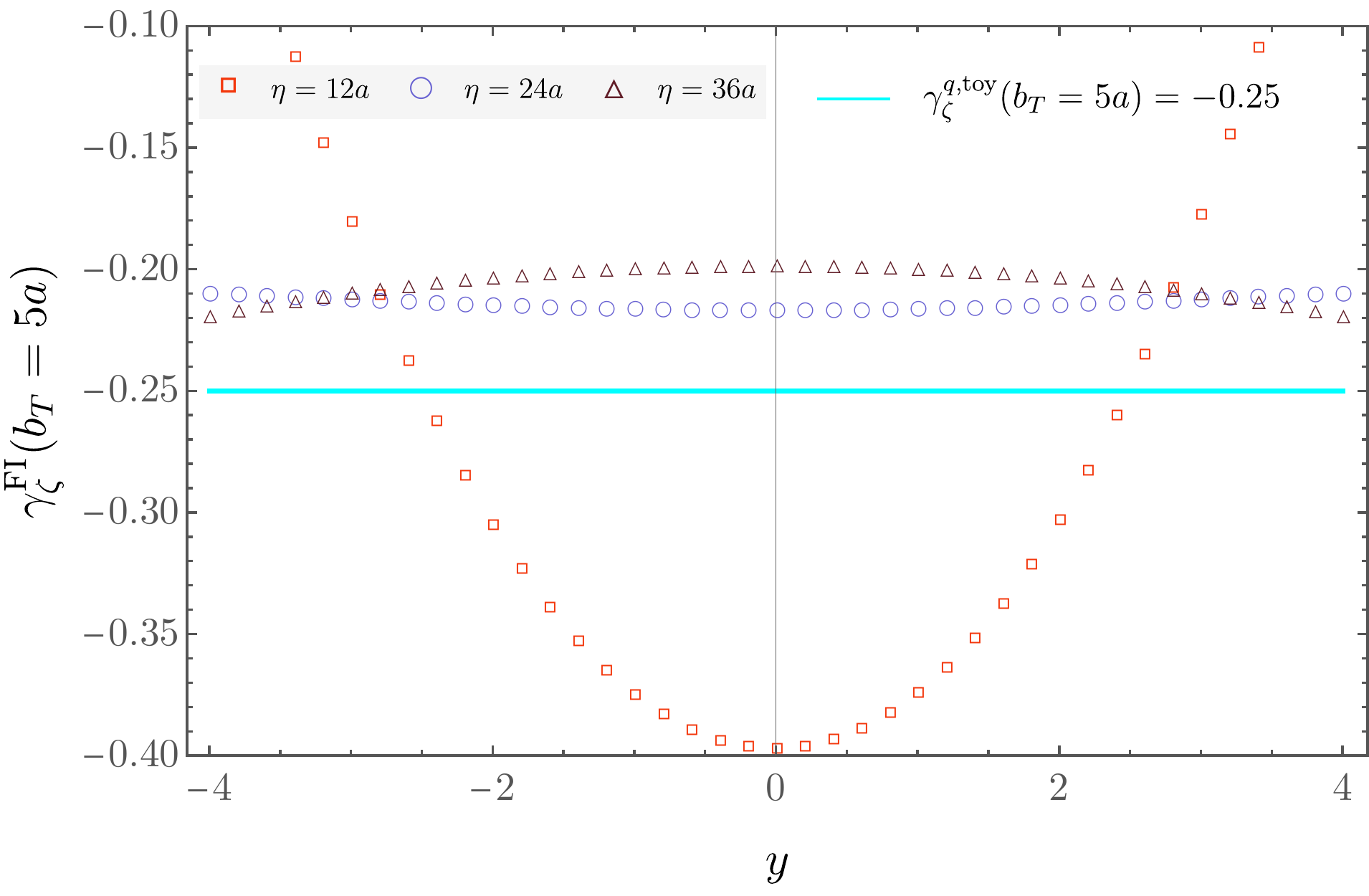}
	\includegraphics[width=0.9\linewidth]{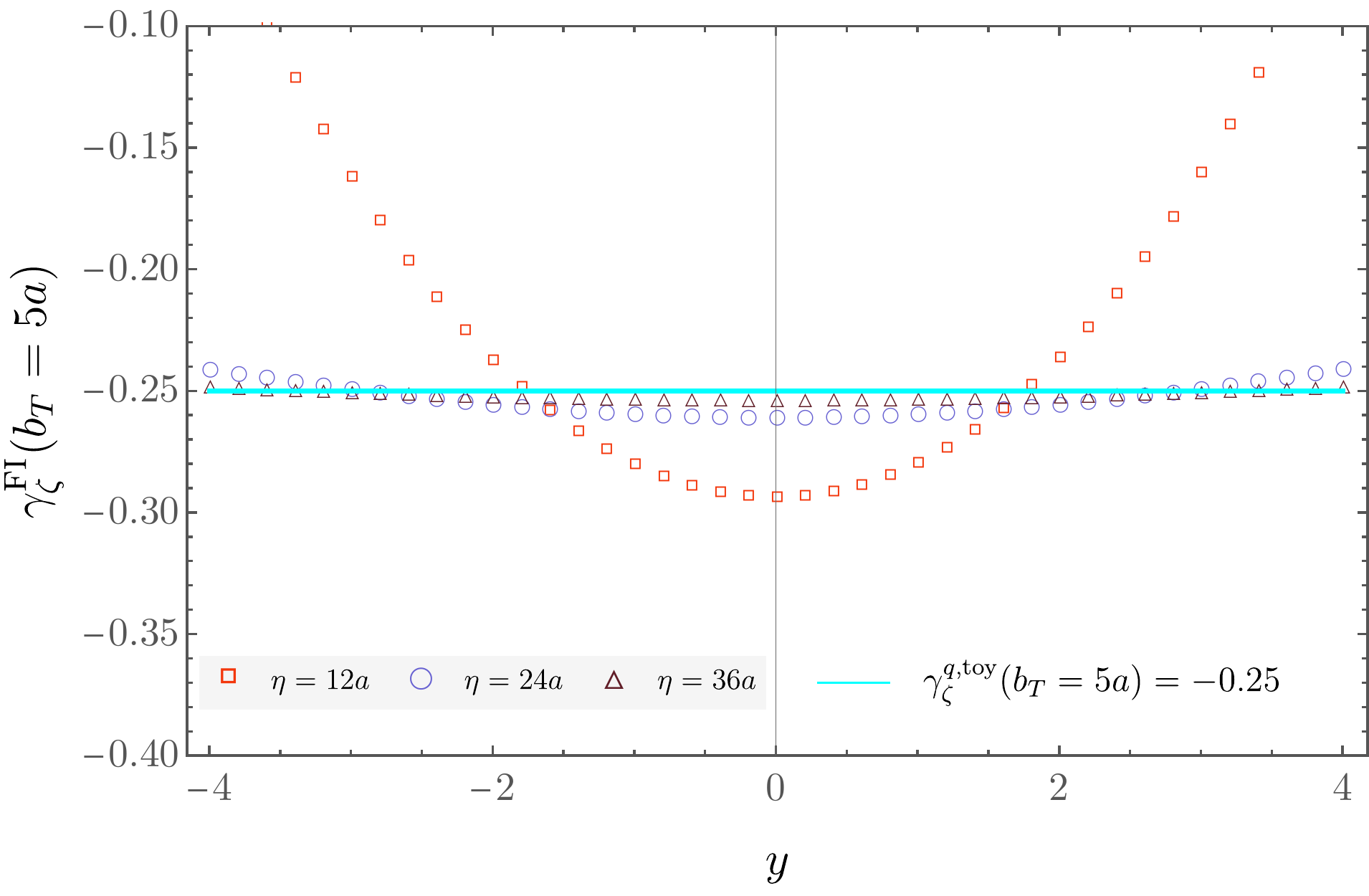}
	\caption{Collins-Soper kernel for the toy model of Eq.~\eqref{eq:toy} determined using the position space approach at $\{P_1^z,P_2^z\}=\{3,4\}\times(2\pi/32)$. Upper panel: extraction with Form I (Eq.~\eqref{eq:gamma_zeta_3}); lower panel: extraction with Form II (Eq.~\eqref{eq:gamma_zeta_4}).
		\label{fig:conv}}
\end{figure}

In this appendix an alternate approach to extract the Collins-Soper kernel in $b^z$-space is investigated, as suggested in Ref.~\cite{Ebert:2019tvc}. By taking the FT of the matching kernel $C_{\rm ns}(\mu, x P^z)$, the Collins-Soper kernel can be expressed as
\begin{align} \label{eq:gamma_zeta_3}
 &\gamma_\zeta^{q,\text{FI}}(\mu, b_T) = \frac{1}{\ln(P^z_1/P^z_2)}\nn\\
 &\times \ln\frac{\int\!d b^z \, \bar C_{\rm ns}(\mu, y - b^z P^z_1, P^z_1) P_1^z B^{\MS}_{\gamma^4}(b^z, \bt, \mu, P^z_1)}
         {\int\!d b^z \, \bar C_{\rm ns}(\mu, y - b^z P^z_2, P^z_2) P_2^z B^{\MS}_{\gamma^4}(b^z, \bt, \mu, P^z_2)}
\,,\end{align}
where
\begin{align} \label{eq:FT_C}
\bar C_{\rm ns}(\mu, b^z P^z, P^z) \equiv \int\!d x \, e^{\img x (b^z P^z)} \, \bigl[ C_{\rm ns}(\mu, x P^z)\bigr]^{-1}
\,,
\end{align}
and the inverse of the matching kernel $C_{\rm ns}(\mu, x P^z)$ is obtained by expanding in $\alpha_s$.
An alternative form is 
\begin{align} \label{eq:gamma_zeta_4}
 &\gamma_\zeta^{q,\text{FII}}(\mu, b_T) = \frac{1}{\ln(P^z_1/P^z_2)}\nn\\
 &\times \ln\frac{\int\!d b^z \, \bar C'_{\rm ns}(\mu, y - b^z P^z_1, P^z_2) P_1^z B^{\MS}_{\gamma^4}(b^z, \bt, \mu, P^z_1)}
         {\int\!d b^z \, \bar C'_{\rm ns}(\mu, y - b^z P^z_2, P^z_1) P_2^z B^{\MS}_{\gamma^4}(b^z, \bt, \mu, P^z_2)}
\,,\end{align}
where 
\begin{align} \label{eq:FT_Cprime}
\bar C'_{\rm ns}(\mu, b^z P^z, P^z) \equiv \int\!d x \, e^{\img x (b^z P^z)} \,  C_{\rm ns}(\mu, x P^z)
\,.
\end{align}
\eqs{gamma_zeta_3}{gamma_zeta_4} are denoted as Form I and Form II, respectively; studying both forms enables a consistency check. Similar to the results obtained via the DFT method outlined in App.~\ref{app:DFT}, the Collins-Soper kernel obtained by either Form I or Form II should not depend on the value of $y$ or on the momentum pair $\{P_1^z, P_2^z\}$, which provides another handle on the relevant systematic uncertainties.

\begin{figure}[tb]
	\centering
	\includegraphics[width=0.9\linewidth]{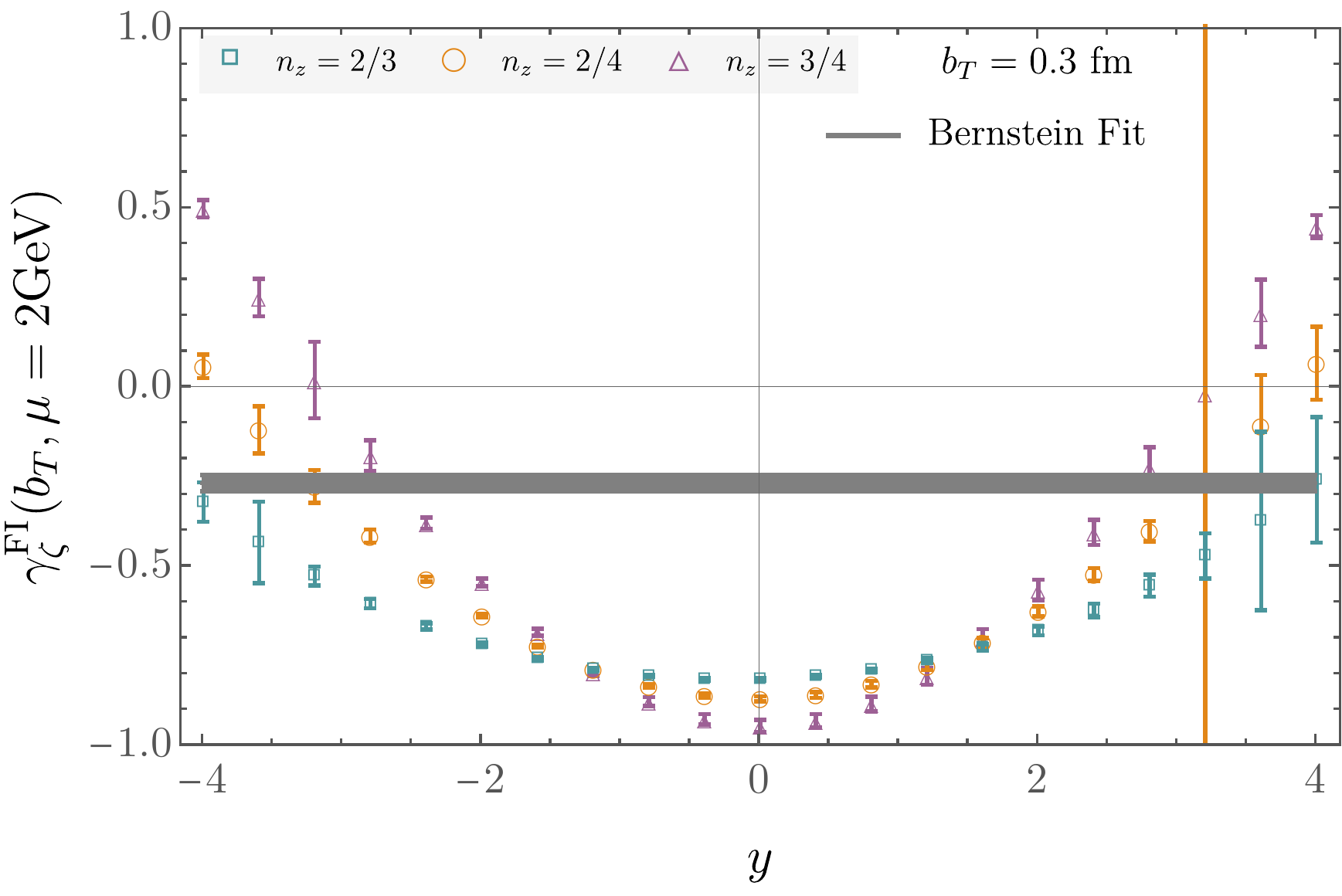}
	\includegraphics[width=0.9\linewidth]{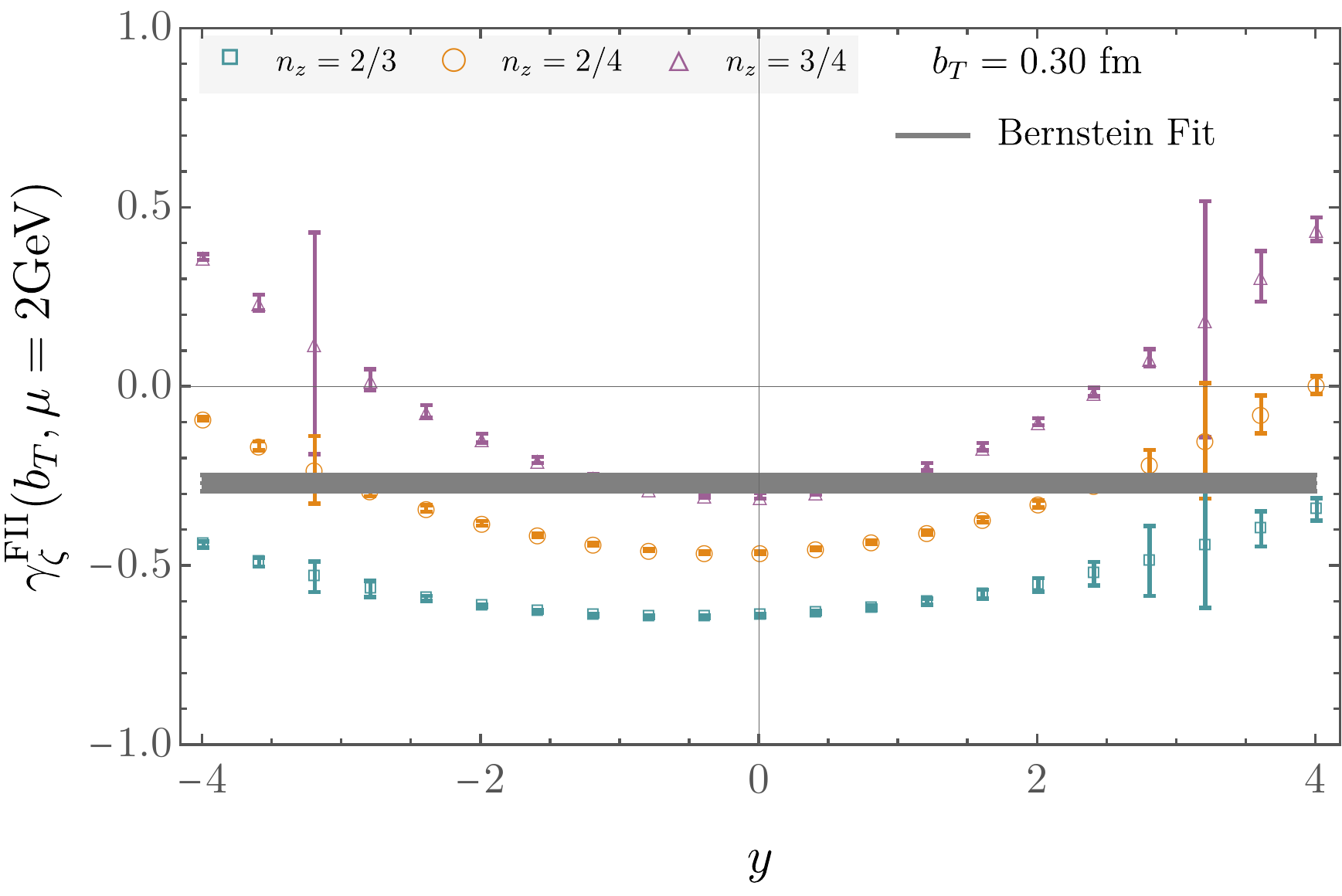}
	\caption{Collins-Soper kernel computed in the position space approach from the lattice data. Upper panel: extraction with Form I (Eq.~\eqref{eq:FT_C}); lower panel: extraction with Form II (Eq.~\eqref{eq:FT_Cprime}).
		\label{fig:csconv}}
\end{figure}

Collins-Soper kernels extracted with the position-space approach for the toy model of Eq.~\eqref{eq:toy} are shown in \fig{conv}. The two forms do not yield consistent answers, which indicates that they are not numerically equivalent and that the perturbative convergence is lost in either or both of the convolution integrals. Nevertheless, it is clear that with Form I the extracted Collins-Soper kernel does not stabilize to the correct result with increasing $\eta$, while with Form II the ratio stabilizes around the true value for sufficiently large $\eta$. With sufficiently large $\eta$ it is possible that this approach may provide a reliable determination of the Collins-Soper kernel, although this will need to be investigated carefully in future work.  

The results of applying the position space approach to the lattice QCD results in this study are shown in \fig{csconv},
 which is compared to the Collins-Soper kernel extracted from the fits with the Bernstein polynomial model, discussed in the main text. With the range of $\eta$ values computed in the numerical study there is no plateau in the $y$-space Collins-Soper kernels, and the different choices of momentum pairs do not appear to converge. Although the result extracted at $\{P_1^z,P_2^z\}=\{3,4\}\times2\pi/L$ is consistent with the results extracted using the model fits at the minimum value, this consistency is not found at different $b_T$ values. Further investigation is needed to confirm whether the position-space approach via Form I or II can provide a valuable consistency check against the DFT approach with larger physical lattice volumes used in calculations.

\bibliography{csrefs}

\end{document}